# EmoStim: A Database of Emotional Film Clips with Discrete and Componential Assessment

Rukshani Somarathna [a], Patrik Vuilleumier [b, c, d], Gelareh Mohammadi [a]

[a] School of Computer Science and Engineering, University of New South Wales, Australia

[b] Laboratory for Behavioral Neurology and Imaging of Cognition, Department of Neuroscience, University of Geneva, Geneva, Switzerland

[c] Neurology Department, University Hospital of Geneva, Geneva, Switzerland

[d] Swiss Center for Affective Sciences, Campus Biotech, University of Geneva

**Abstract**
Emotion elicitation using emotional film clips is one of the most common and ecologically valid methods in Affective Computing. However, selecting and validating appropriate materials that evoke a range of emotions is challenging. Here we present EmoStim: A Database of Emotional Film Clips as a film library with a rich and varied content. EmoStim is designed for researchers interested in studying emotions in relation to either discrete or componential models of emotion. To create the database, 139 film clips were selected from literature and then annotated by 638 participants through the CrowdFlower platform. We selected 99 film clips based on the distribution of subjective ratings that effectively distinguished between emotions defined by the discrete model. We show that the selected film clips reliably induce a range of specific emotions according to the discrete model. Further, we describe relationships between emotions, emotion organization in the componential space, and underlying dimensions representing emotional experience. The EmoStim database and participant annotations are freely available for research purposes. The database can be used to enrich our understanding of emotions further and serve as a guide to select or create additional materials.

## 1   Introduction

Affective Computing (AC) is an interdisciplinary domain that focuses on understanding emotions and enhancing machines with emotional intelligence [1]. Therefore, there has been a growing interest in developing various experimental settings to better probe and characterize emotions. Those experimental settings vary depending on the theoretical model used to define emotions, the material used, and the measures collected. An abundant literature has mainly assumed either discrete or dimensional models of emotion, and more rarely considered appraisal theories to represent emotions. Although they are not mutually exclusive, their emphasis and scope differ. For example, the discrete model explains emotions as distinct entities, such as happiness and sadness that underlie more complex emotions [2]. The dimensional theory describes emotions according to different levels of arousal and valence elements [3], and the appraisal model invokes a process-based generation [4]. The latter hypothesizes that different components are jointly engaged and interact to trigger specific emotions. The appraisal model can help explain how emotions are triggered by specific circumstances and how they influence behaviour and decision-making [4]. The appraisal model can also help explain why different people may have different emotional reactions to the same situation, as they may have different appraisals of the situation based on their personal experiences, beliefs, and values [5, 6]. Therefore, this model assumes that emotions are highly subjective. Further, some research considered appraisal model to explain the neural activation of the brain [7]. Moreover, the constructivist theory of emotions proposes that emotions arise from conceptual categorization processes that rely on core affective dimensions, such as valence and arousal, which are further integrated with other sources of knowledge determined by perception, attention, and past experiences [8, 9]. Furthermore, existing studies [10, 11], suggest that emotion categories are not confined to a single region or system within the brain. Instead, they are represented as configurations spanning across multiple interconnected brain networks. These findings align with the theories of appraisal and constructivism, which proposes that emotions are distinguished by a combination of perceptual, mnemonic, prospective, and motivational elements [10].

Different stimuli, such as films [12, 13], images [14, 15], and sounds [16, 17], have been used to induce emotions and are generally pre-assigned to different categories according to one given model. Among these stimuli, films have gained wider attention and a reputation in AC as a useful medium to effectively induce diverse emotions [12, 18, 19]. Features of films, such as cost-effectiveness [20], the possibility to elicit intense emotions [21], easy laboratory setup [19], and triggering of subjective and objective changes [12, 13, 22, 23], can be listed as some advantages.

Here we present a database of emotional clips that have also been analyzed to understand emotion formation assuming the Component Process Model (CPM) [6], a variant of the appraisal model where emotions are proposed to result from interactive effects of different sub-processes. Full CPM focuses on five main components: Appraisal, Motivation, Physiology, Expression, and Feeling, which have rarely been studied using data-driven approaches. We further explore the relationships between the CPM and the discrete model of emotion using this database.

In this paper, we first present a film library named EmoStim, which was assessed in terms of both discrete and componential models of emotion. We then explain our main analysis of the dataset features. The EmoStim dataset and the presented analysis can be used in future research as a material or guide to study emotions using different approaches. Our annotated dataset can also be further examined to study emotions using state-of-the-art techniques and algorithms in order to extend the current result. The EmoStim film database and participant annotations are freely available for research purposes upon request.

The rest of the paper is structured as follows. The following section introduces the Component Process Model and related works. In section 3, we discuss our approach. Then we present our analysis and results in section 4. Section 5 includes our discussion and conclusion.

## 2    Background

Discrete theories of emotions postulate that a limited number of distinct entities represent main or "basic" emotions. However, many issues remain disputed; hence, there is no consensus on the exact number of discrete emotions [20, 24]. Nevertheless, the six-emotion model proposed by Ekman, et al. [25] (anger, happiness, fear, surprise, disgust, sadness), as well as the six-emotion model by Frijda, et al. [26] (desire, happiness, interest, surprise, wonder, sorrow), or the 11 emotion model by Izard [27] (joy, interest, fear, sadness, disgust, anger, guilt, shame, contempt, love, attachment) are commonly used in research. Here we consider a wider range of 14 categories (fear, anxiety, anger, shame, warm-hearted, joy, sadness, satisfaction, surprise, love, guilt, disgust, contempt, and calm), using a subset of emotions adapted from Differential Emotion Scale (DES) [28, 29].

### 2.1    Component Process Model (CPM)

The Component Process Model (CPM) is a variant of appraisal theories seeking to explain the unfolding of emotions from event perception and interpretation to subjective feeling. CPM hypothesizes emotions as a process of synchronized components including physiological changes, behaviour, and cognition over a limited time [30]. As shown in Figure 1, the five main components of the CPM can be defined as Appraisal, Motivation, Physiology, Expression, and Feeling, which are interconnected [6].

According to a standard CPM framework, an event is first assessed by the appraisal component at four levels: 1) Relevance: "Is the incident relevant for me?", 2) Implications: "What are the consequences?", 3) Coping potential: "How well can I overcome the consequence?" and 4) Normative significance: "Is this important for my self-concept and norms?".

Following such appraisals, the response will affect the other four components. This can then trigger the motivation component activating action tendencies and formulating appropriate responses modifying the previous motivational state. The physiological component incorporates bodily changes based on appraisal and motivational changes (e.g., faster heartbeat), and the expression component activates expressive motor behaviour (e.g., smiling). These components are integrated and constantly fused, merging into a coherent conscious experience defined as the feeling component. Reflection on the resulting experience can be described by categorical and verbal labels (sad, happy, and fear). The layout of these main five components sets up a recursive and parallel functioning of the CPM.

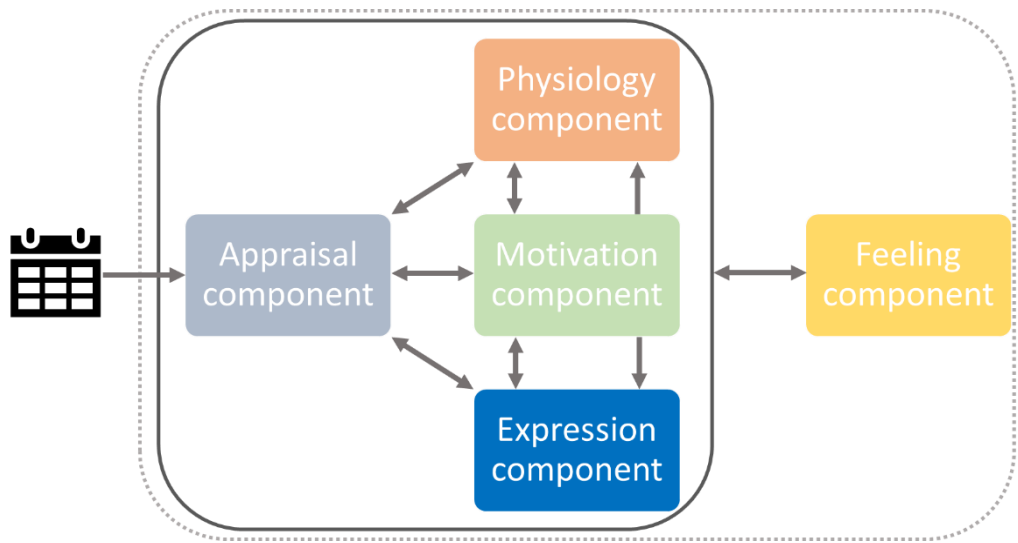

Figure 1 Component Model of emotion with five components.

Neuroscience studies show that emotions have distributed representations in multiple brain regions [10, 31, 32]. Moreover, emotion patterning relies on differential neocortical involvements together with various patterns of cortical-subcortical interactions [10]. Furthermore, there is no one-to-one correspondence of emotions with specific brain regions [11, 33]. Hence, neuroscience findings are generally consistent with the brain's representation of emotion according to the componential model associated with appraisal, pleasantness, novelty, goal-relevance, action tendencies, and social norms rather than discrete or dimensional views [10, 32, 34].

## 2.2 Related Works

According to surveys [20, 35], film clips are the most prominent paradigm for emotion elicitation. In line with this, various studies have assessed the efficacy of film clips in inducing emotions represented by dimensional and discrete theories. Moreover, annotated emotional film clips based on the dimensional model [12, 22, 36] and the discrete model [19, 37-40] are publicly available. However, these approaches focus on the feeling component and ignore the more complex process by which emotions are triggered. Furthermore, they do not allow studying the full emotional model with the five components described by the CPM.

The effectiveness of films has been verified by independent researchers [19, 41] and previous works [42, 43] to provide accounts based on appraisal theory. But those datasets are not publicly available. Therefore, here we present EmoStim as an open film library to study both componential and discrete models of emotion together. Our previous experimental work has already shown the effectiveness of the EmoStim library in understanding emotional features. For example, Mohammadi, et al. [42] uncovered six dimensions defined by the CPM principles to interpret emotional experience, beyond the traditional aspects of arousal and valence. This high-dimensional outcome shows the limitations of previous works ignoring full CPM. In another case, but using a subset of the EmoStim library, Menétrey [43] collected both subjective and physiological measures from twenty participants. The results demonstrated the potential of EmoStim film clips in inducing a range of emotions as defined by discrete theory and triggering distinctive physiological variations. Similarly, EmoStim has been used to investigate emotion organization within brain networks and their link with different components [32].

Overall, these observations emphasize the importance of the CPM perspective for better understanding emotion and the potential of EmoStim for studying the richness of human affective experiences. They also highlight the usefulness of developing annotated multimedia and dataset for future research. Therefore, we present EmoStim as a dataset of emotional film clips and subjective measures that can be shared across researchers and fields.

## 3   Approach

We first present the process of selecting emotional film clips, assessments, and data collection setup.

### 3.1   Material selection and assessment

Film clips provide naturalistic events due to their dynamic nature while being easily implemented in the laboratory [19]. They can induce powerful emotions theorized by both discrete [18, 19] and dimensional [12, 22] models, including complex emotions such as tenderness or compassion [19] beyond basic categories such as fear or disgust. Their effectiveness for studying componential appraisal models [18, 43] is also supported by their impact on physiological bodily states such as heart function [13, 43], skin conductance [13, 18], and brain activity [12, 13, 22, 23]. Our study embraced a wide range of discrete theory-based emotions that were induced using film clips to assess the corresponding CPM descriptors.

As the initial step, we selected 139 emotional film clips from affective literature [19, 37, 39, 44] based on their availability. Film clips were presented in a randomized order and rated on discrete emotional categories and CPM descriptors. The discrete scale was extracted from the Differential Emotion Scale (DES) to evaluate 14 affective states: fear, anxiety, anger, shame, warm-hearted, joy, sadness, satisfaction, surprise, love, guilt, disgust, contempt, and calm [29, 45]. To assess CPM descriptors, we used a modified version of the GRID instrument, CoreGRID [2], and selected a subset of 39 features depicting the five main components of CPM (appraisal, motivation, expression, physiology and feeling). CoreGRID descriptors were chosen based on applicability for a passive emotion elicitation during movie watching, rather than active first-person engagement as in real events. This CoreGRID selection strategy appropriately reduced the noninformative data from the absence of variance. For example, descriptors such as "it was caused by my own behavior" or "it was important and relevant for my goals" were removed because they would consistently rate as "No". Supplementary material Table S1 provides the full list of CoreGRID items, and supplementary material Table S2 summarizes the 14 discrete theory-based emotions questions.

### 3.2   Experimental setup

This work was approved by the Geneva Cantonal Research Committee and followed their guidelines in accordance with the Helsinki declaration. We used the crowdsourcing web platform: CrowdFlower to recruit participants for the study. The selected participants were limited to native English speakers from USA or UK. The estimated duration for each task was set at 10 minutes with a 110¢(USD) reward per task (an effective hourly wage of $6.6 (USD)). However, the actual average time for completing the task was around 6 minutes. We ensured that participants had watched the film clip using six quality control questions (e.g., *did you see any animal, see anyone crying, hear any gunshot, see sports activity, see corpse*). We discarded the data if a rater made two wrong answers or more out of six questions. The distribution of the correct quality control answers is given in Supplementary Figure S1.

We advised participants to experience emotions freely, recall feelings, and evaluate their experiences on their own real feelings, not what they believed people should feel while watching the clip. Participants started by self-reporting personality traits using the Big Five Inventory 10 (BFI-10) questionnaire [46] (Supplementary material Table S3) and then watched the film clips in randomized order. They were instructed to mark their emotional experience on the CoreGRID and discrete emotion questionnaires along a 5-point Likert scale, from "not at all" to "strongly". We randomized the order of each questionnaire's items.

First, in a pilot experiment, we collected five assessments per film clip (n=139). We then chose a subset of 99 clips from these based on the emotional intensity, and the emotion discreetness observed among pilot raters. Next, to achieve more statistical power for reliable inferences on emotionality, we collected ten more assessments per selected clip. The final film dataset duration is about 3.7 hours, with an average of 133.8 seconds per clip. It comprises 1792 validated assessments from 638 workers (358 males, mean age = 34, SD = 11). However, in the current paper, we analyze 1567 assessments belonging to the 99 selected clips with at least 15 assessments per video clip, gathered from 617 participants (workers with a unique ID), which ensured good statistical power for inferences. The participants' average age is 33.78 years (standard deviation = 11.76) for the used dataset with 1567 assessments. The gender distribution consists of 940 males and 627 females, while the geographic distribution comprises 1272 individuals from the USA and 295 from the UK. Supplementary material "Dataset_FilmClipsDetails" (https://tinyurl.com/FilmClipsDetails) lists the 139 film clips used in the full experiment, with details of the 99 clips selected for this paper, and supplementary material section 1 illustrates the average intensity of each emotion.

## 4 Analysis and Results

### 4.1 Discrete emotion categories

We first analyzed the score distribution computed from the 1567 ratings across 99 clips to explore the effectiveness of selected stimuli in inducing a wide range of emotions. Figure 2 represents results for 14 discrete theory-based emotions along the 1-5 points Likert scale. Although all emotions were experienced to some level, the distribution of ashamed, warm-hearted, joy, love, guilt and contemptuous was reported less frequently.

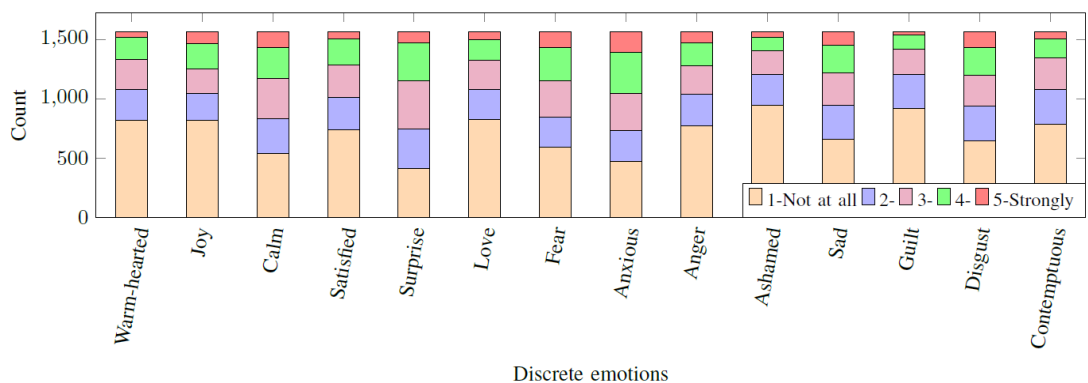

Figure 2 Distribution of scorings in 14 emotions. Colour scales correspond to the proportion of samples.

To test for any similarity or confusion across different emotions, we calculated Spearman correlation matrices per movie and then computed the average to obtain the concatenated matrix. The results are given in Figure 3 and show clear high-level differences between positive and negative emotions, with more varying correlations within these high-level families. Accordingly, warm-hearted, joy, love, and satisfaction are related with each other to different degrees. Calm is also moderately correlated with joy, satisfaction, and love, whereas surprise does not show a high correlation with any other emotion. Among negative categories, fear is correlated with anxious, and anger correlates mostly with ashamed, contempt, and disgust, followed by sad and guilt. Guilt and ashamed are better correlated and show a similar association with anger and contemptuous. Sad is correlated with anger, guilt, contemptuous, and disgust. Contemptuous is correlated with disgust, anger, sad, ashamed, and guilt. Comparing the average correlation among all positive (warm-heartedness, joy, calm, satisfaction, surprise, love) with that among all negative emotions (fear, anxious, anger, ashamed, sad, guilt, disgust, contemptuous), without the diagonal identity values, showed similar values of 0.43 and 0.48 for positive and negative categories, respectively. Hence, valence did not affect the relative similarity/dissimilarity between discrete emotions in our dataset.

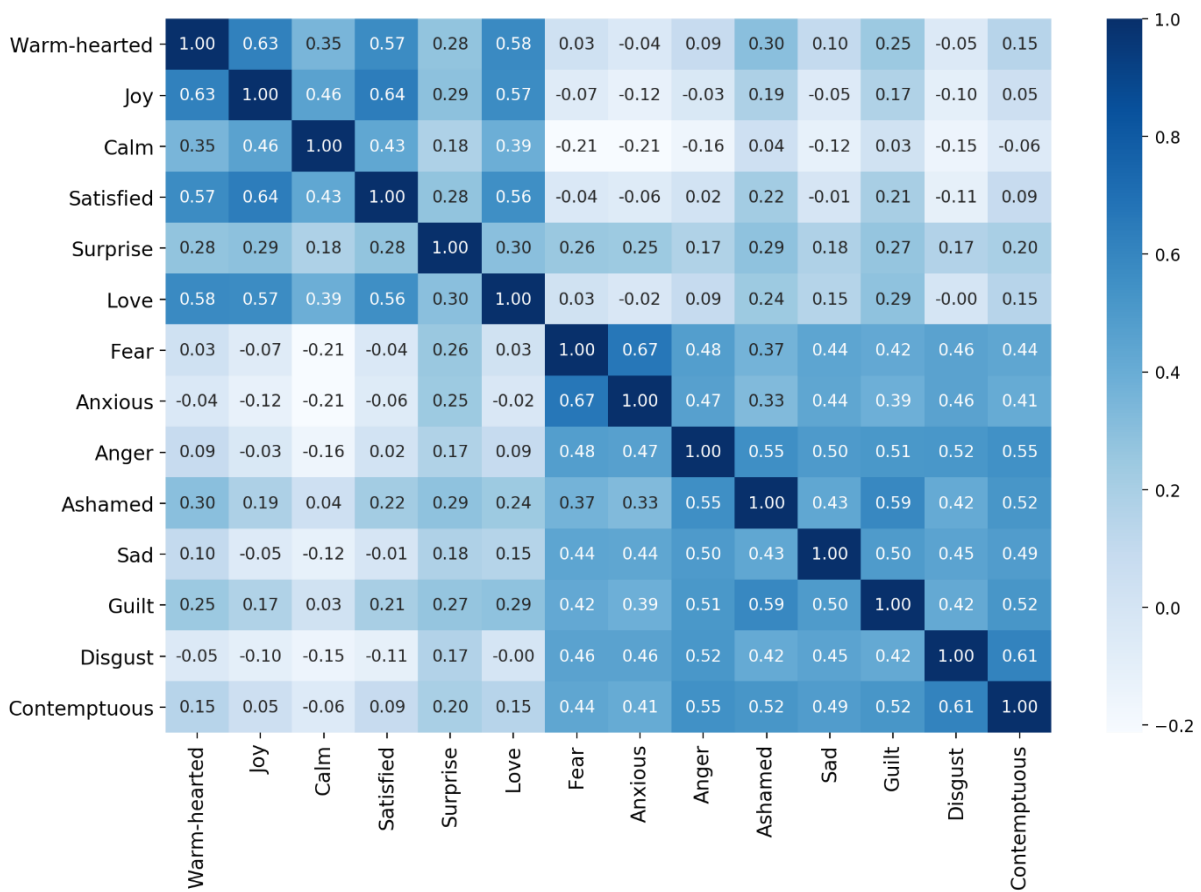

Figure 3: Average emotion correlations at movie level. Scores represent the Spearman correlation values.

To further explore similarities and differences between emotions, we applied hierarchical clustering to the correlation matrix obtained above. As shown in Figure 4, this analysis confirms a high-level differentiation between positive and negative emotions, as classically observed in the literature. At the lower level, six clusters are observed that can be interpreted as embarrassment (guilt, ashamed), threat (anxious, fear), irritation (sad, disgust, anger, contemptuous), surprise, serenity (calm), and happiness (joy, satisfaction, warm-hearted, love). Moreover, looking at the overall functional and situational relationships of these clusters suggests a distinction possibly opposing self-directed negative (threat and embarrassment) or positive (joy/satisfaction) affective responses, against other-directed negative (anger/sad/contempt) or positive (love/warm-hearted) responses. These results accord with a hierarchical organization of emotion terms and their similarity, putatively reflecting differences in underlying components.

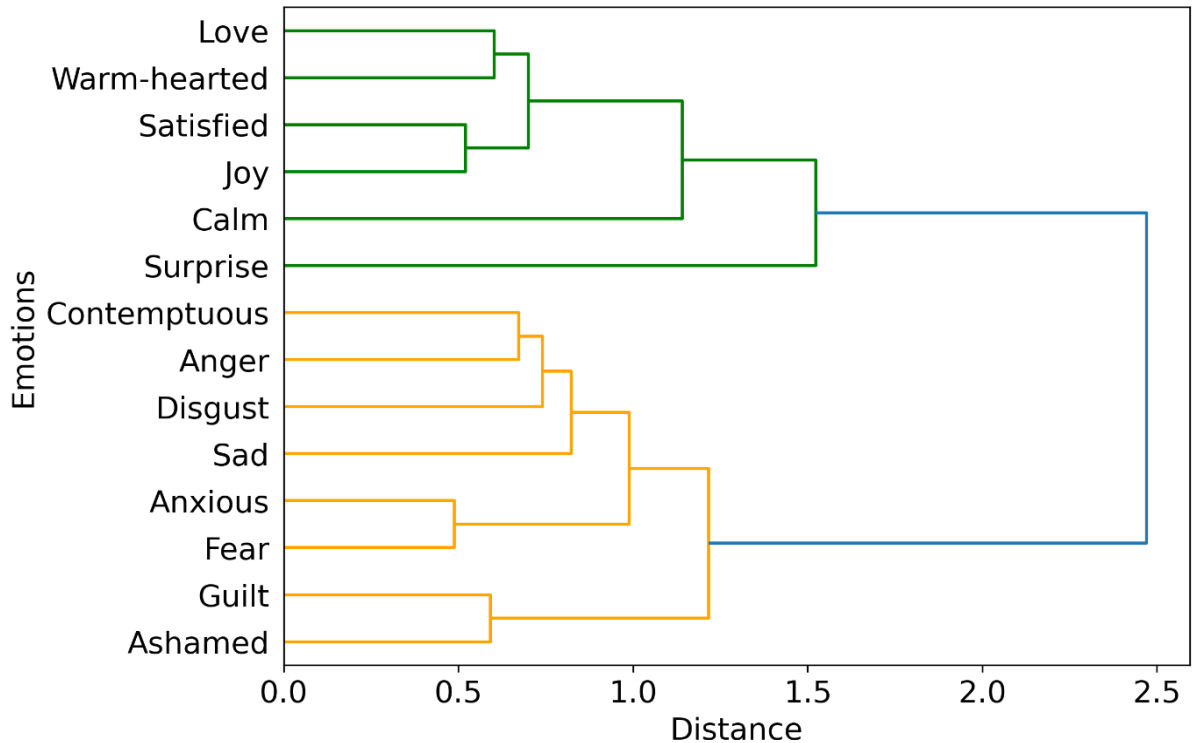

Figure 4: Results of the correlation-based hierarchical clustering of discrete emotion ratings across movies.

We also investigated gender effects on emotion rating averages over all film clips using a Wilcoxon Rank Sum with Holm's correction for family wise error. As shown in Table 1, men rated film clips significantly higher on all emotions except for anxious and disgust, where there is a significant difference. Interestingly, the largest difference concerned a lesser degree self-related positive emotion (satisfied) to a self-related negative emotion with social significance (guilt/shame), higher in males than females, whereas the least whereas the least difference was observed for more basic, sensory-driven, and non-social emotions (disgust, fear, anxious).

Table 1 Scores of the Wilcoxon Rank Sum with Holm's correction for family wise error for gender effects for emotion rating averages. (SD: Standard deviation). For effect size calculation male group was considered as the first group and female group was considered as the second group.

|  | Male |  | Female |  | Test statistic | p-value for gender analysis | Effect size (First group is Male) |
| --- | --- | --- | --- | --- | --- | --- | --- |
|  | Mean | SD | Mean | SD |  |  |  |
| **Warm-hearted** | 2.10 | 1.22 | 1.79 | 1.16 | 5.228 | p<0.001 | 0.257 |
| **Joy** | 2.21 | 1.36 | 1.88 | 1.27 | 4.803 | p<0.001 | 0.247 |
| **Calm** | 2.59 | 1.36 | 2.28 | 1.29 | 4.363 | p<0.001 | 0.237 |
| **Satisfied** | 2.27 | 1.29 | 1.87 | 1.16 | 6.094 | p<0.001 | 0.322 |
| **Surprise** | 2.71 | 1.20 | 2.40 | 1.28 | 5.011 | p<0.001 | 0.257 |
| **Love** | 2.12 | 1.28 | 1.78 | 1.13 | 5.028 | p<0.001 | 0.276 |
| **Fear** | 2.45 | 1.36 | 2.41 | 1.40 | 0.843 | p<0.001 | 0.034 |
| **Anxious** | 2.67 | 1.35 | 2.68 | 1.45 | 0.070 | p<0.001 | -0.007 |
| **Anger** | 2.17 | 1.33 | 1.97 | 1.23 | 2.666 | p<0.100 | 0.155 |
| **Ashamed** | 1.90 | 1.15 | 1.56 | 1.03 | 6.204 | p<0.100 | 0.312 |
| **Sad** | 2.34 | 1.33 | 2.17 | 1.35 | 2.780 | p<0.100 | 0.128 |
| **Guilt** | 1.90 | 1.10 | 1.54 | 0.97 | 6.322 | p>0.500 | 0.336 |
| **Disgust** | 2.28 | 1.30 | 2.36 | 1.46 | -0.340 | p>0.500 | -0.061 |
| **Contemptuous** | 2.06 | 1.20 | 1.90 | 1.21 | 2.926 | p>0.500 | 0.135 |

## 4.2 Componential emotion features

Next, we examined the distribution of CoreGRID ratings across emotional clips. As shown in Figure 5, all CPM descriptors were experienced to some level. However, a few items were reported less frequently, such as *"want to destroy something", "want to damage something", "close your eyes", "show tears", "have stomach troubles", "sweat", "produce abrupt body movements", "produce speech disturbances", "breathing is slowing down"*, and *"have the jaw drop"*.

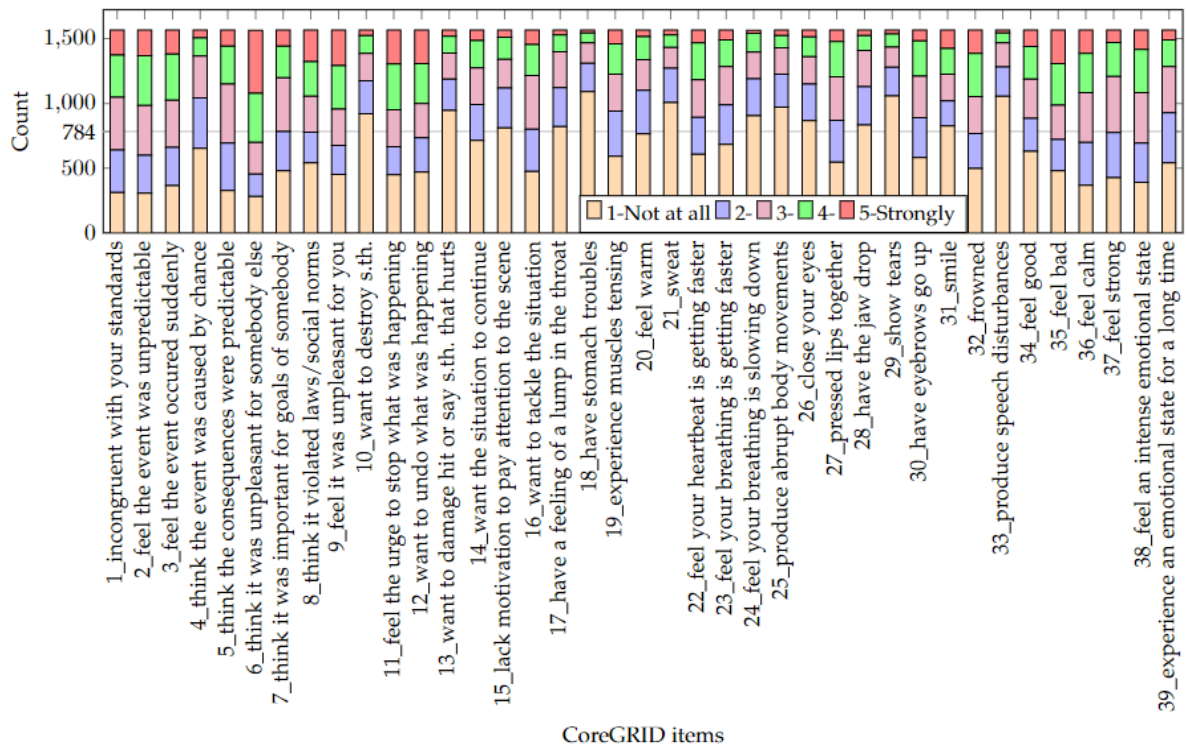

Figure 5 Distribution of scorings for 39 CoreGRID descriptors. Colour scales correspond to the proportion of samples.

We then performed a cluster analysis on the CPM profile of different emotion categories as defined by the discrete model to identify similarities or differences among them in the componential space. As shown in the following formula, the emotion profile was calculated using the weighted average of the normalized CoreGRID item scores, where the weight of each sample was proportional to the scores of each emotion term [42, 47].

$$CP_j = \frac{\sum_{i=1}^{n} w_{ij} x_i}{\sum_{i=1}^{n} w_{ij}}$$

In the above formula, $w_{ij}$ is the score of emotion term j in sample i, $x_i$ is the CoreGRID score vector for sample i, $CP_j$ is the weighted average CPM profile for emotion term j, and n is the total number of samples. We conducted a Ward hierarchical clustering using the Euclidean distance metric [42, 47]. We opted to utilize the Ward algorithm for clustering analysis as we aim to find clusters based on variance with a relatively large number of real-valued variables with similar scales [48]. In addition, we did not want to fix the number of clusters and also some studies have used Ward algorithm to identify clusters, and this allows for a better comparison of cluster structures [18, 49]. However, it is important to note that Ward's algorithm may be sensitive to outliers and noise, potentially leading to suboptimal results. Additionally, this algorithm can be computationally expensive and requires visual inspection of dendrograms to determine the optimal number of clusters. Despite these limitations, the use of Ward's hierarchical clustering method can be justified in this context, given the need to identify groups of similar emotions based on multiple variables and uncover potential subgroups of interest through dendrogram analysis.

Results are depicted in Figure 6 and again show a high-level difference between positive (green cluster), and negative (orange cluster) affect. At the lower level, we observed seven clusters that resembled clustering based on discrete terms (Figure 4), but with some notable differences. These seven clusters can be identified as embarrassment (guilt, ashamed), surprise, irritation (contemptuous, anger, sad), distress or aversion (anxious, fear, disgust), happiness (satisfaction, joy), affection (love, warm-hearted), and serenity (calm). This similarity to the hierarchical clustering obtained from the discrete emotion ratings supports the validity of CPM in capturing the functional organization of emotional experience across a range of categories. Among emotions whose relationships changed between the two analyses, surprise shifted to the negative valence side, while calm remained on the positive side. More interestingly, disgust now clustered with fear/anxiety rather than sad/anger, whereas sadness was still grouped with the sad/anger cluster, disclosing a clearer distinction between more sensory-driven and more socially-determined sources of aversion, respectively.

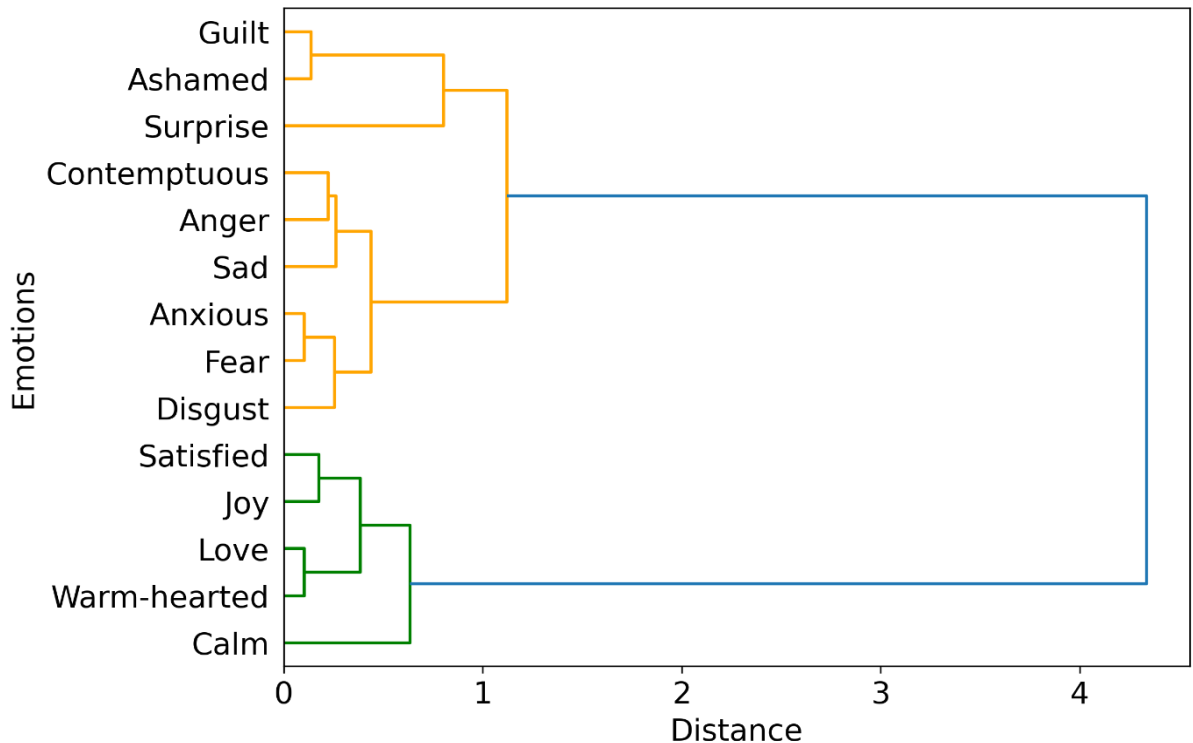

Figure 6 Results of the hierarchical clustering of discrete emotion terms using their CPM representation. Two general clusters of negative (orange cluster) and positive (green cluster) emotions with meaningful sub-clusters can be distinguished.

Supplementary material Figure S2 illustrates each emotion's within-subject normalized average profile on the 39 CoreGRID descriptors. This figure might be used as a guide by researchers interested in selecting stimuli that trigger certain CoreGRID items. Clear differences between positive and negative emotions are noted not only for *"feel good", "feel calm", "feel bad", "smile", "feel warm",* but also for *"feel your breathing is getting faster", "feel your heartbeat is getting faster", "have stomach troubles", "have a feeling of a lump in the front", "experience muscles tensing", "want the situation to continue", "want to undo what was happening", "feel the urge to stop what was happening", "want to damage, hit or say something that hurts", "want to destroy something", "think it violated laws/social norm",* and *"feel it was unpleasant for you"*.

To complement previous analyses, we ran two new clustering analyses at the movie-level for the mean values of 14 emotions and CPM. Results from the discrete emotion ratings revealed five main emotion clusters differentially induced by the film clips, shown in supplementary material Figure S3. These clusters appear to reflect specificities in clip content associated with calm, happiness, anxious, distress, and irritation. Detailed graphs are available in supplementary material Figures S5-S9.

The second clustering analysis using the average GRID item values at the movie level is shown in supplementary material Figure S10. It identified six-film clip clusters that could be correlated with CPM features. Further analysis revealed that the first cluster is dominated by GRID items such as *"feel good", "feel calm," "smile", and "feel strong"*, which are related to valence. The second cluster was suggestive of appetitive motivations/action tendencies as it is dominated by items such as *"situation to continue", "smile", "feel good", "feel calm",* and *"feel warm"*. The third cluster was loaded mainly with items belonging to norm violation *("incongruent with your standards", "think it violated laws/social norms")*, defensive *("feel the urge to stop", "want to undo", "want to tackle the situation"),* and autonomic features *("muscles tensing", "heartbeat is getting faster")*, as well as expression or feeling ("*did you frowned?*", "*did you feel bad?*"), which can potentially be interpreted in terms of goal and norm significance. The fourth cluster is loaded with items belonging to norm violation, unexpected, defensive, and negative expression, potentially defined as expectation violations or obstructiveness. The fifth cluster is mainly dominated by items belonging to unexpected and valence features, which can be interpreted as novelty. The sixth final cluster is loaded with items belonging to valence appraisal, calm, expected, unexpected, and defensive ratings, suggesting that it can be categorized as coping ability. Detailed graphs are available in supplementary material Figures S11-S16.

We also compared the movies associated with each cluster to find similar relationships between the CoreGRID clustering and the discrete emotion clustering at the movie level. Correspondingly, emotion cluster 1 (calm) is related to CoreGRID 1 (valence) and 6 (coping ability), emotion cluster 2 (happiness) is associated with CoreGRID 2 (motivations/action tendencies), emotion cluster 3 (anxious) is related to CoreGRID 4 (obstructiveness) and 5 (novelty), emotion cluster 4 (distress) is associated with CoreGRID 3 (norms), and emotion cluster 5 (irritation) is related to CoreGRID 4 (obstructiveness).

Finally, we conducted an exploratory factor analysis with a varimax rotation to find the latent dimensions explaining emotional experience using the z-scored normalized 39 CoreGRID items. We identified six factors explaining 51.6% total variance. Then, using the z-score normalized 39 CoreGRID items and 14 emotions, we performed a factor analysis and applied a varimax rotation [32, 50]. We chose to use varimax orthogonal rotation because it optimizes the distribution of the squared loadings within each factor by maximizing their variance and reduces the presence of cross-factor loadings (e.g., orthogonal dimensions). This results in clearer identification and interpretation of the underlying factors in the analysis. We identified four factors based on scree plot results. The four factors explain 37.3% total variance as shown in supplementary material Table S5. These factors can be interpreted as norms (10.99%), valence (12.55%), novelty or suddenness (6.52%), and action tendencies (7.18%). As shown in Figure 7, we plotted the loading values of each factor for 14 emotions. For better representation, we reversed the sign of the action tendency and norms values to show positive tendencies. As expected, valence shows the highest positive loadings for joy, satisfaction, love, warm-hearted, and calm. In contrast, disgust, anxious, fear, anger, sadness, and contempt all load negatively on valence. Joy is the most positive and fear the most negative. Interestingly, surprise had an intermediate valence load. On the other hand, for norm values, joy, satisfaction, love, warm-hearted, and calm show positive loadings; while other emotions (disgust, anxious, fear, surprise, guilt, ashamed, anger, contemptuous and sad) are negatively loaded on norms, with disgust showing the highest (negative) load. For novelty and suddenness, fear and surprise show high loading, unlike calm, that is the lowest. Lastly, guilt, anger, contempt, and sad show high action-tendency loadings, whereas disgust and fear show moderate loadings, and calm and satisfaction are the lowest on action tendency.

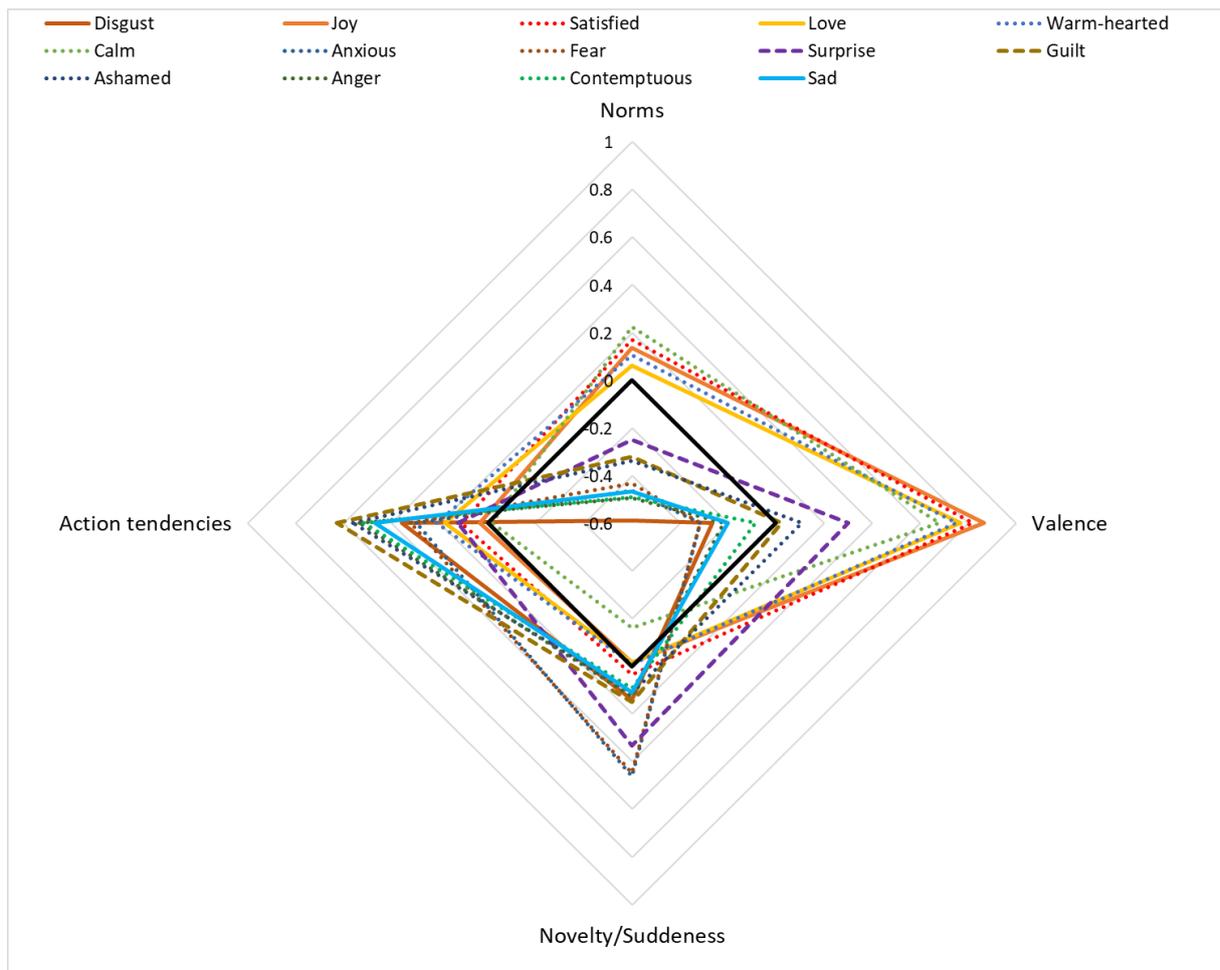

Figure 7: Spider plot for the factor analysis loading values of each factor for 14 emotions. Zero level is indicated by the black line. Scales of all dimensions in this plot are between –0.6 and 1, where norm violation vs compliance ranges from -0.6 to 0.2, negative vs positive valence ranges from -0.3 to 0.9, no novelty vs novelty ranges from -0.2 to 0.4, and appetitive vs defensive action tendencies ranges from -0.02 to 0.6.

## 5   Potential uses of EmoStim

Our EmoStim dataset, which includes emotion annotations and film clip details, has numerous potential use cases. These include:

- To further explore the emotion understanding from a componential perspective and using techniques such as Machine Learning to explore the feasibility of computational modelling.
- Serving as a benchmark for developing and evaluating automatic emotion recognition algorithms, which could be applied to domains such as psychology, human-computer interaction, and AC.
- Providing insights into audience reactions to different types of films, which could inform decision-making in film production, marketing, and distribution.
- Facilitating teaching on emotion recognition and analysis in courses such a psychology, emotion, and AC.
- Supporting the design of more emotionally engaging experiences in interactive media, such as films, video games and virtual reality, by understanding how users emotionally respond to different visual and audio stimuli.
- Enabling research on mental health and emotional well-being, such as studying the effects of different films on people's mood and emotions.

- Selecting subset of the collected clips according to the emotional content for eliciting emotions in emotion studies.

## 6 Discussion and Conclusion

Although emotional film clips are widely used in affective computing, finding appropriate material for a range of emotions can be challenging. In parallel, classical methods of emotion modelling based on discrete and dimensional theories are widely used in the literature but do not appear sufficient to account for explaining the whole range of emotion mechanisms. More recently, other models, such as "constructivist" theory and appraisal theory were also proposed, which are better aligned with neuroscience findings that show multiple brain processes subserving emotional experience [11, 33]. In this study we focused on CPM because it is easier to operationalise and measure. Moreover, in this study we assumed that emotions are highly subjectively pertaining to the principles of the CPM.

In the current work, we present EmoStim as a database of emotional film clips validated with both discrete and CPM-based labels that may be further refined in future research. Specifically, we report measures based on 99 emotion clips which have at least 15 rating items per film clip. A repository of these film clips and rating data collected is accessible to researchers upon their request through this link https://tinyurl.com/EmoStimDataset.

The first goal of this study was to determine the efficacy of selected film clips in inducing a wide range of emotions. Results demonstrate that all targeted emotions were experienced to some level. However, the distribution of warm-heartedness, joy, love, guilt, and contempt is relatively low. A possible explanation for this might be the complex nature of these emotions, which imply social or self-conscious processing (guilt, contemptuous) [51]. Similarly, Izard [51] interpreted love and attachment feelings as related to human evolution, normative development, and adaptation, that involve a high level of self/self-other representation and cultural cognition. Therefore, the selected film clips and duration may not be sufficient to reliably trigger such emotions for some participants. A similar explanation concerning attachment-related emotions (love, warm-hearted) underscores the inadequacy and difficulty of current lab experiments to obtain valid emotional stimuli as they result from biological tendencies [19]. In any case, in further analysis (see below), we found evidence for a distinction between emotions with self-directed and other-directed features, which does not only support the role of self-conscious and social processing in emotion elicitation, but also indicates that EmoStim material and CPM descriptors could successfully capture these aspects despite their low incidence during movie watching.

A second goal of our study was to inspect the organization of emotion space covered by EmoStim. To do so, we computed a correlation matrix across emotions at the movie level (see Figure 3) that revealed a clear high-level differentiation between positive and negative emotions, demonstrating the efficacy of selected clips to induce different valence levels and replicating previous research with other stimuli [21]. Furthermore, results showed a high association between attachment-related (love, warm-heartedness) and happiness-related (joy, satisfaction) emotions, in line with their shared positive valence. Calm was also distinctively associated with positive emotions, particularly happiness-related, which may reflect the pleasantness characteristic of calm but with a lower arousal level. In contrast, fear, anxiety, anger and disgust demonstrated strong correlations with each other, reflecting their shared negative valence and perhaps other aspects, such as uncertainty or unpredictability associated with these different emotion terms [52]. Likewise, the correlation between guilt and ashamed and their similar general relationships with other emotions is presumably due to sharing negative valence as well as potency, unpredictability, and social aspects [52]. This correlation accords with the literature, which considers shame, guilt and contempt as social or self-conscious emotions that require higher-order cognition and cultural cognition [51]. On the other hand, surprise did not correlate with any other particular emotion, suggesting specific characteristics of this experience putatively based on its ambiguous valence [53] and the role of novelty or suddenness in its elicitation [52].

These complex relationships among different emotions were further dissected by our clustering analysis that sought to uncover their structural organization in the CPM space. Consistent with the literature [18, 24], our results (see Figure 6) again underscored the main high-level differences between positive and negative emotions. However, at the lower level, we observed distinct and meaningful subclusters: embarrassment, surprise, irritation, distress, happiness, affection, and serenity. Interestingly, while calm remained a distinct state of positive affect, in this analysis surprise appeared more similar to the negative than the positive subclusters, indicating that it arose in more negative contexts in the selected film clips. A similar pattern for surprise was already reported in film-based literature [18], although this emotion is classically considered neutral or ambiguous in terms of valence [53, 54]. By comparing these CPM clustering results with the analysis of emotion terms at the movie level (see Figure 4), we could also observe a better functional similarity among emotions in the former than in the latter case. Accordingly, while serenity, embarrassment, threat, and social irritation (anger/contempt) showed similar patterns in both analyses, disgust was closer to threat (rather than embarrassment) and sad closer to embarrassment (rather than disgust) when considering CPM features. This would be consistent with disgust involving more appraisal of harm and avoidance responses similar to threat, whereas sadness implies more other-directed social cognitions shared with embarrassment. However, at the movie-level, the happiness cluster was associated with four terms (joy, satisfaction, warm-hearted, and love), whereas in the CPM space, these terms were clustered in only two lower levels interpreted as happiness (positive self-directed) and affection (positive attachment-related). Moreover, a similar dissociation between other-regarding and self-conscious emotions appeared to emerge more systematically among negative and positive emotions in the CPM clustering analysis, suggestive of an important role of social factors for certain kinds of emotions. Overall, these results confirm the capacity of film clips to trigger varying similar and dissimilar emotions, as well as the value and reliability of film feature annotations to capture their underlying functional organization. Furthermore, this validates the validity of our experimental paradigm and the EmoStim database.

Finally, we conducted an exploratory factor analysis to reveal the latent dimensions underpinning emotion formation. We found four factors: norms, valence, novelty/suddenness, and action tendencies. This study supports evidence from previous observations [18, 32, 50, 52], where more than two dimensions beyond valence and arousal are required to fully explain the emotional experience. Furthermore, these four factors seem to converge with a role of different component processes that are assumed to mediate differences between particular emotions, as discussed above. Here we also examined how the 14 discrete emotions mapped against the loading values of these four factors (see Figure 7). Besides a clear valence factor, we observed that positive emotions (joy, satisfaction, love, warm-hearted), as well as calm, loaded strongly on the first factor corresponding to behaviours aligned with social values and norms. In contrast, negative emotions loaded negatively on this factor, highlighting their link to disruptive events and norm violation. On the other hand, emotions associated with uncertainty and urgency are loaded high on novelty and suddenness factors, in contrast to calm. Specifically, high loadings of surprise [55] and fear [49] for novelty and suddenness accords with previous reports. Interestingly, the last fourth factor related to action-tendency is manifested mainly by emotions associated with embarrassment and irritation (including guilt, shame, anger, contempt, and sadness). This resonates with the previous view of Scherer [56], whereby guilt, shame, anger, and sadness are defined as emotions that provide adaptive functions that drive action, recovery, and motivational needs. Previous research has shown that variations of anger are related to action and action tendencies [10, 27].

Taken together, our data demonstrate the efficacy of selected film clips in the EmoStim database to induce a variety of emotions for empirical studies in affective sciences. Nevertheless, films are passive stimuli where a participant observes emotional content [21] rather than being directly implicated him/herself by ongoing events. Therefore, unlike more active procedures with Virtual Reality (VR) and video games, using films also imply potential limitations due to passive participation [57, 58] and lack of ecological validity [59]. These limitations can be mitigated by using passive stimuli in combination with active stimuli. For instance, recent affective research has shown the efficacy of VR paradigms in studying emotions using games [47, 59] or immersive films [60, 61] and images [62]. The effectiveness of VR can be explained by this immersion capacity, allowing a feeling of presence, interaction, and easy installation while still permitting controlled experiments [24]. However, while VR is a growing technology, films are backed by more robust research and more easily combined with other research tools, including neuroimaging techniques and advanced physiology sensors. Consequently, a validated film dataset such as ours can be valuably used for future research in behavioral sciences and may guide the design of future studies investigating emotional unfolding in more highly immersed and ecological valid conditions including VR. Moreover, to fully address the five components assumed in CPM, future research should also measure physiological and facial signals to better characterize the differential patterns of emotion experiences.

While the film dataset used in this study provided valuable insights into the trends and patterns in the emotions, it is important to acknowledge the constraints of the dataset. Firstly, despite the smaller scale of EmoStim in comparison to other datasets such as BoLD [63] we chose to annotate film clips using discrete emotions and CoreGRID items in accordance with our approach to understanding emotions from a CPM perspective. As a result, generating and utilizing short clips that may not effectively induce emotions as required by the CPM would not be practical or feasible for our study. Moreover, we acknowledge that the dataset was collected solely from participants in the USA and UK and that, therefore, it may not be representative of other cultural or linguistic groups. However, we ensure the diversity within the population by ensuring a mix of ages, genders, and regions. Therefore, we ensure that the dataset is inclusive and culturally sensitive to USA and UK regions. Additionally, the dataset does not include the test-retest reliability [64] which is used in research to measure reliability. Because it may not be practical for our crowdsourcing applications in emotion research. This is due to the potential difficulty in finding willing participants to take part in the study multiple times, as well as the possibility that external factors beyond the study could influence their emotional experiences. However, we have collected more than 15 assessments per video clip from 617 participants in our dataset, providing a sufficient level of rater reliability.

In sum, emotions play an important role in human development and everyday life. Therefore, understanding emotion evolution is critical in different domains and applications, but the availability of reliably validated material is still challenging and often limited to only a restricted set of emotions. In this paper, we present EmoStim as a database of emotional film clips built from a selection of 99 film clips combined with rich participant annotation and multistep validation. These clips cover a large emotional distribution with reliable induction. They allow a robust differentiation of affective states defined by the discrete model as well as by more elementary features derived from a CPM theory-based approach. Our results show the efficacy of our database in studying emotional experience effectively. Future research can use this database to extend emotion-related research with new measures and/or analyze the collected evaluations made available with EmoStim in order to enhance our understanding of emotion generation and experience in humans.

## 7    Acknowledgments
The authors would like to thank the Swiss National Science Foundation (SNF Sinergia No. 180319) and the National Centre of Competence in Research (NCCR) Affective Sciences (under grant No. 51NF40-104897) for funding the data collection.

**EmoStim: A Database of Emotional Film Clips with Discrete and Componential Assessment**

Rukshani Somarathna [a], Patrik Vuilleumier [b, c, d], Gelareh Mohammadi [a]

[a] School of Computer Science and Engineering, University of New South Wales, Australia

[b] Laboratory for Behavioral Neurology and Imaging of Cognition, Department of Neuroscience, University of Geneva, Geneva, Switzerland,

[c] Neurology Department, University Hospital of Geneva, Geneva, Switzerland

[d] Swiss Center for Affective Sciences, Campus Biotech, University of Geneva

**Supplementary material**
Table S1 CoreGRID questionnaire

| GRID Questionnaire | Component |
|---|---|
| **While watching this movie, did you...** | |
| 1. think it was incongruent with your standards/ideas? | Appraisal |
| 2. feel that the event was unpredictable? | Appraisal |
| 3. feel the event occurred suddenly? | Appraisal |
| 4. think the event was caused by chance? | Appraisal |
| 5. think that the consequence was predictable? | Appraisal |
| 6. feel it was unpleasant for someone else? | Appraisal |
| 7. think it was important for somebody's goal or need? | Appraisal |
| 8. think it violated laws/social norms? | Appraisal |
| 9. feel in itself was unpleasant for you? | Appraisal |
| 10. want to destroy s.th.? | Motivation |
| 11. feel the urge to stop what was happening? | Motivation |
| 12. want to undo what was happening? | Motivation |
| 13. want to damage, hit or say s.th. that hurts? | Motivation |
| 14. want the situation to continue? | Motivation |
| 15. lack the motivated to pay attention to the scene? | Motivation |
| 16. want to tackle the situation and do s.th.? | Motivation |
| 17. have a feeling of lump in the throat? | Physiology |
| 18. have stomach trouble? | Physiology |
| 19. experience muscles tensing? | Physiology |
| 20. feel warm? | Physiology |
| 21. sweat? | Physiology |
| 22. feel heartbeat getting faster? | Physiology |
| 23. feel breathing getting faster? | Physiology |
| 24. feel breathing slowing down? | Physiology |
| 25. produce abrupt body movement? | Expression |
| 26. close your eyes? | Expression |
| 27. press lips together? | Expression |
| 28. have the jaw drop? | Expression |
| 29. show tears? | Expression |
| 30. have eyebrow go up? | Expression |
| 31. smile? | Expression |
| 32. frown? | Expression |
| 33. produce speech disturbances? | Expression |
| 34. feel good? | Feeling |
| 35. feel bad? | Feeling |
| 36. feel calm? | Feeling |
| 37. feel strong? | Feeling |
| 38. feel an intense emotional state? | Feeling |
| 39. experience an emotional state for a long time? | Feeling |

Table S2 Discrete emotions questionnaire

| While watching this movie, did you feel… |
|---|
| 1. fearful, scared, afraid? |
| 2. anxious, tense, nervous? |
| 3. angry, irritated, mad? |
| 4. warm-hearted, gleeful, elated? |
| 5. joyful, amused, happy? |
| 6. sad, downhearted, blue? |
| 7. satisfied, pleased? |
| 8. surprised, amazed, astonished? |
| 9. loving, affectionate, friendly? |
| 10. guilty, remorseful? |
| 11. disgusted, turned off, repulsed? |
| 12. disdainful, scornful, contemptuous? |
| 13. calm, serene, relaxed? |
| 14. ashamed, embarrassed? |

Table S3 BFI questionnaire

| I see myself as someone who… |
|---|
| 1. is reserved? |
| 2. is generally trusting? |
| 3. tends to be lazy? |
| 4. is relaxed and handles stress well? |
| 5. has few artistic interests? |
| 6. is outgoing sociable? |
| 7. tends to find fault with others? |
| 8. does a thorough job? |
| 9. gets nervous easily? |
| 10. has an active imagination? |

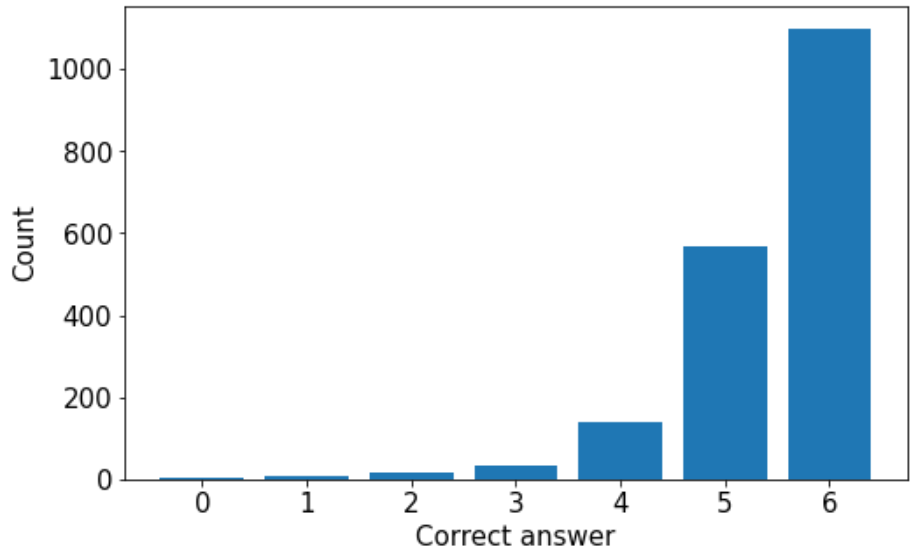

Figure S1  Count of correct quality control answers for the six questions.

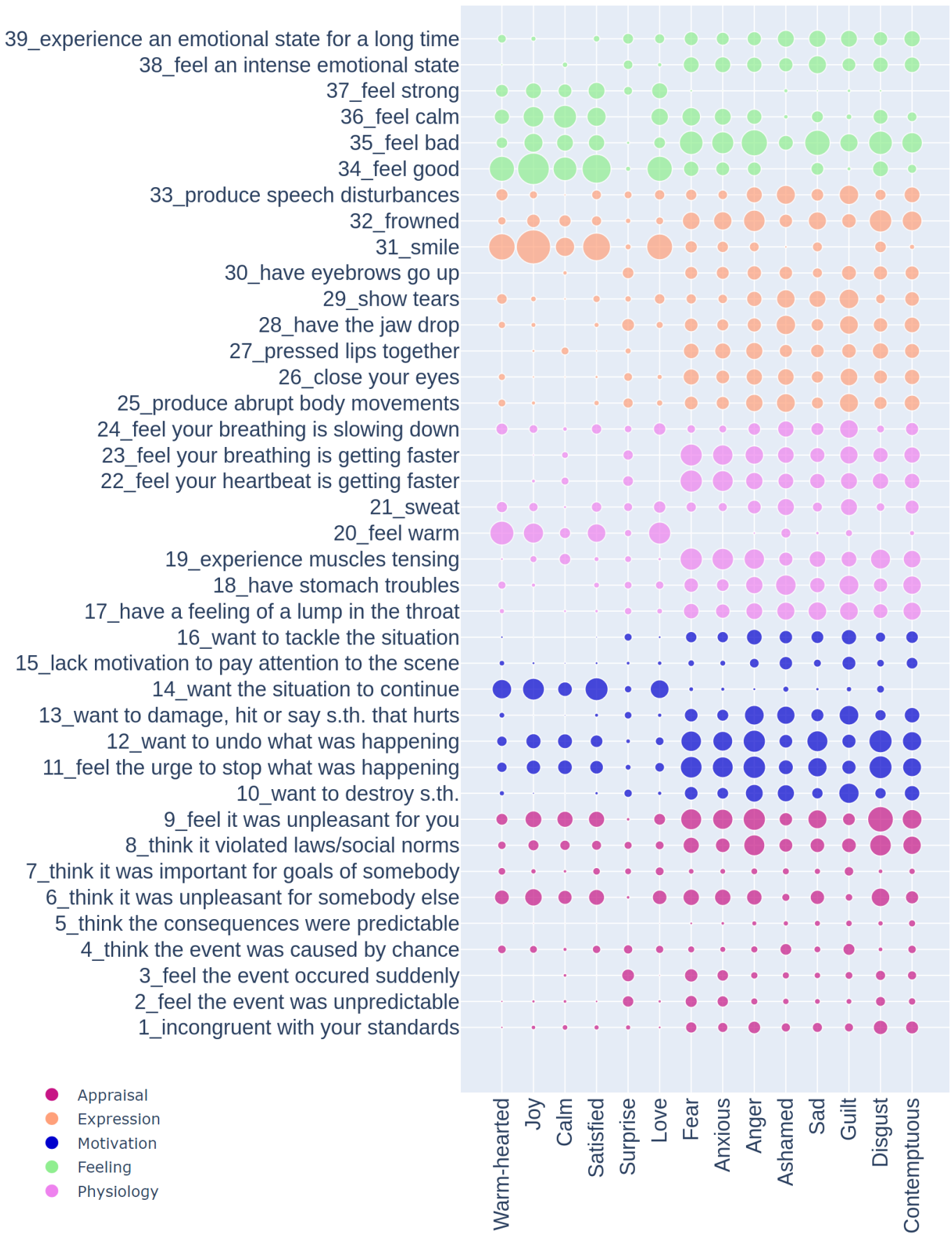

Figure S2 Discrete theory-based emotion profiles in CoreGRID space derived from hierarchical clustering. The size of the bubble is proportional to the value of the weighted average CPM profile for each emotion term.

**Clustering at movie level for average emotion ratings**

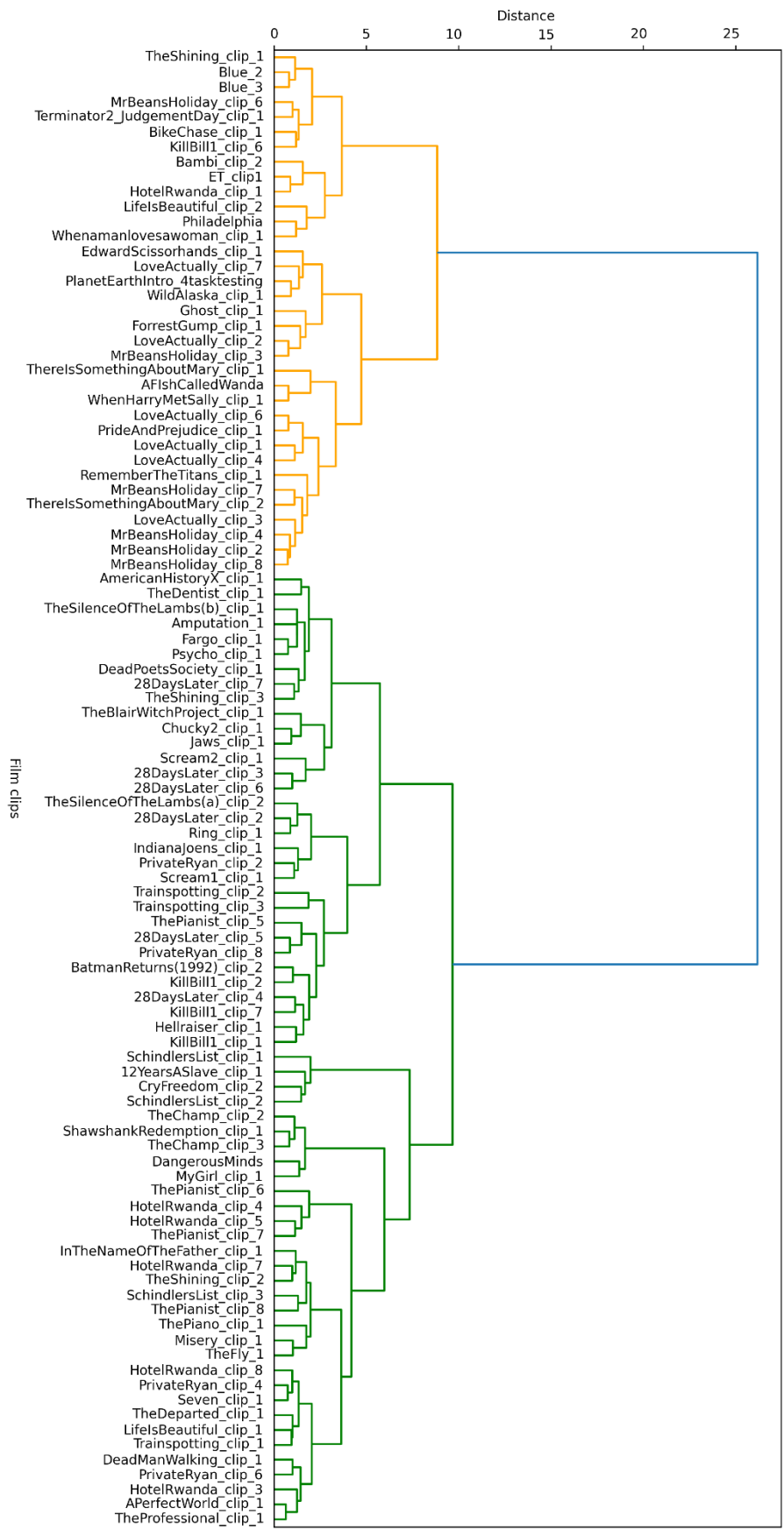

Figure S3: Hierarchical clustering at movie-level for average emotion and CPM ratings.

To find similar movies that trigger emotions, we run a clustering analysis at the movie level by taking the average of emotions for each film. Similar to the previous hierarchical clustering (see main text), we could identify five groups including calm, happiness, anxious, distress, and irritation, as shown in figures 4-8.

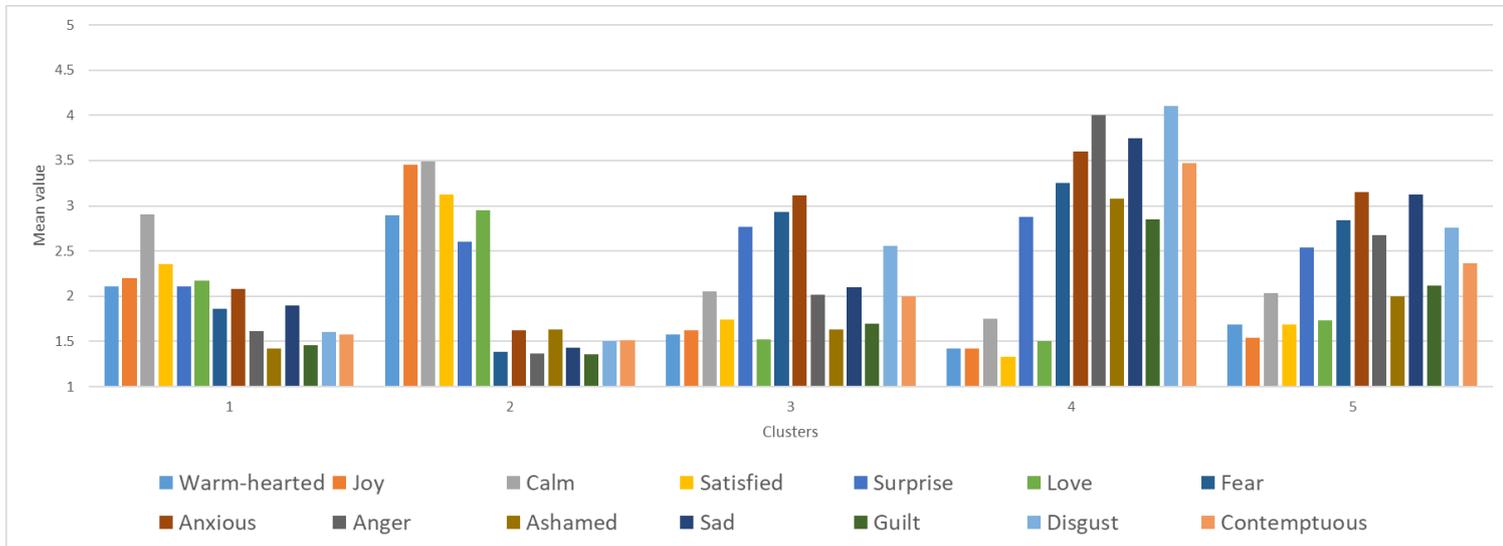

Figure S4: Emotion rating mean values for each cluster from 1-5

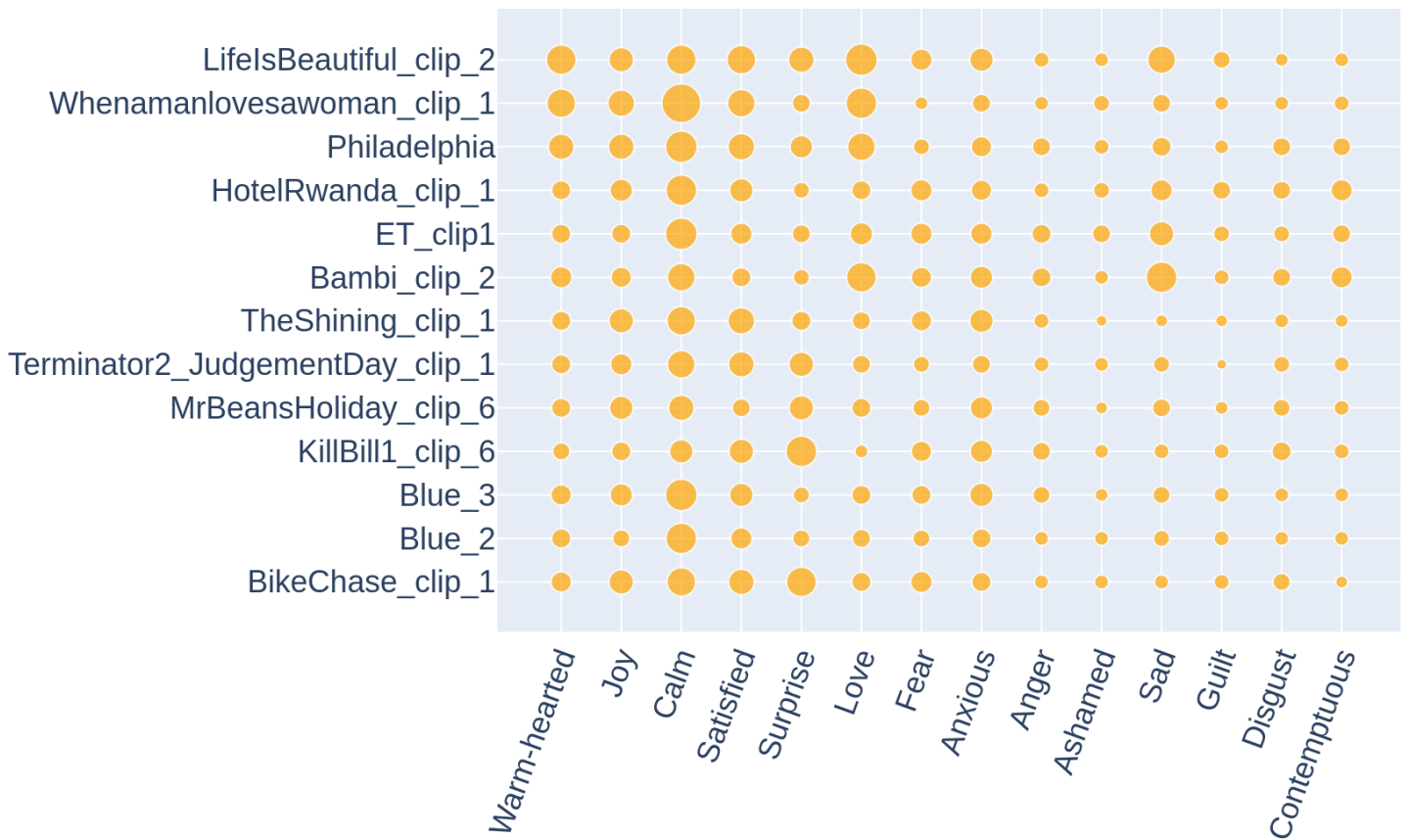

Figure S5: Emotion rating values for cluster 1

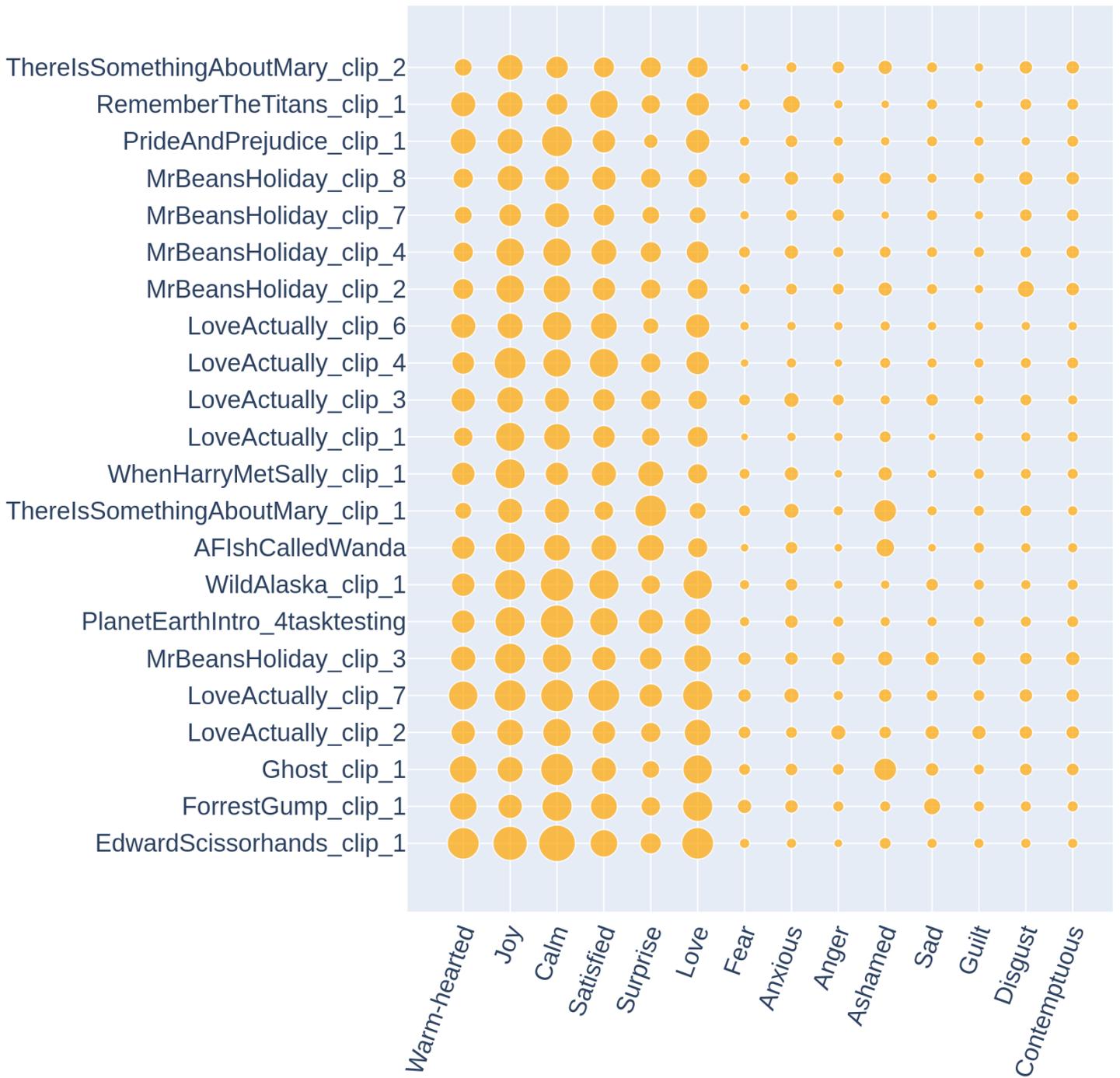

Figure S6: Emotion rating values for cluster 2

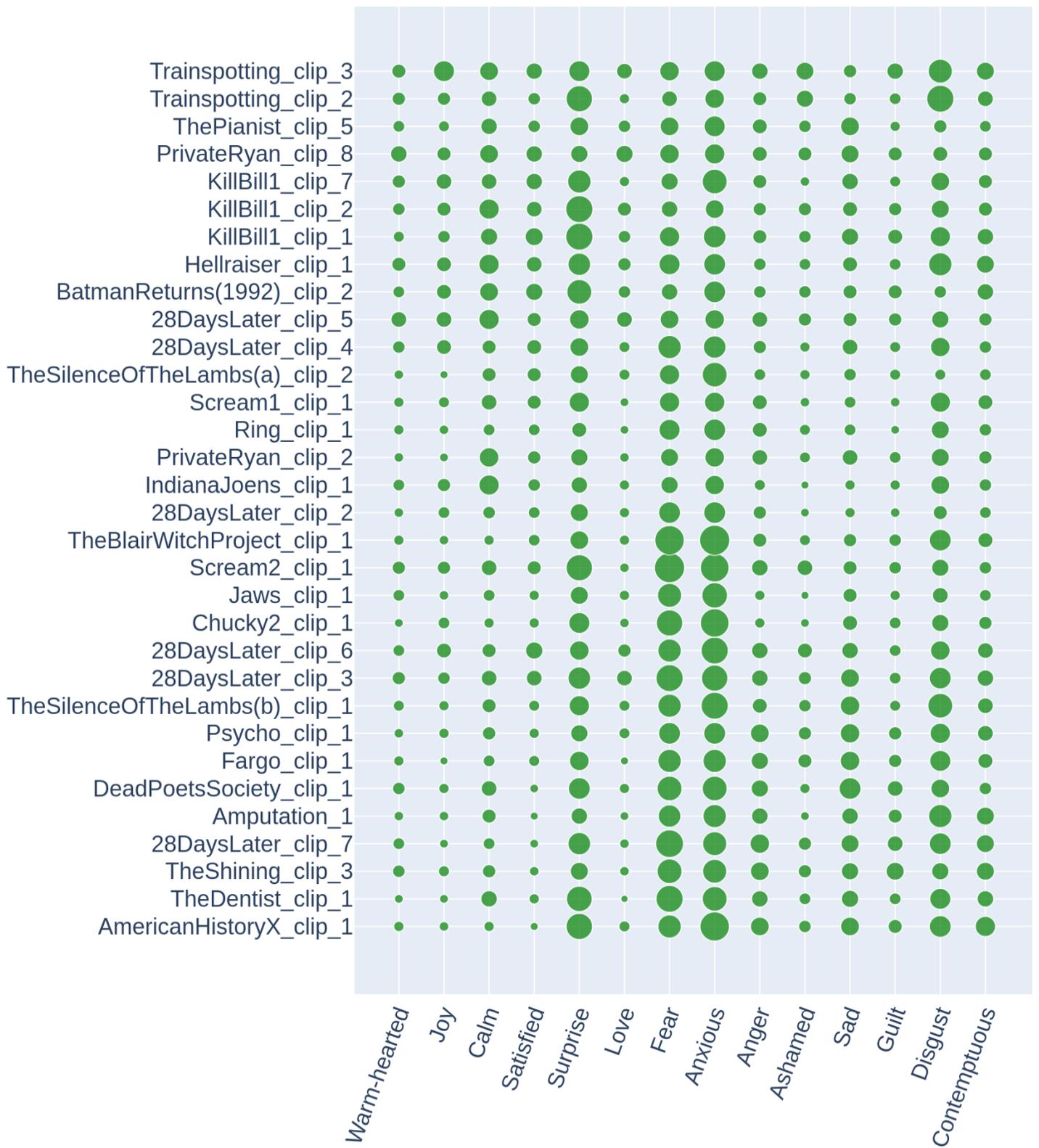

Figure S7: Emotion rating values for cluster 3

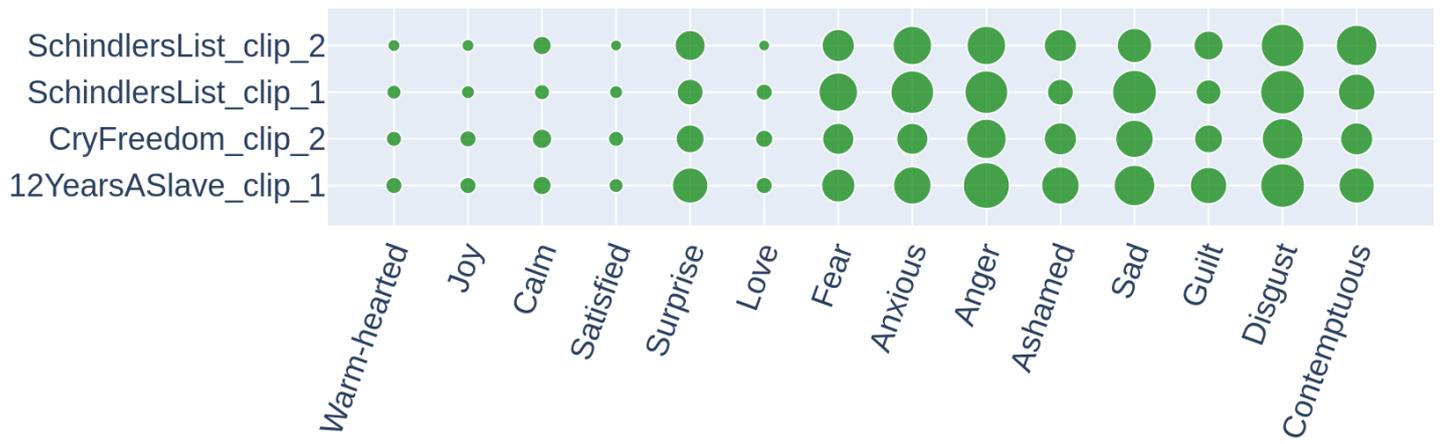

Figure S8: Emotion rating values for cluster 4

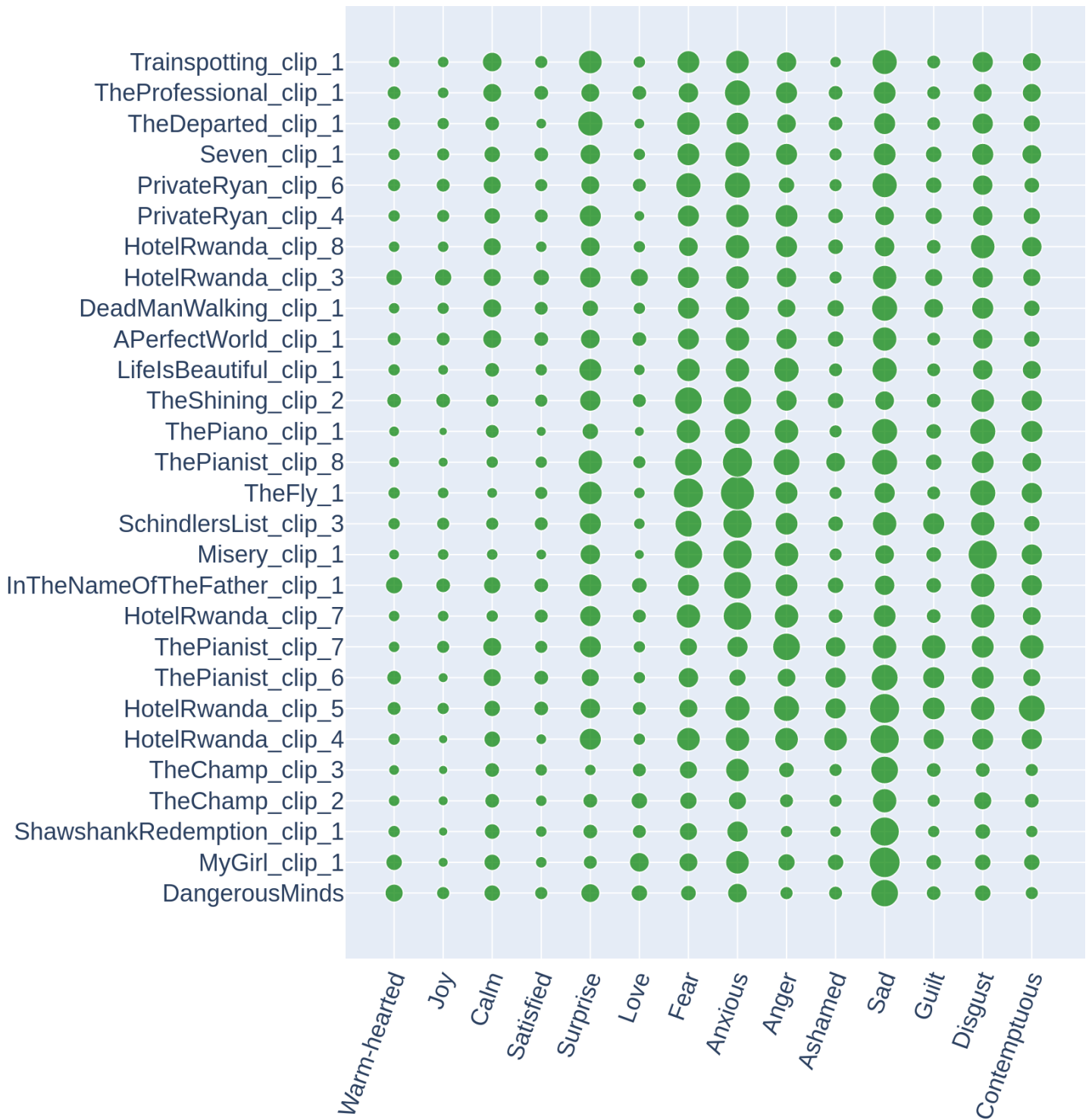

Figure S9: Emotion rating values for cluster 5

**Clustering at movie level for average CPM ratings**

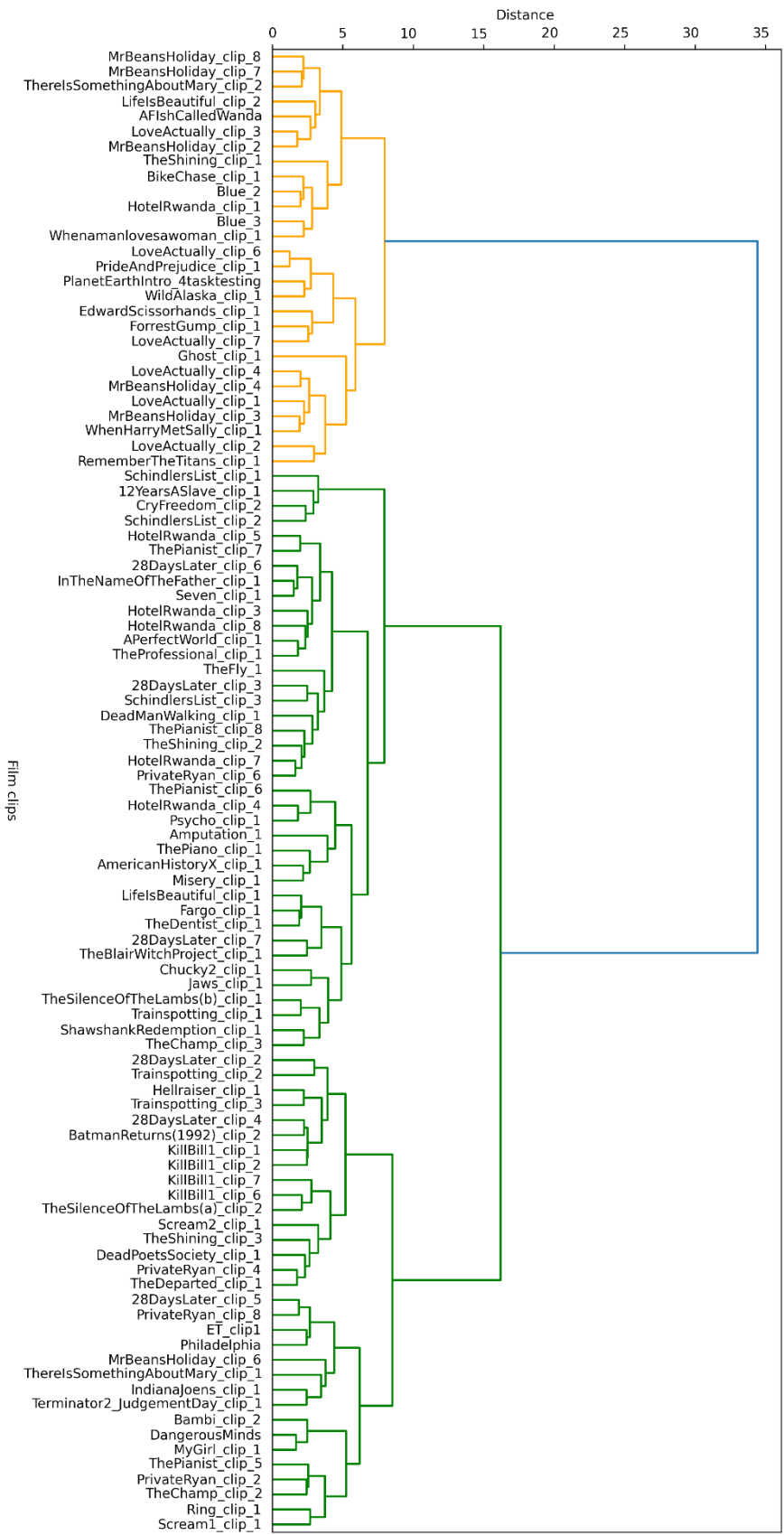

Figure S10: Hierarchical clustering of CPM items at movie-level TODO caption.

To find movies that trigger similar componential patterns, we ran a clustering analysis at the movie level by taking the average of CoreGRID items. We could identify six sources of similarity including valence, appetitive motivations/action tendencies, norms, obstructiveness, novelty appraisal, and coping potential, as shown in figures 10-15. The results also suggest that our film database has globally more negative than positive content. Nevertheless, our findings demonstrate that the selected emotional clips can induce a wide range of discrete emotions and CPM features at several intensities.

Figure S11: CoreGRID rating values for cluster 1

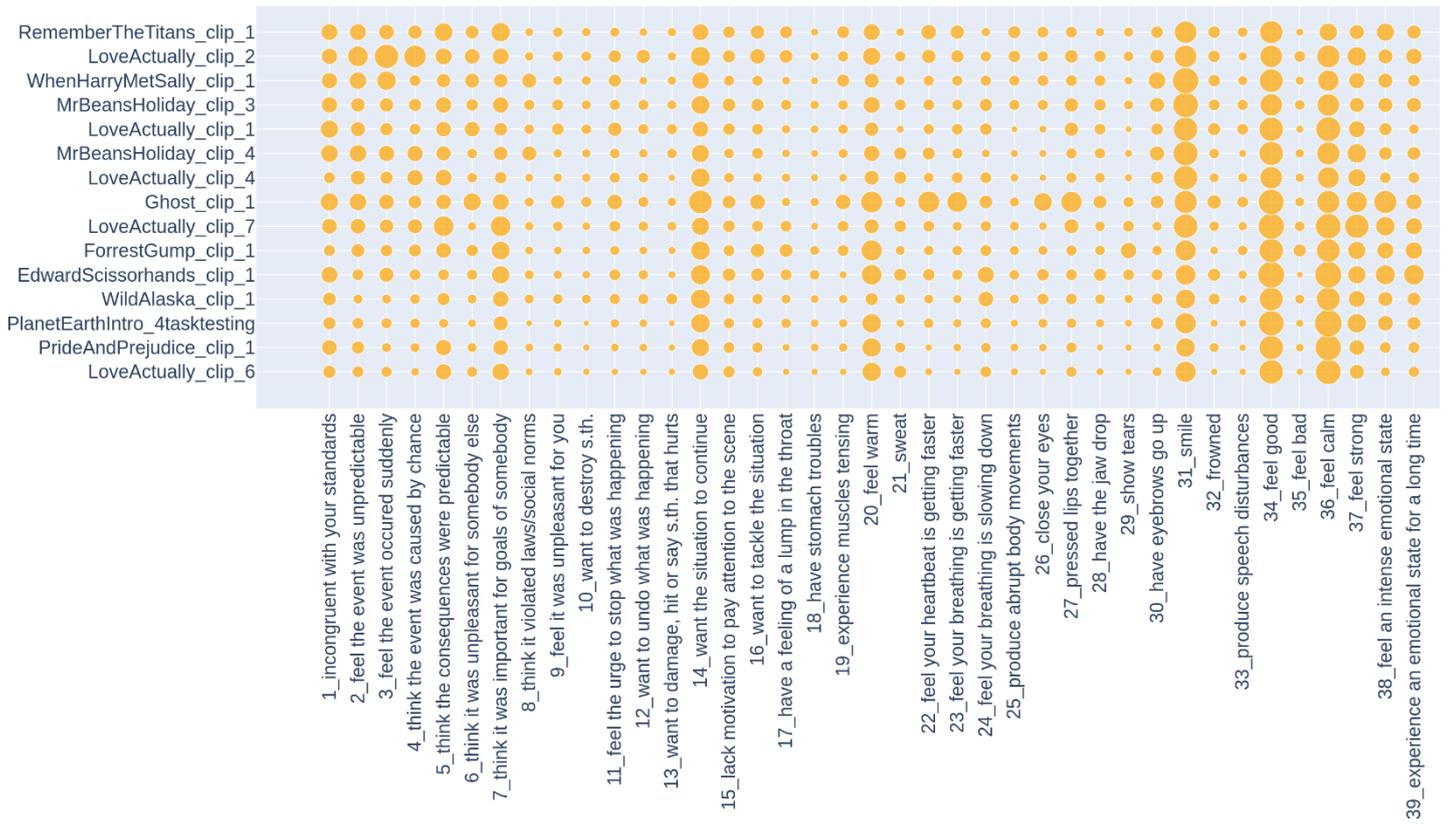

Figure S12: CoreGRID rating values for cluster 2

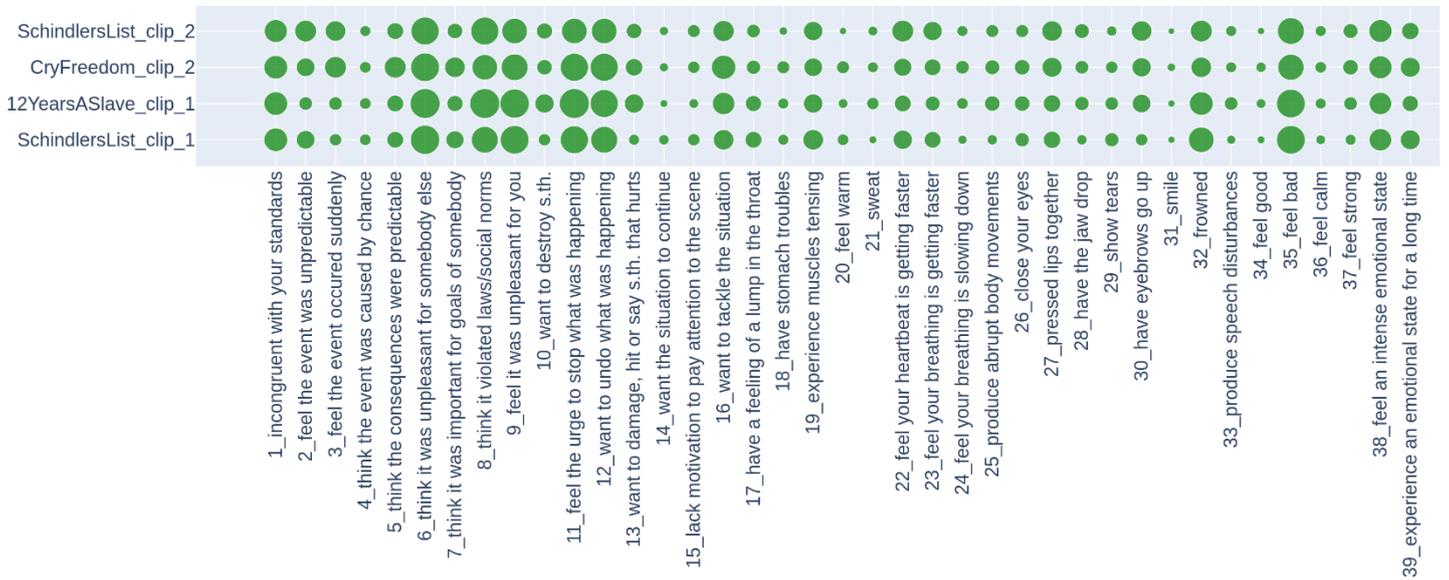

Figure S13: CoreGRID rating values for cluster 3

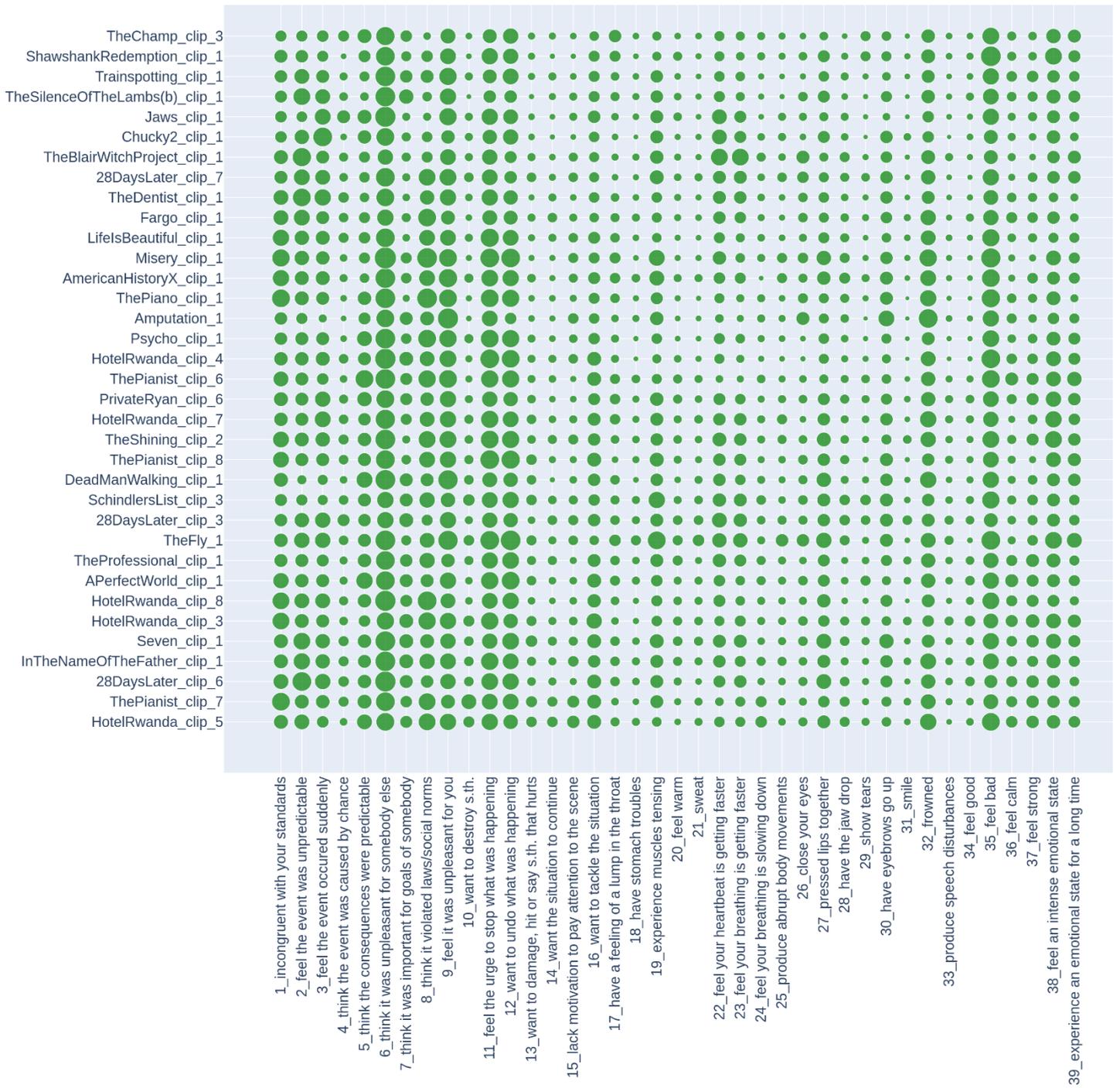

Figure S14: CoreGRID rating values for cluster 4

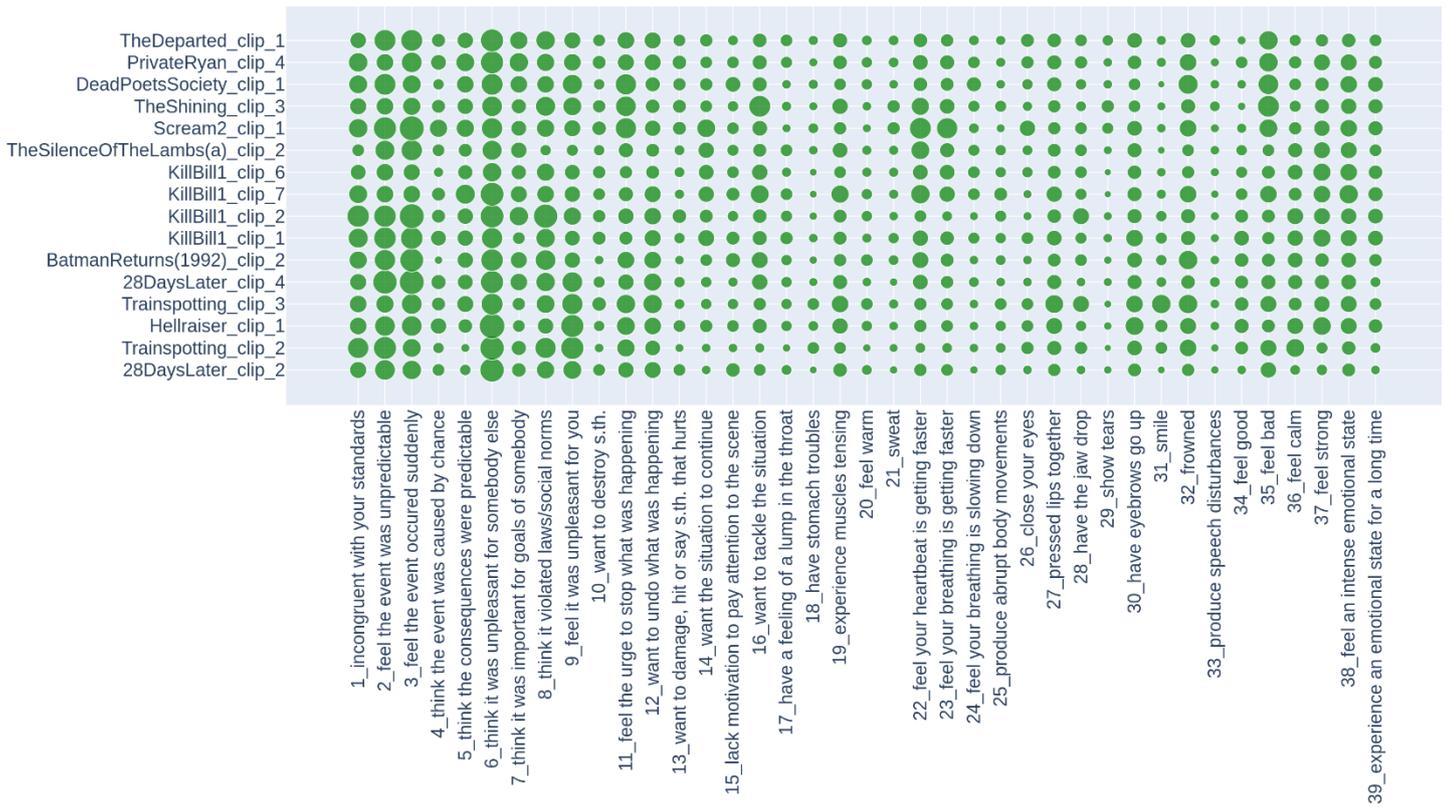

Figure S15: CoreGRID rating values for cluster 5

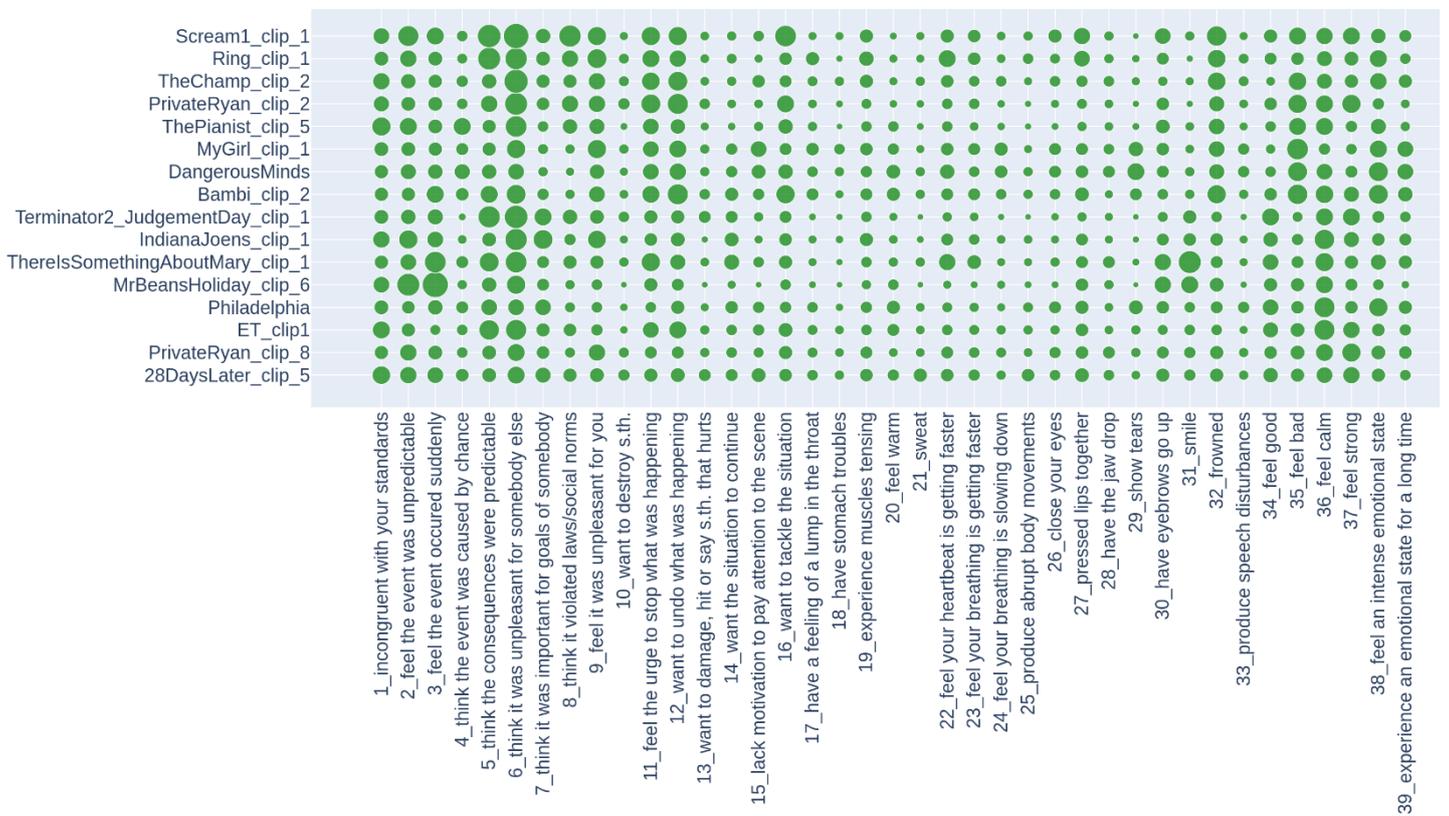

Figure S16: CoreGRID rating values for cluster 6

Table S5: Results of the exploratory factor analysis using the z-score normalized 39 CoreGRID items and 14 emotions. F1 can be interpreted as norms (10.99%), F2 as valence (12.55%), F3 as novelty and suddenness (6.52%) and F4 as action tendencies (7.18%). The bolded values indicate the maximum coefficient for each row.

| Item | Component | F1 | F2 | F3 | F4 |
|---|---|---|---|---|---|
| feel the urge to stop what was happening | Motivation | **0.75** | -0.27 | 0.24 | -0.16 |
| want to undo what was happening | Motivation | **0.74** | -0.27 | 0.20 | -0.17 |
| feel it was unpleasant for you | Appraisal | **0.69** | -0.36 | 0.16 | -0.19 |
| think it was unpleasant for somebody else | Appraisal | **0.66** | -0.34 | 0.15 | 0.03 |
| feel bad | Feeling | **0.64** | -0.40 | 0.18 | -0.30 |
| think it violated laws/social norms? | Appraisal | **0.61** | -0.20 | 0.15 | -0.19 |
| Disgust | | **0.59** | -0.27 | 0.14 | -0.36 |
| frowned | Expression | **0.55** | -0.25 | 0.22 | -0.27 |
| Anxious | | **0.46** | -0.32 | 0.46 | -0.30 |
| incongruent with your standards | Appraisal | **0.44** | -0.06 | 0.11 | -0.11 |
| feel an intense emotional state | Feeling | **0.44** | 0.05 | 0.42 | -0.24 |
| want to tackle the situation | Motivation | **0.38** | 0.09 | 0.26 | -0.25 |
| pressed lips together | Expression | **0.37** | -0.04 | 0.35 | -0.32 |
| feel the event was unpredictable | Appraisal | **0.33** | -0.00 | 0.32 | 0.01 |
| think the consequences were predictable | Appraisal | **0.17** | 0.04 | 0.06 | -0.15 |
| Joy | | -0.14 | **0.86** | -0.02 | -0.03 |
| Satisfied | | -0.17 | **0.82** | 0.03 | -0.11 |
| feel good | Feeling | -0.28 | **0.80** | -0.05 | 0.02 |
| smile | Expression | -0.21 | **0.78** | -0.00 | -0.04 |
| Love | | -0.06 | **0.77** | -0.02 | -0.18 |
| Warm-hearted | | -0.10 | **0.75** | -0.08 | -0.21 |
| Calm | | -0.22 | **0.68** | -0.16 | 0.02 |
| want the situation to continue? | Motivation | -0.22 | **0.56** | 0.19 | -0.19 |
| feel calm | Feeling | -0.33 | **0.54** | -0.21 | 0.08 |
| feel warm | Physiology | -0.04 | **0.53** | 0.17 | -0.33 |
| feel strong | Feeling | 0.04 | **0.50** | 0.16 | -0.06 |
| think it was important for goals of somebody | Appraisal | 0.20 | **0.24** | 0.20 | -0.19 |
| feel your heartbeat is getting faster | Physiology | 0.25 | -0.06 | **0.81** | -0.29 |
| feel your breathing is getting faster | Physiology | 0.21 | -0.03 | **0.79** | -0.36 |
| Fear | | 0.43 | -0.32 | **0.45** | -0.32 |
| experience muscles tensing | Physiology | 0.44 | -0.15 | **0.45** | -0.36 |
| feel the event occurred suddenly | Appraisal | 0.36 | 0.04 | **0.37** | -0.02 |
| have eyebrows go up | Expression | 0.31 | 0.05 | **0.34** | -0.30 |
| Surprise | | 0.25 | 0.30 | **0.33** | -0.13 |
| produce speech disturbances | Expression | 0.10 | 0.20 | 0.21 | **-0.71** |
| have stomach troubles | Physiology | 0.18 | 0.12 | 0.23 | **-0.69** |
| show tears | Expression | 0.10 | 0.15 | 0.18 | **-0.68** |
| Guilt | | 0.32 | 0.02 | 0.15 | **-0.63** |
| sweat | Physiology | 0.06 | 0.22 | 0.30 | **-0.60** |
| produce abrupt body movements | Expression | 0.19 | 0.12 | 0.35 | **-0.60** |
| feel your breathing is slowing down | Physiology | 0.09 | 0.24 | 0.24 | **-0.58** |
| want to damage, hit or say s.th. that hurts | Motivation | 0.25 | 0.08 | 0.24 | **-0.57** |
| Ashamed | | 0.34 | 0.11 | 0.11 | **-0.56** |
| close your eyes | Expression | 0.20 | 0.07 | 0.34 | **-0.54** |
| Anger | | 0.49 | -0.22 | 0.13 | **-0.53** |
| want to destroy s.th. | Motivation | 0.25 | 0.07 | 0.28 | **-0.52** |
| Contemptuous | | 0.49 | -0.09 | 0.09 | **-0.52** |
| have a feeling of a lump in the throat | Physiology | 0.29 | 0.07 | 0.32 | **-0.50** |
| have the jaw drop | Expression | 0.26 | 0.17 | 0.34 | **-0.47** |
| Sad | | 0.47 | -0.20 | 0.11 | **-0.47** |
| lack motivation to pay attention to the scene | Motivation | 0.11 | 0.09 | 0.06 | **-0.42** |
| experience an emotional state for a long time | Feeling | 0.36 | 0.21 | 0.36 | **-0.37** |
| think the event was caused by chance | Appraisal | 0.13 | 0.25 | 0.19 | **-0.29** |

# 1 Average intensity of each emotion

This section consists of average intensity ratings bar plots for each of the 99 movie clips. (WARM: Warm-hearted, JOY: Joy, CALM: Calm, SATIS: Satisfied, SURP: Surprise, LOVE: Love, FEAR: Fear, ANX: Anxious, ANG: Anger, ASHAM: Ashamed, SAD: Sad, GUILT: Guilt, DISG: Disgust, CONTEMPT: Contemptuous)

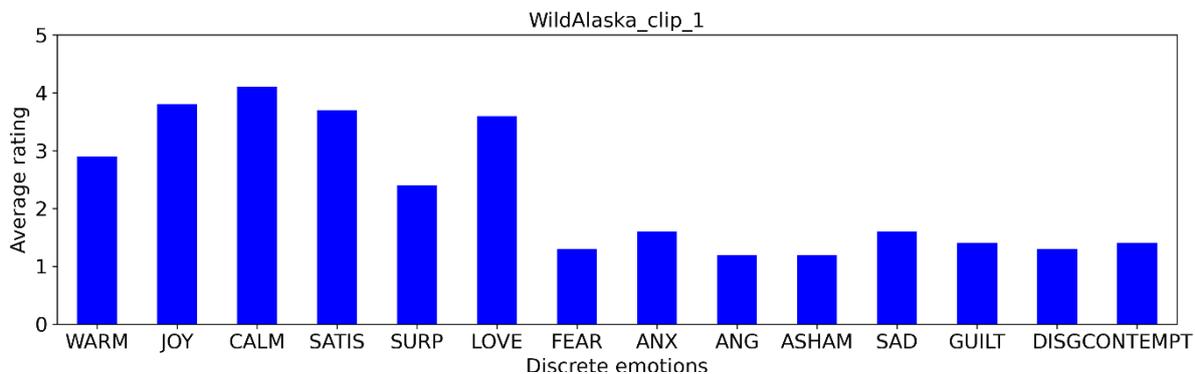
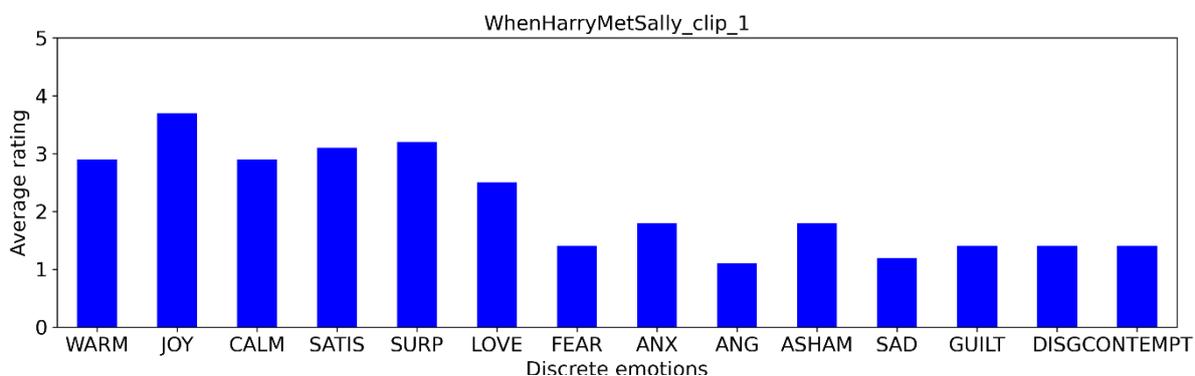
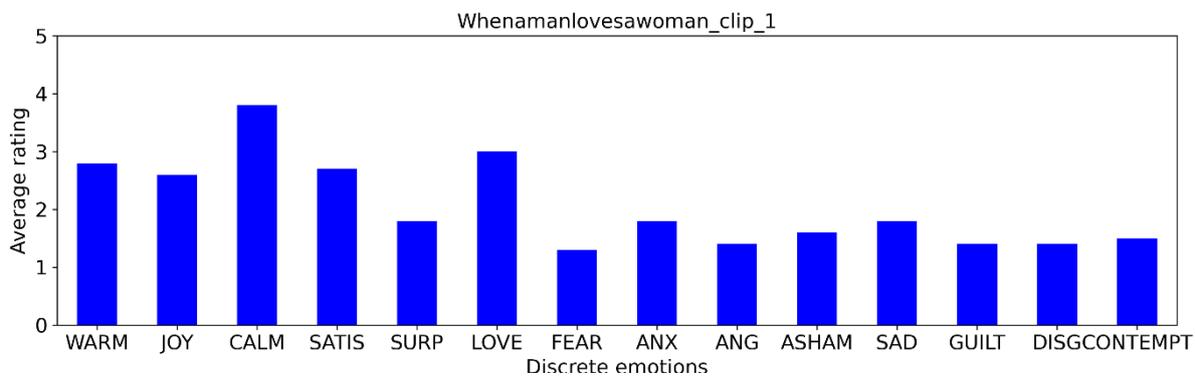
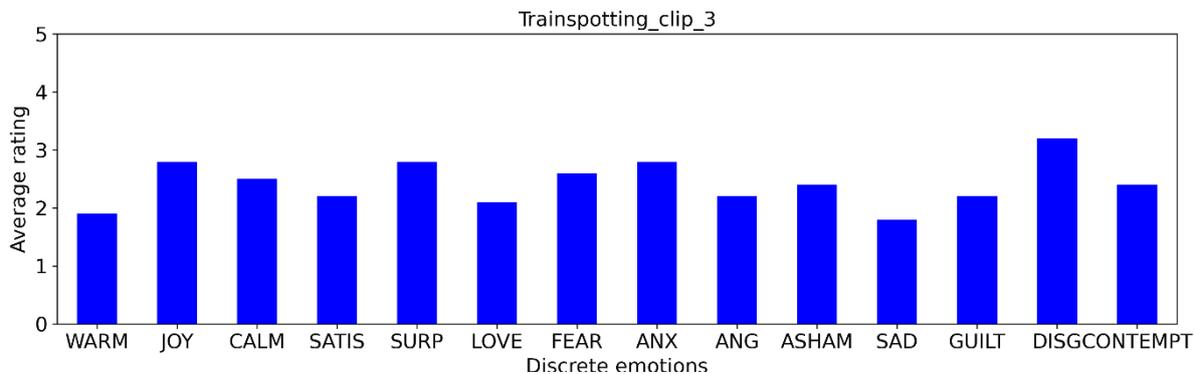

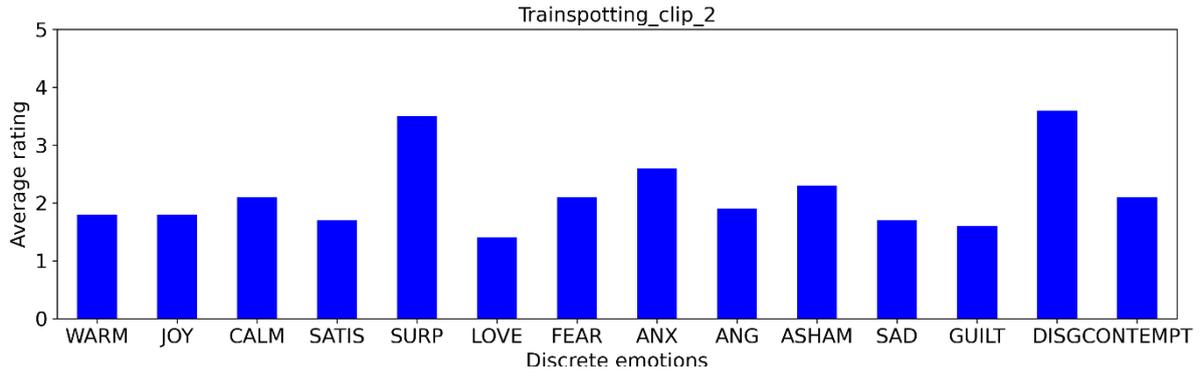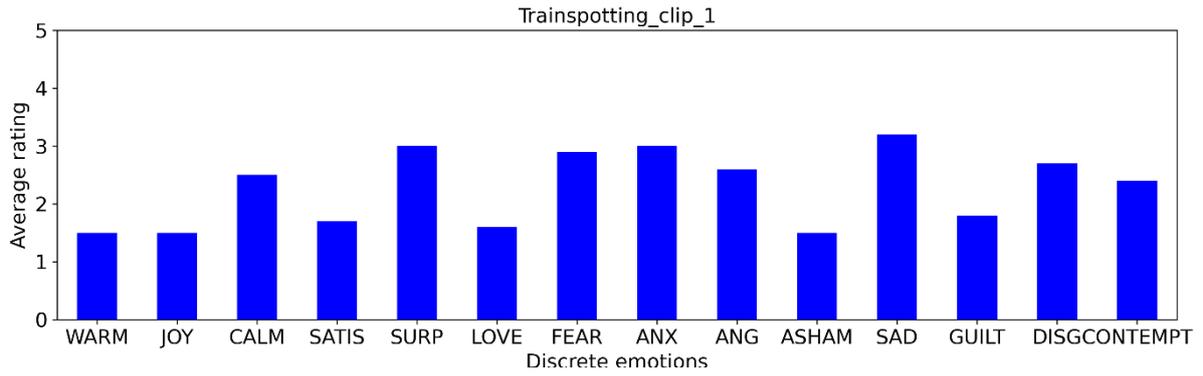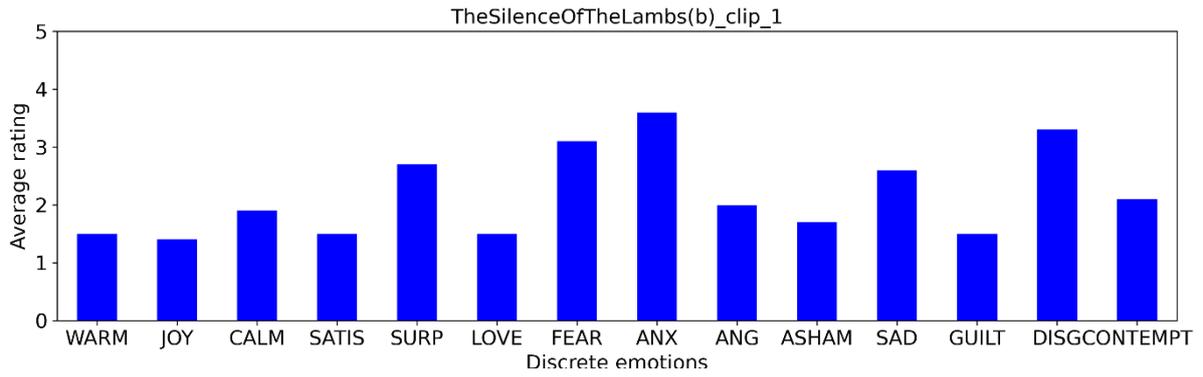

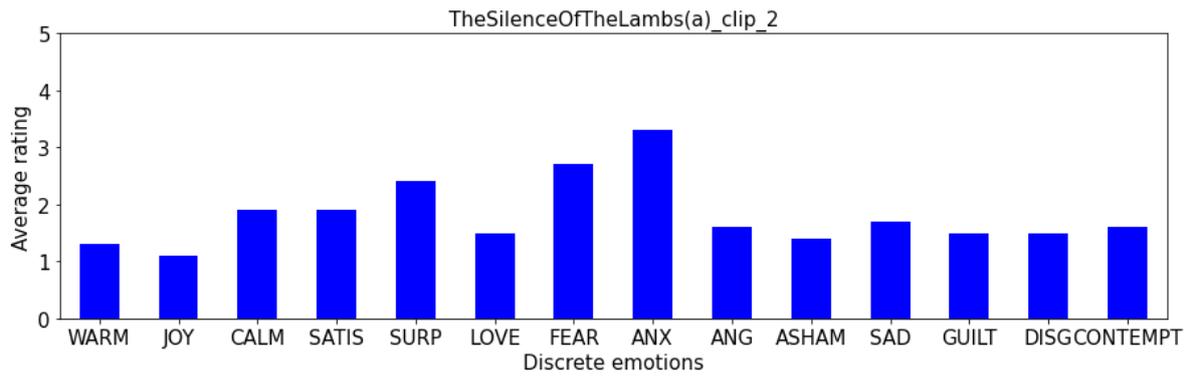
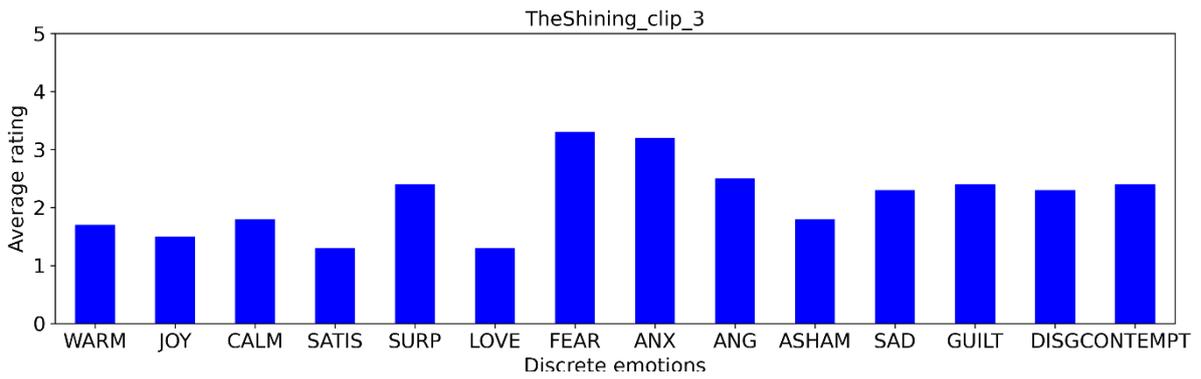
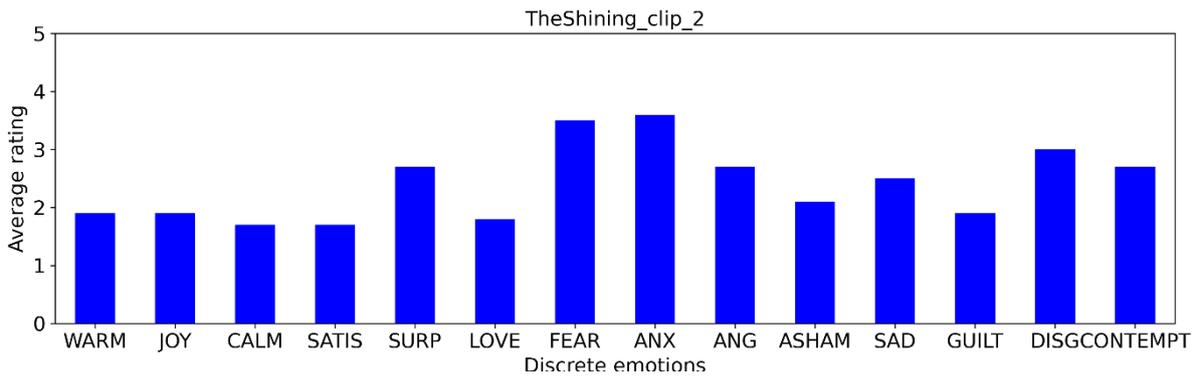
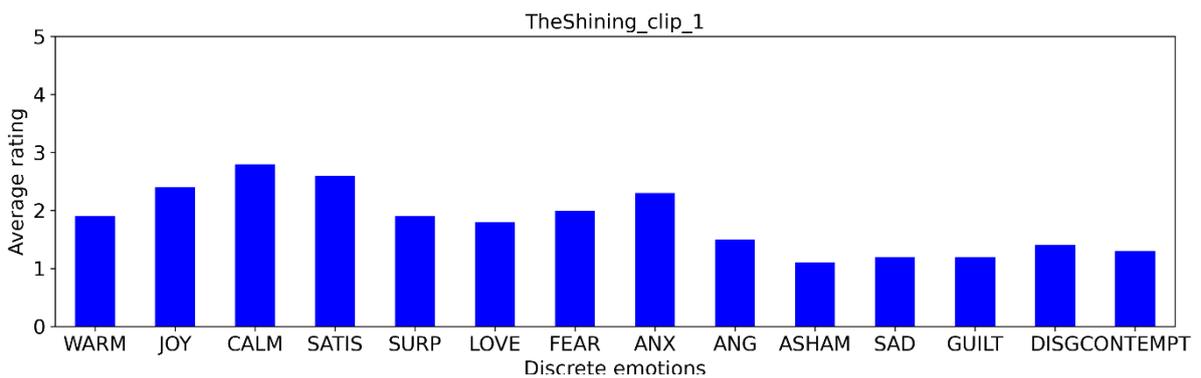

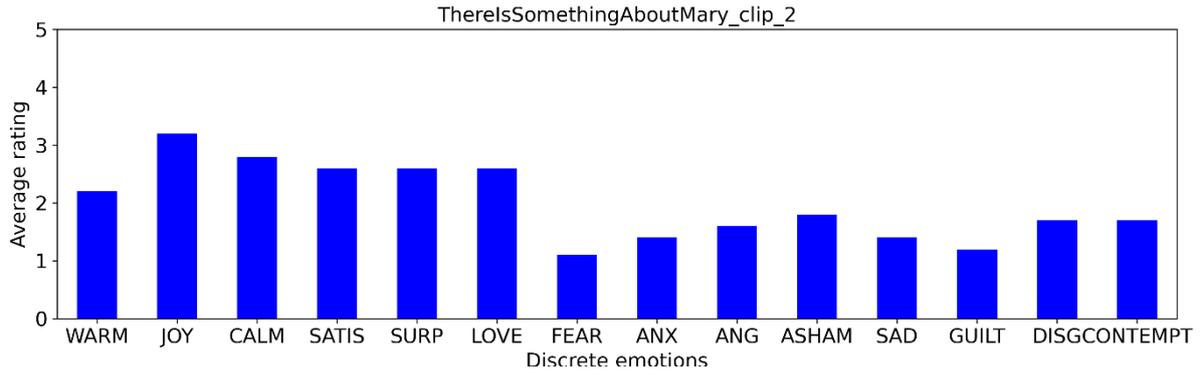
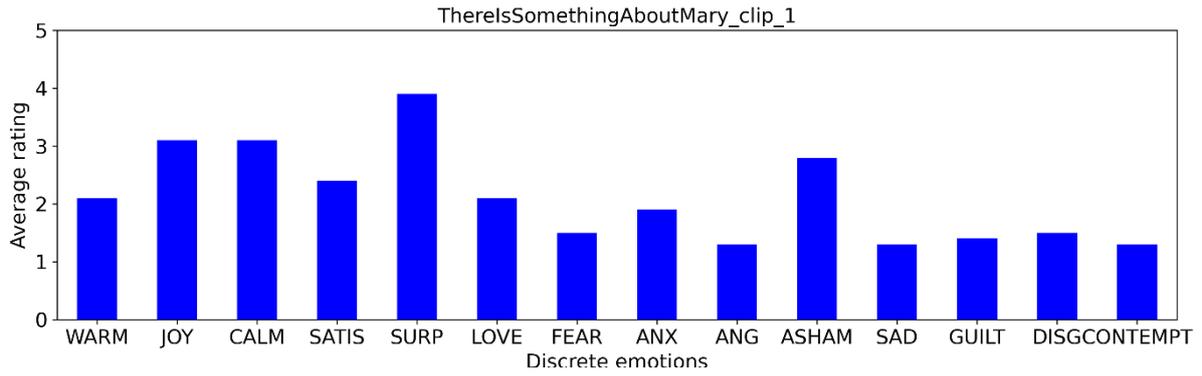
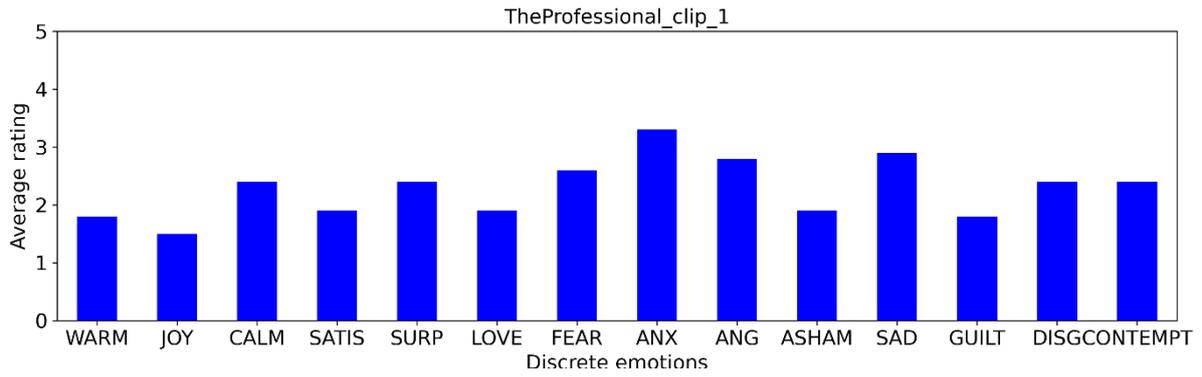
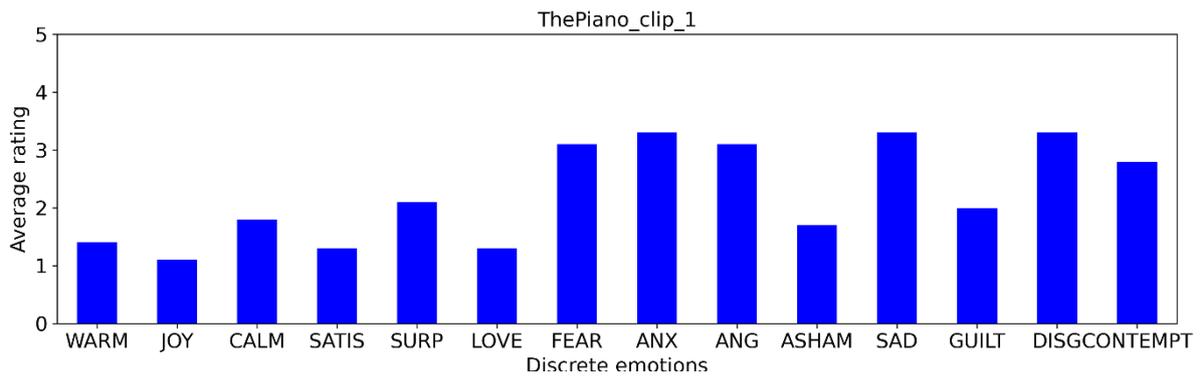

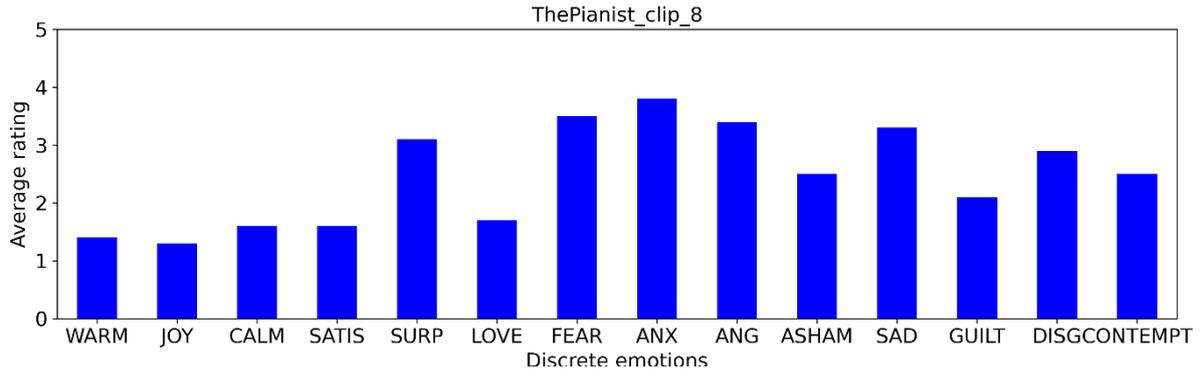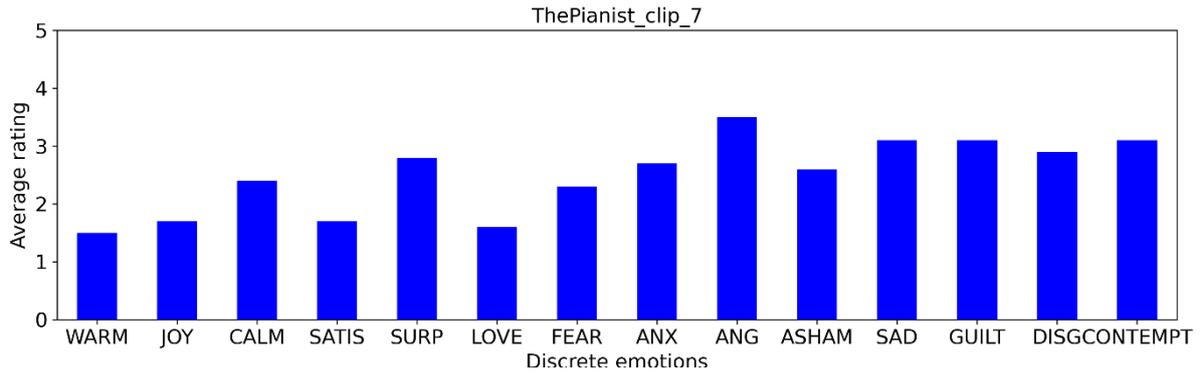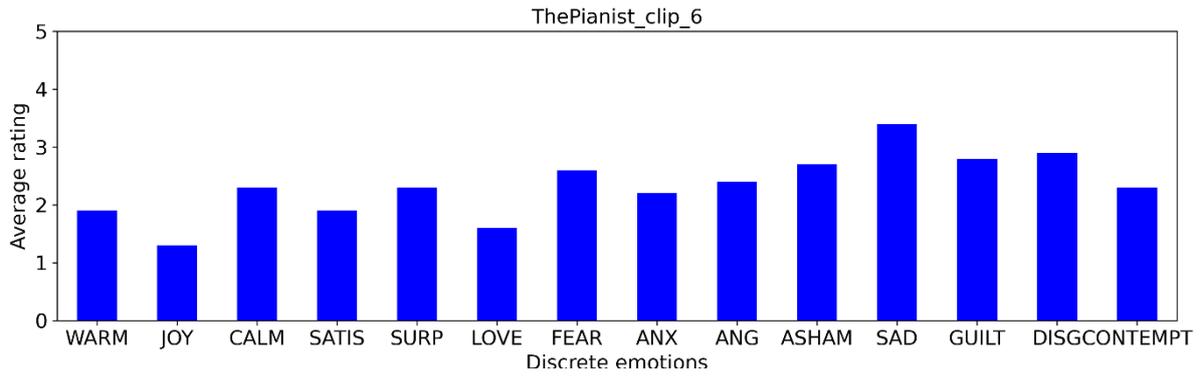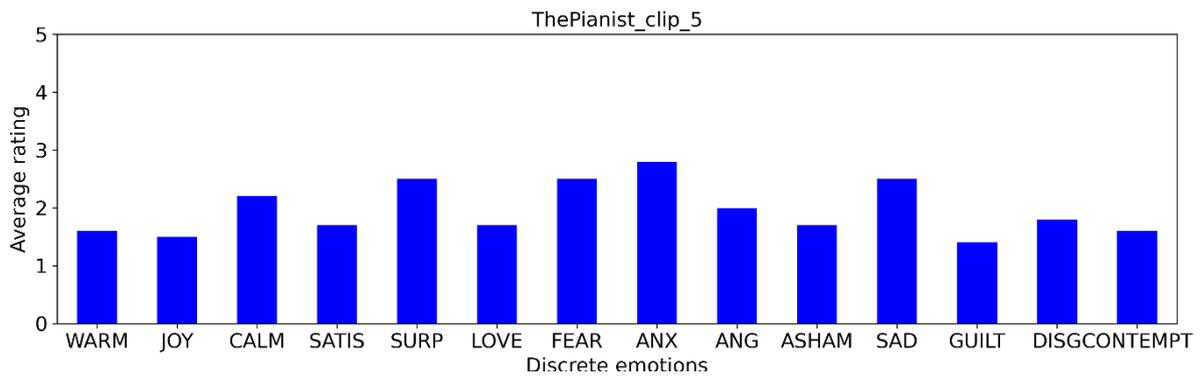

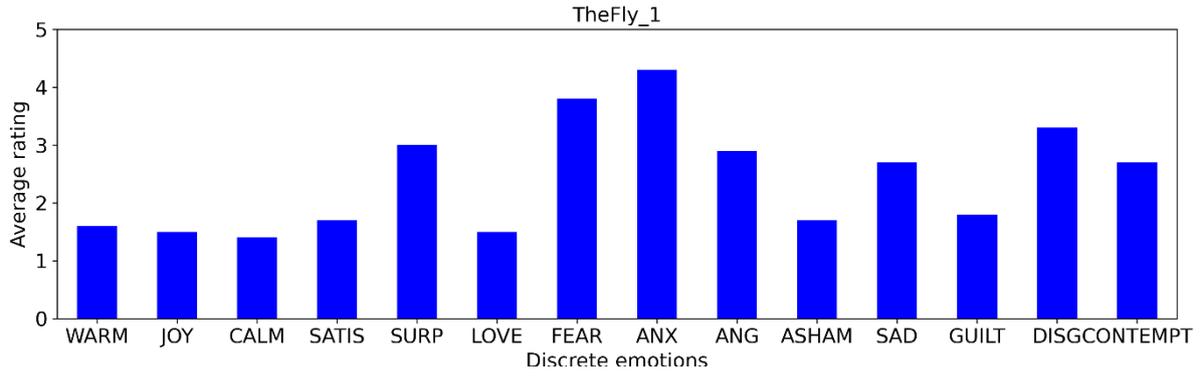
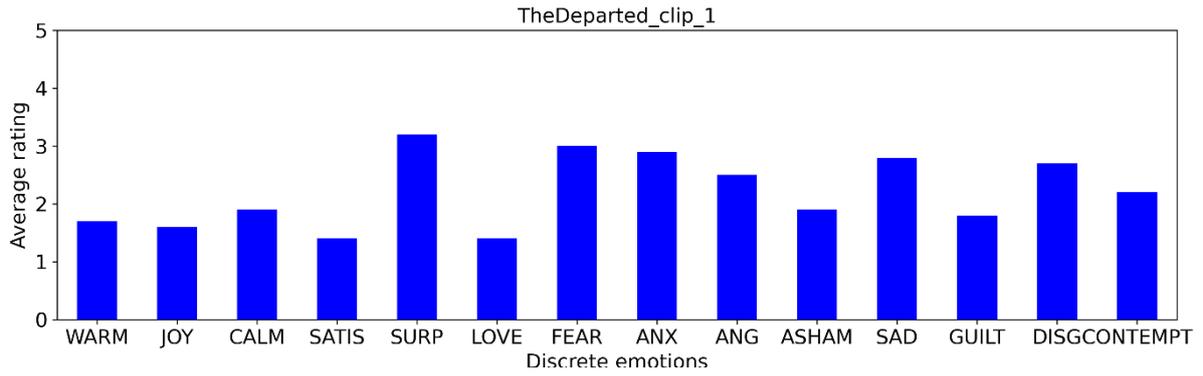
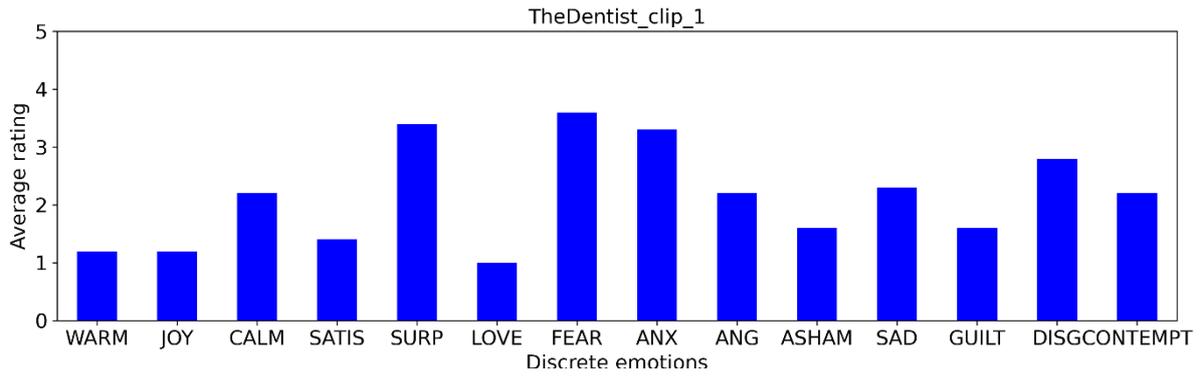
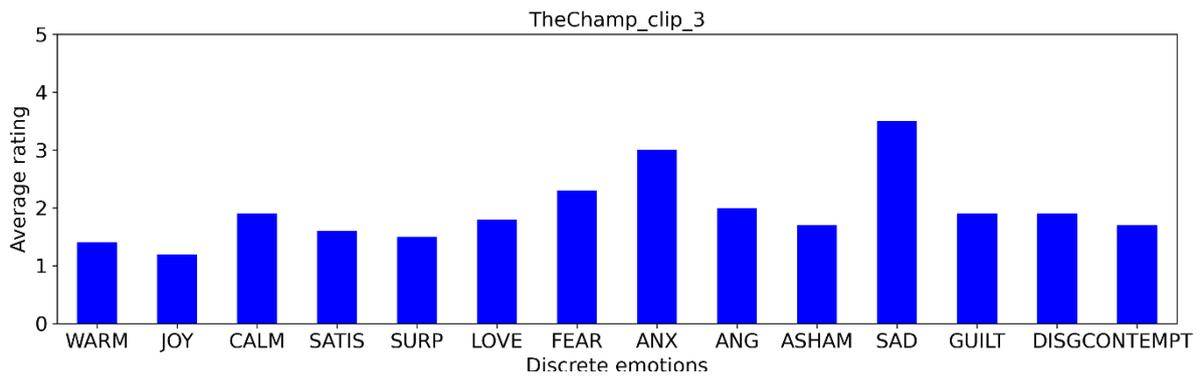

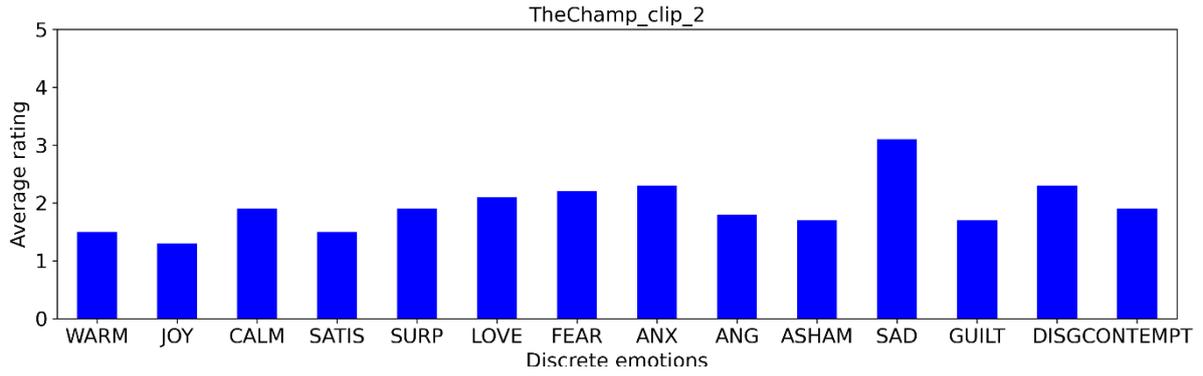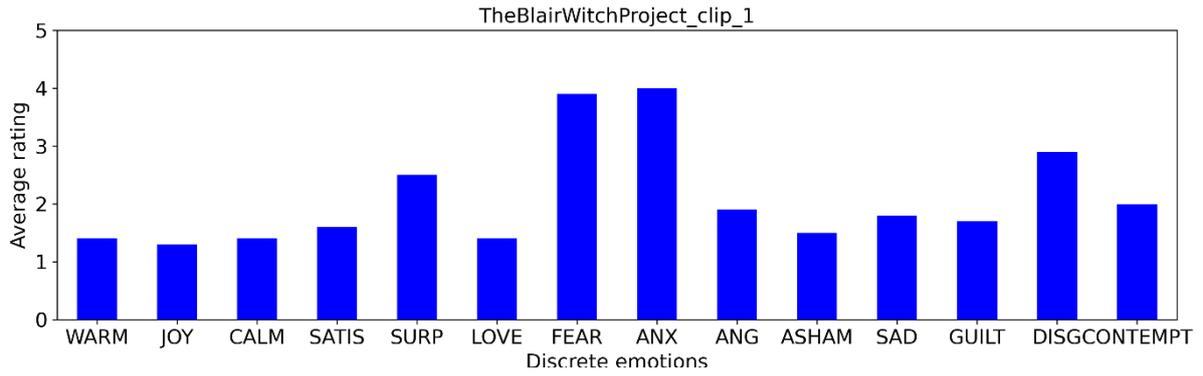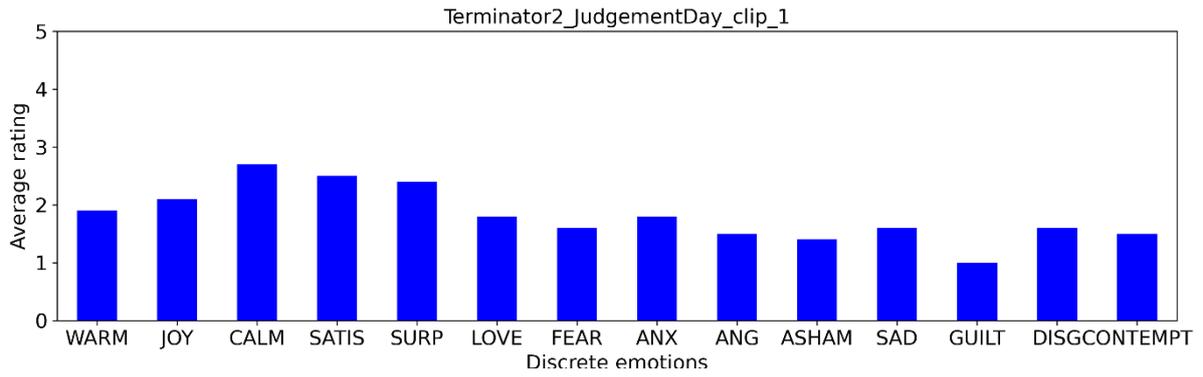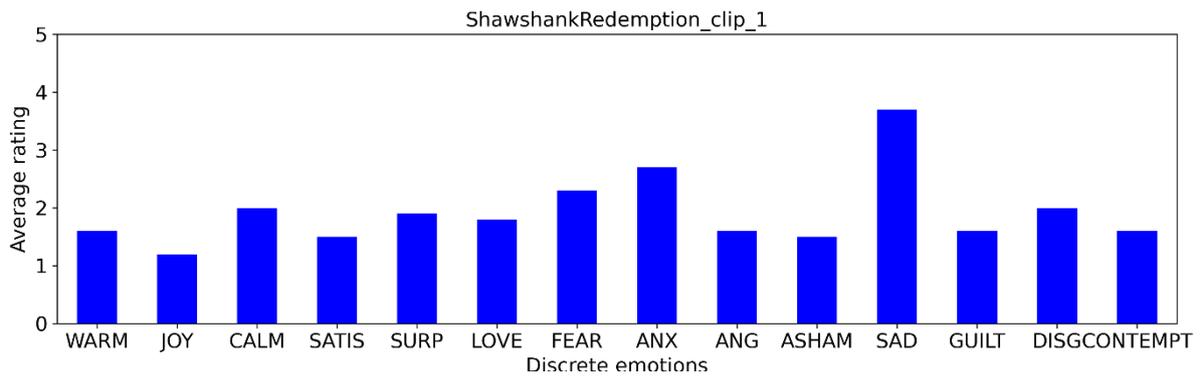

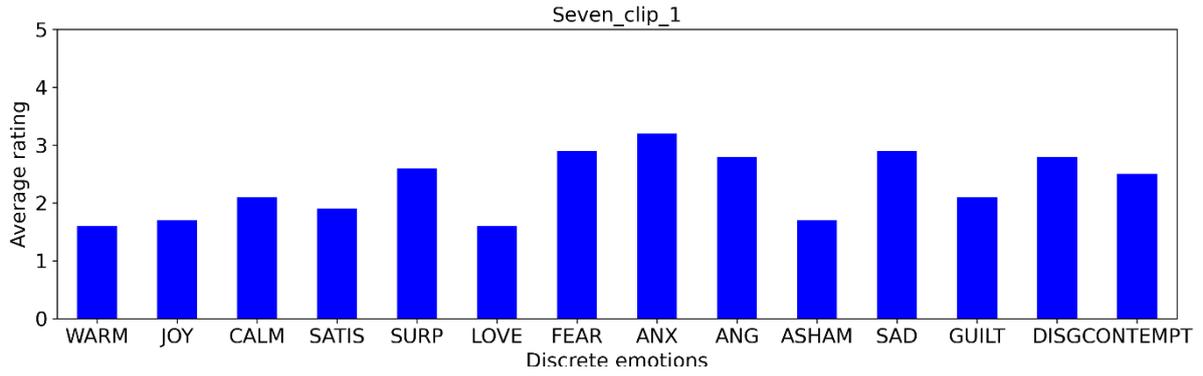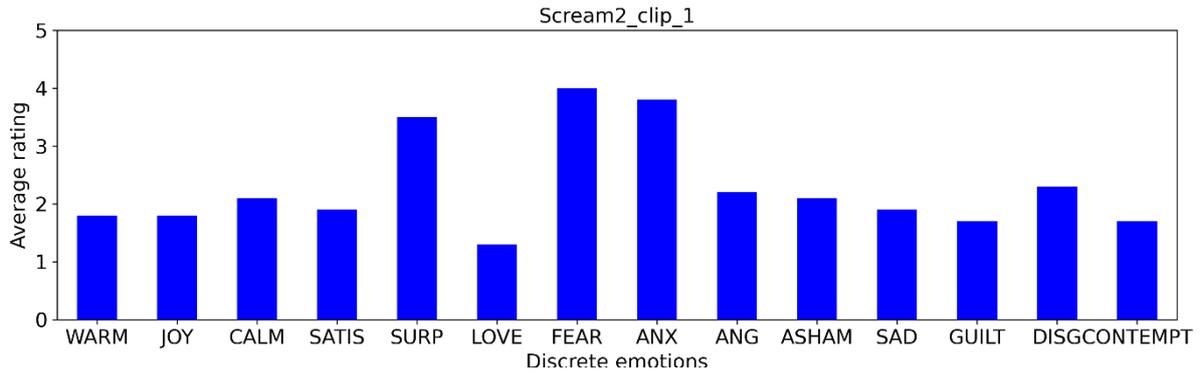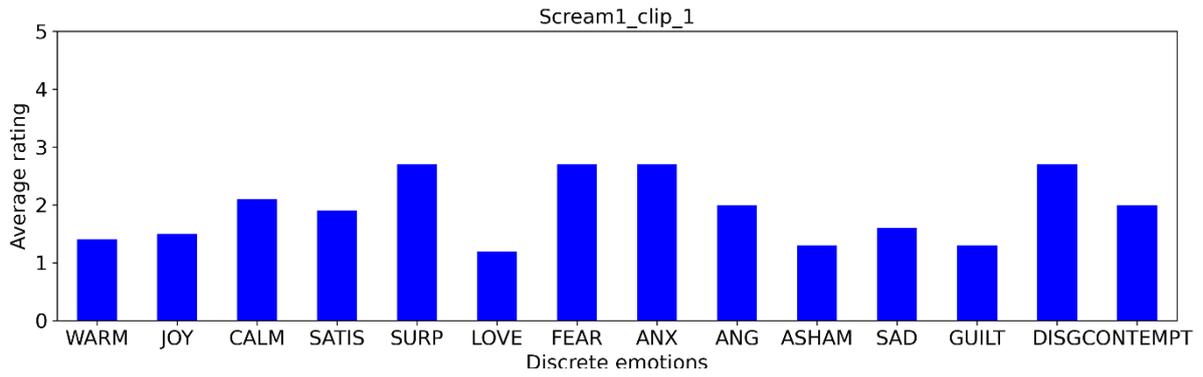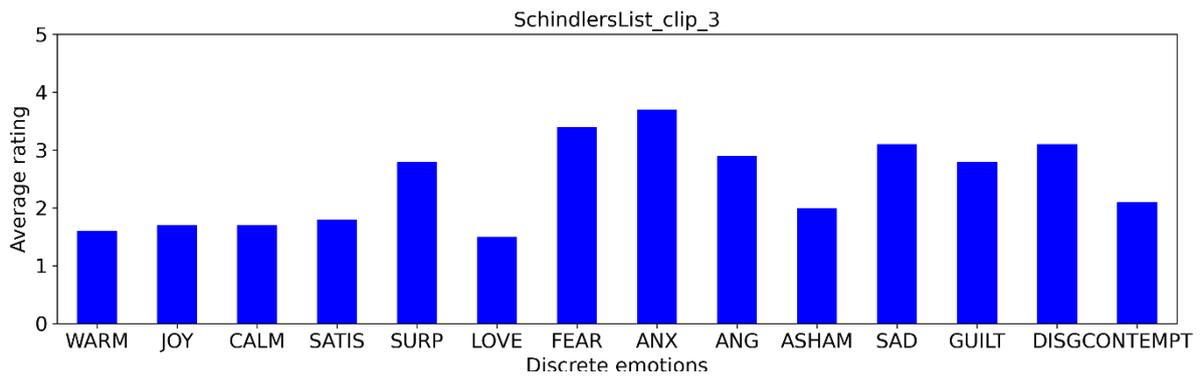

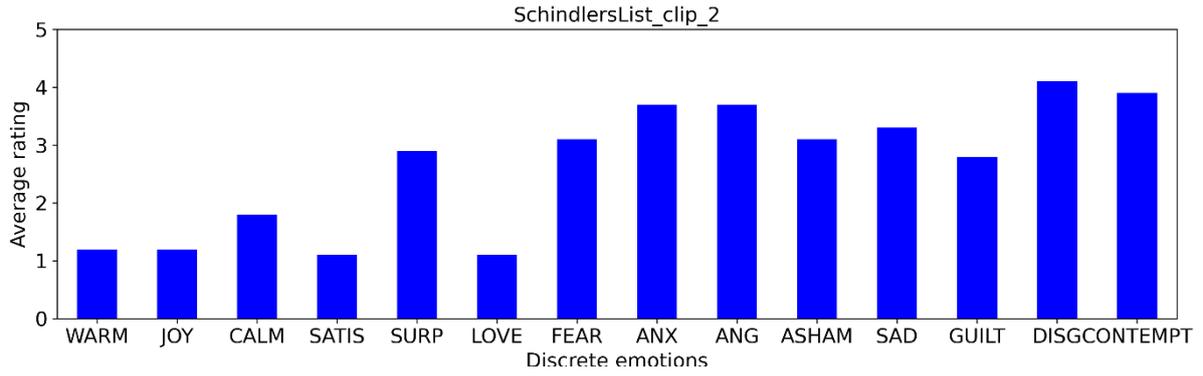
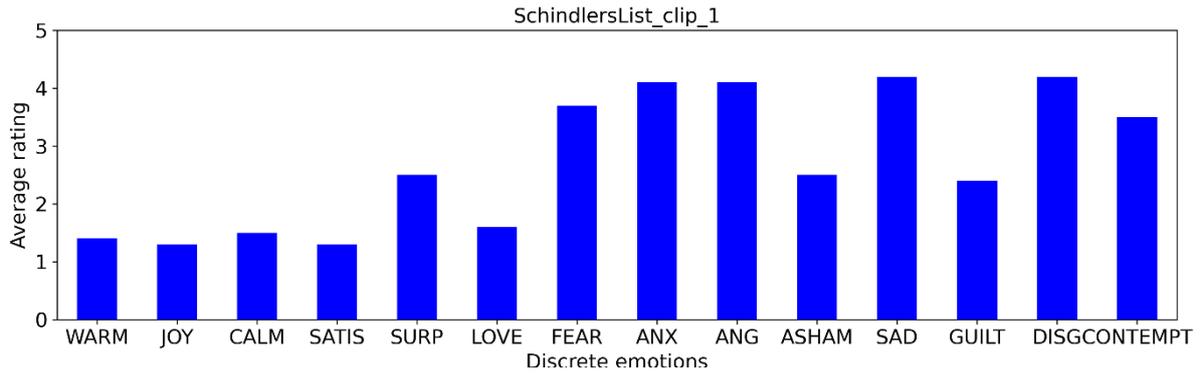
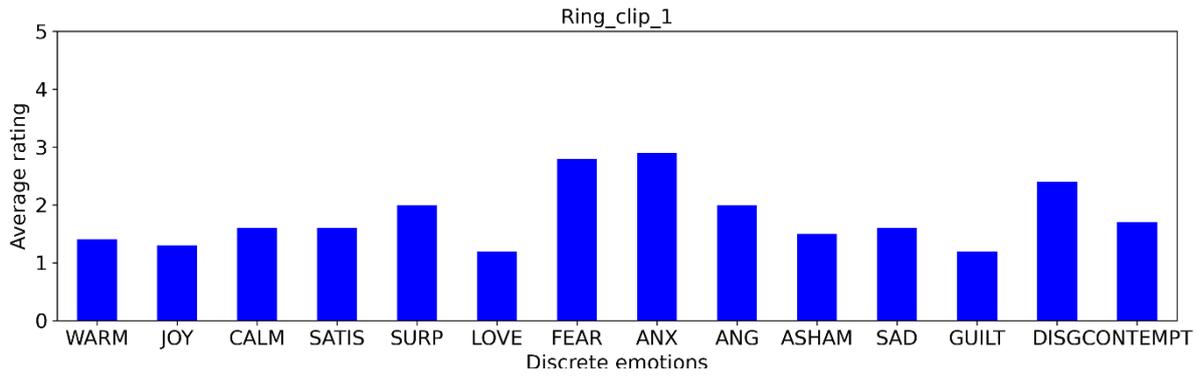
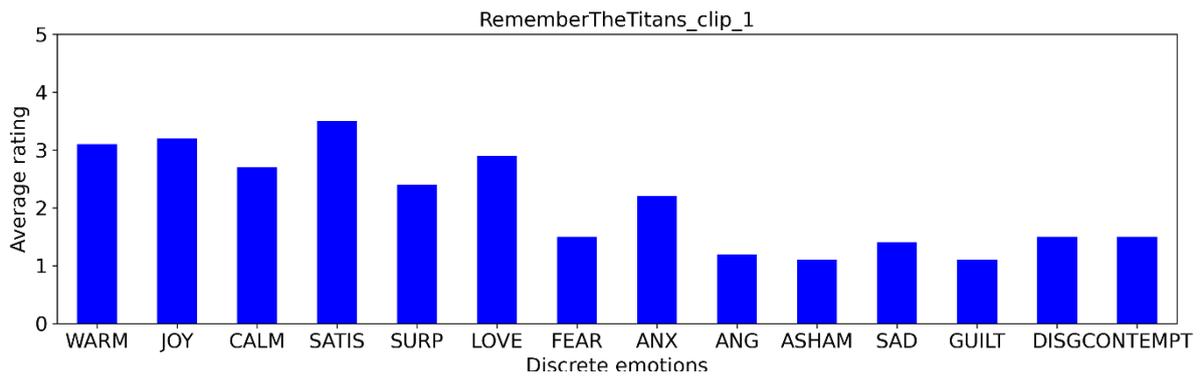

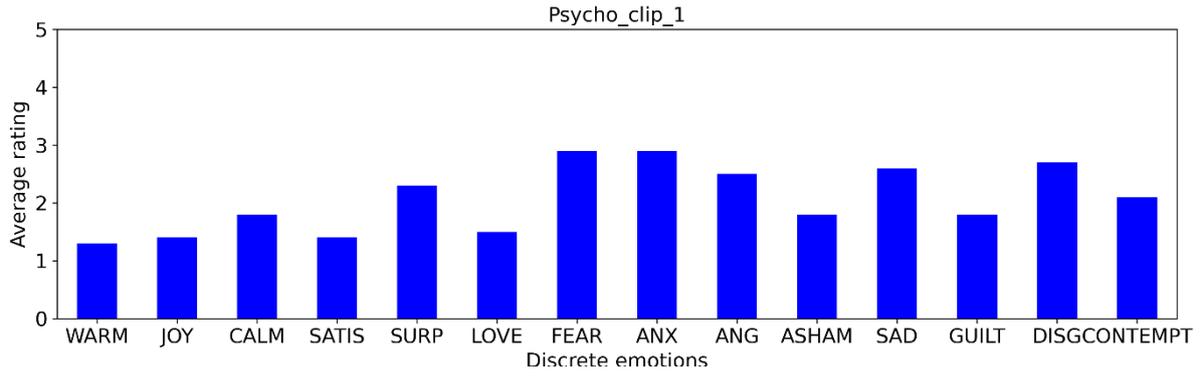
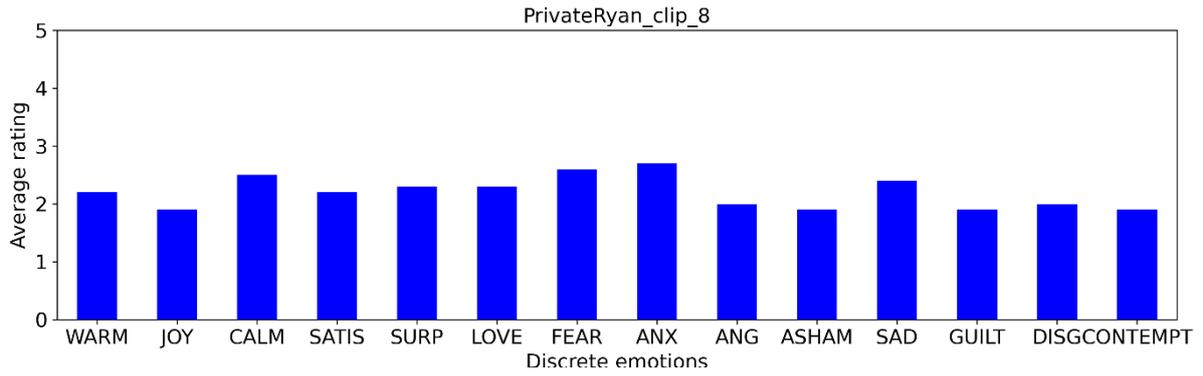
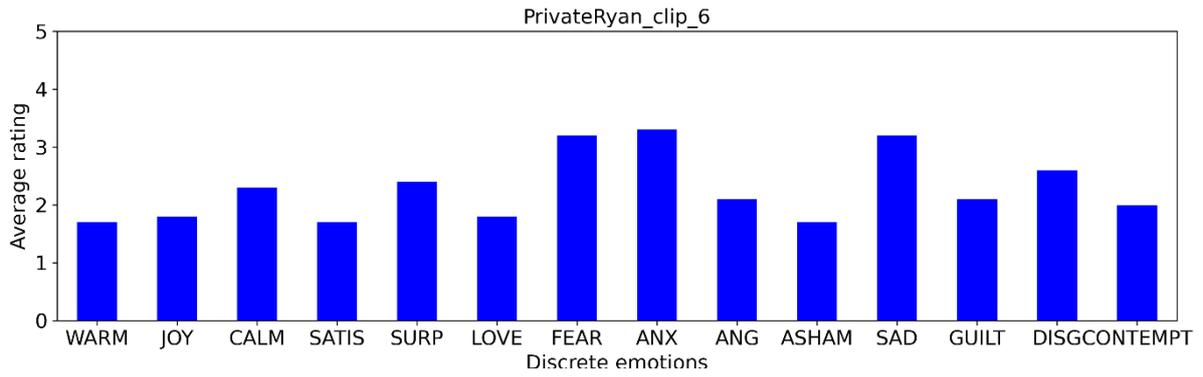
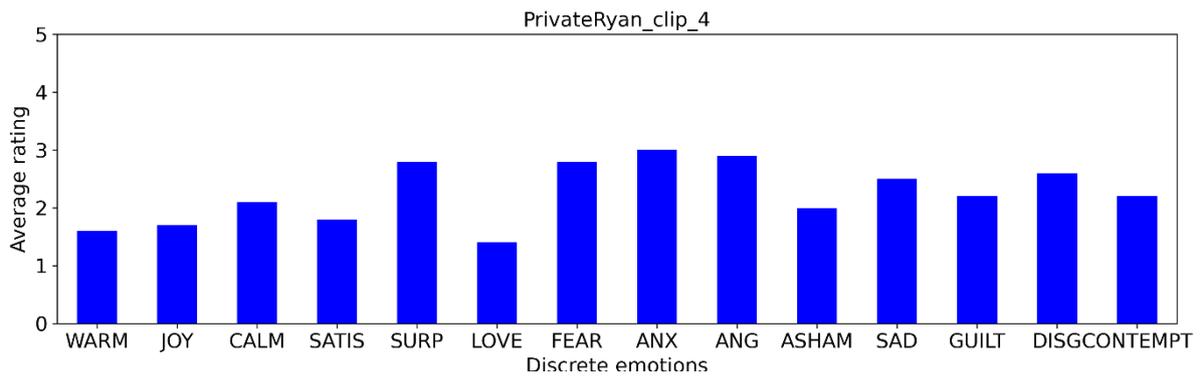

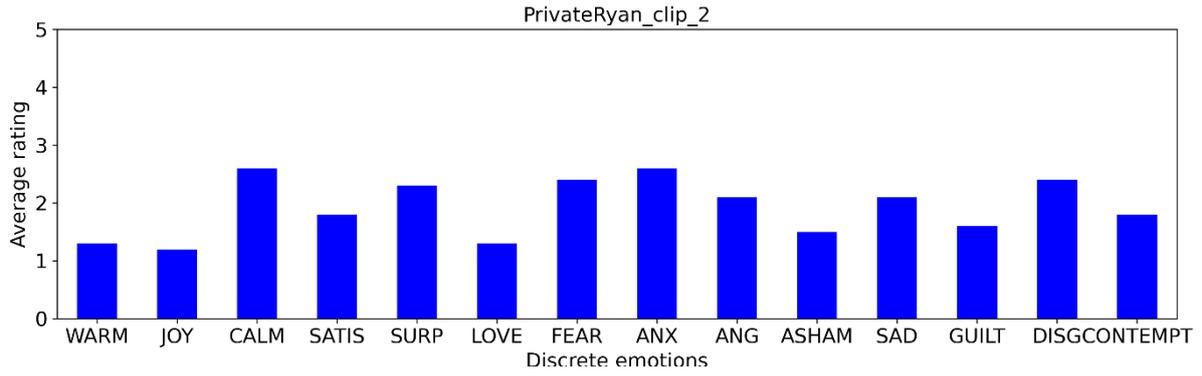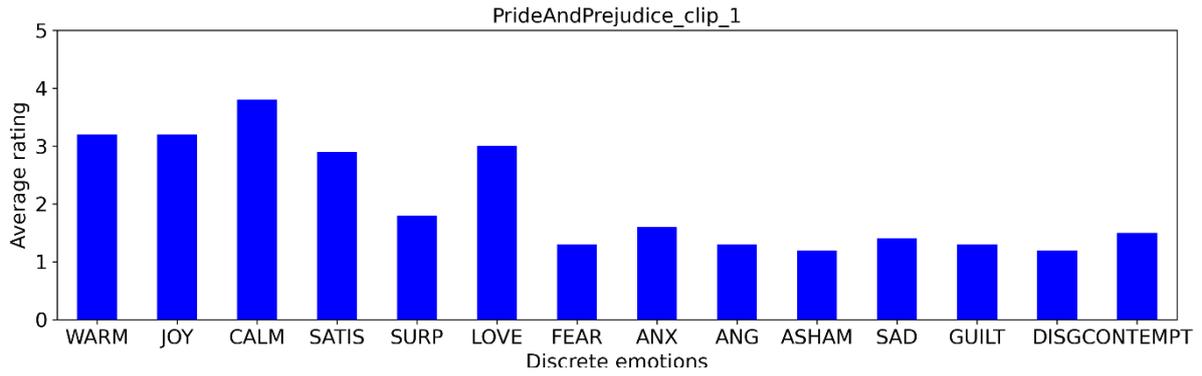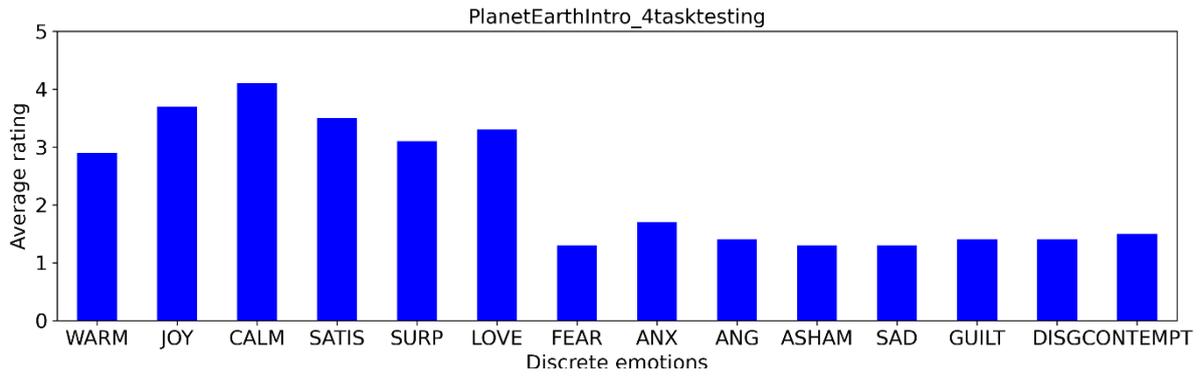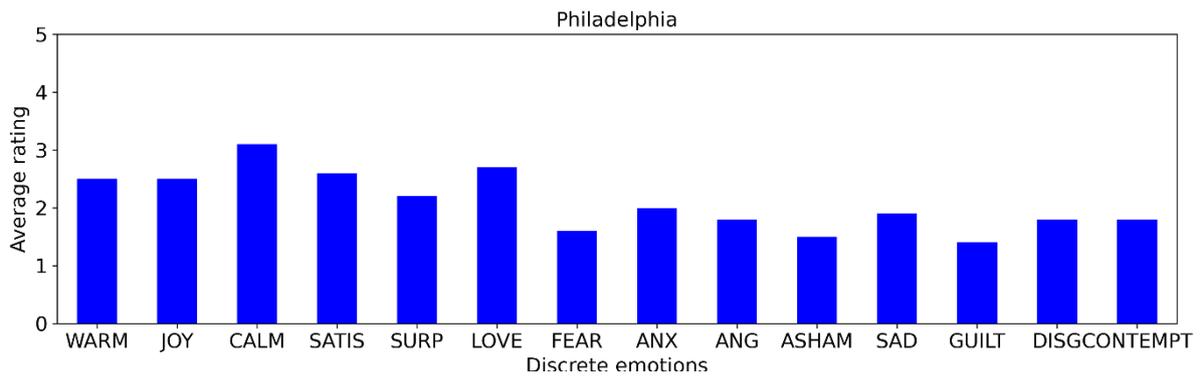

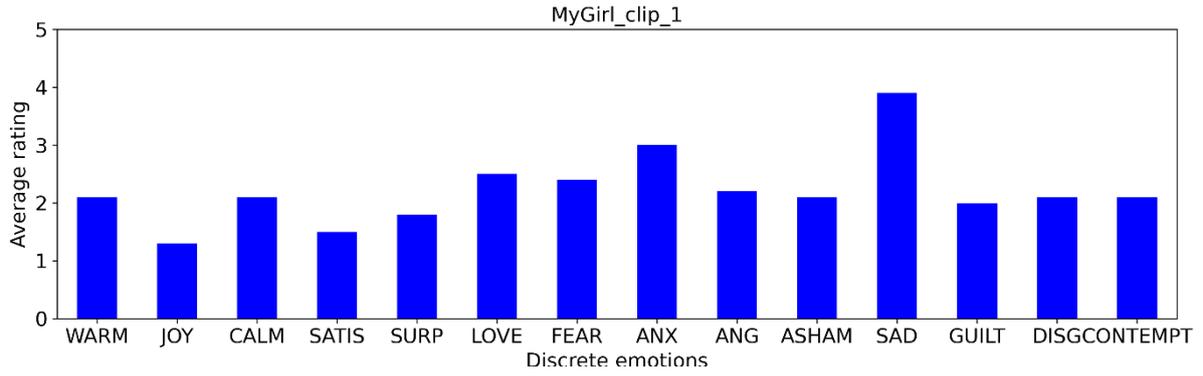
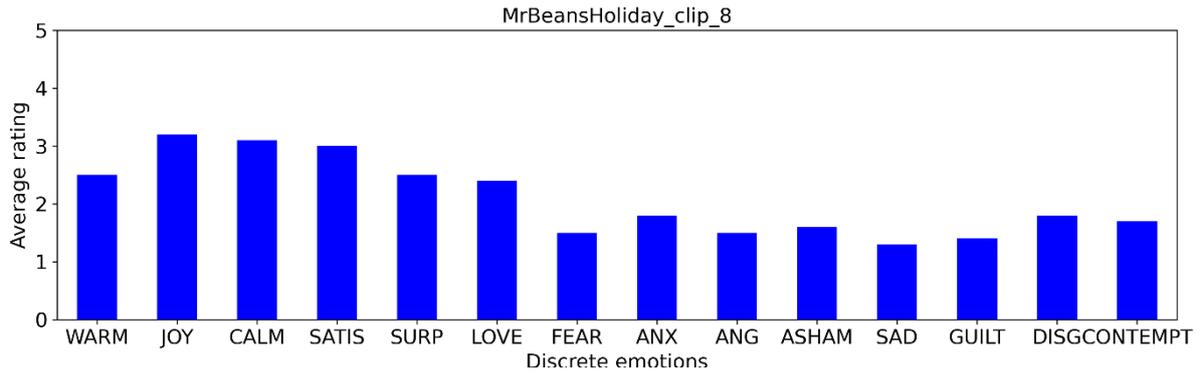
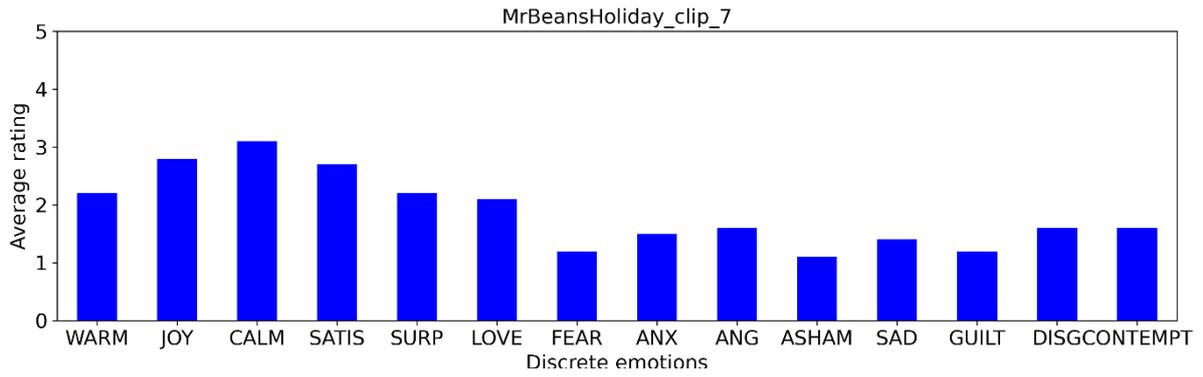
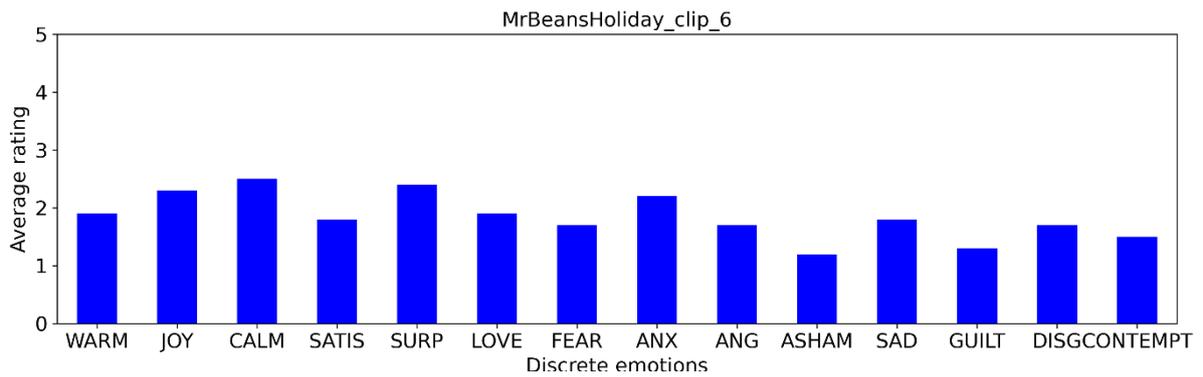

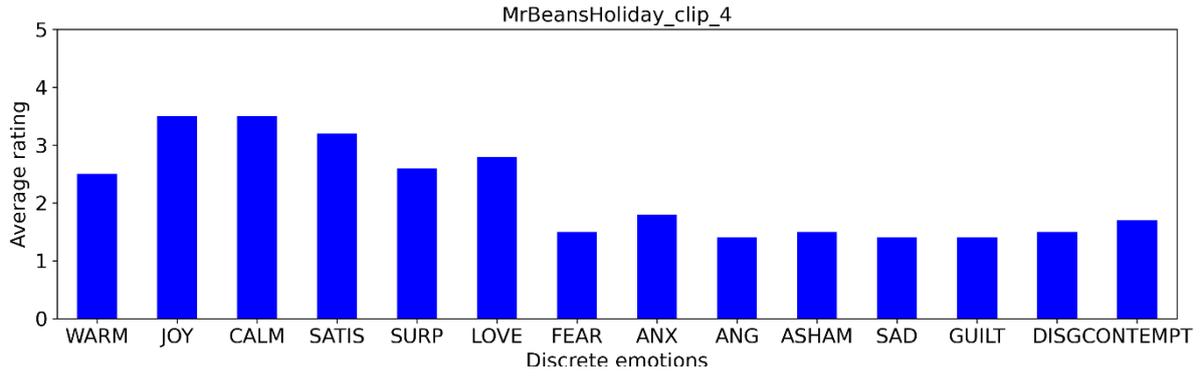
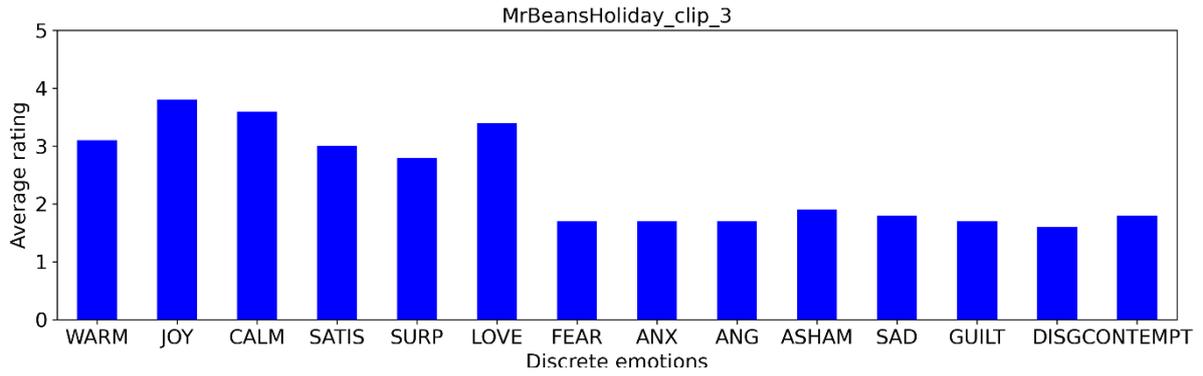
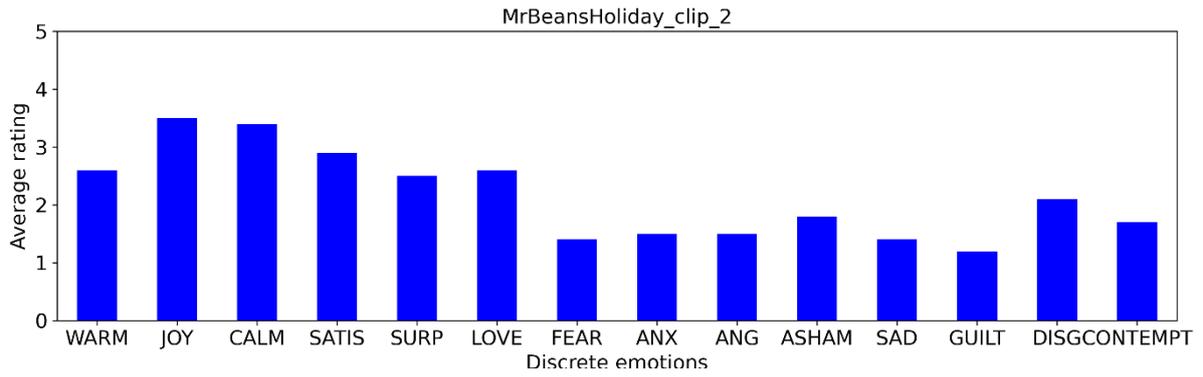
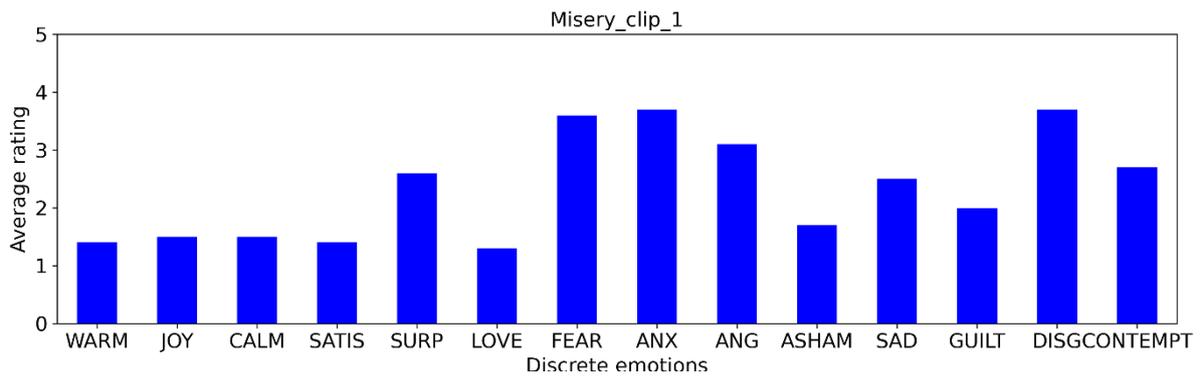

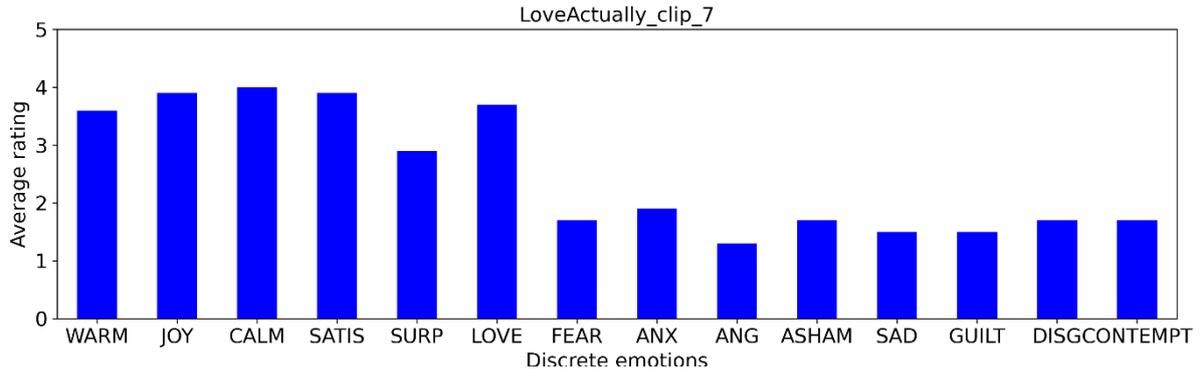
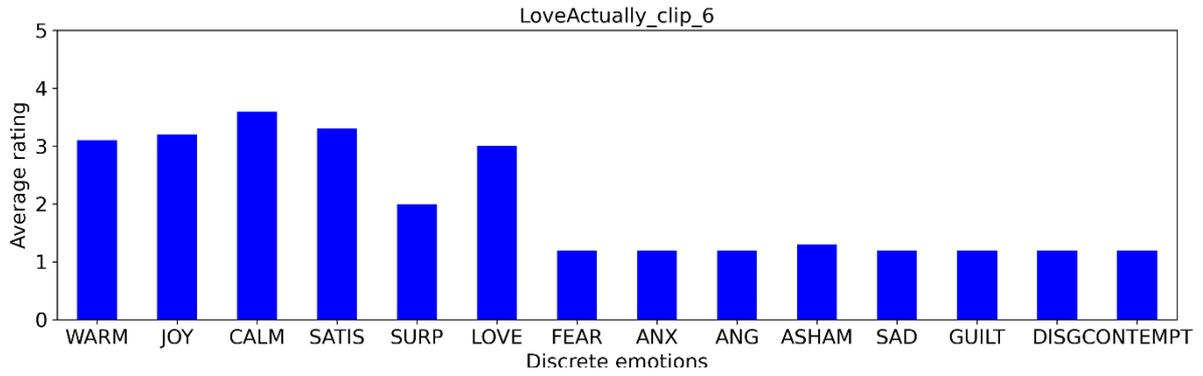
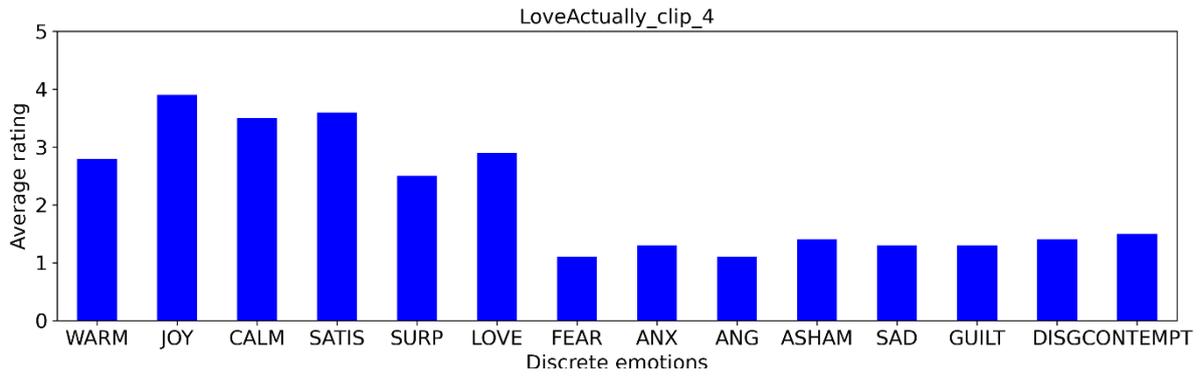
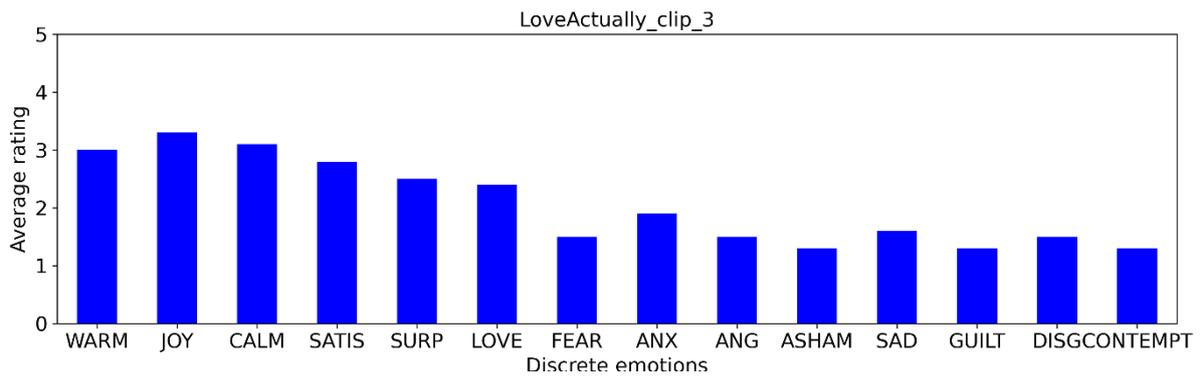

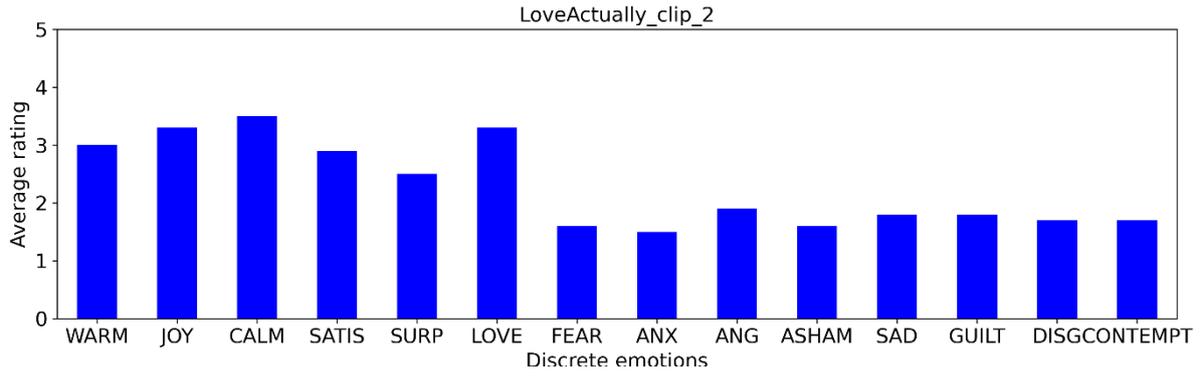
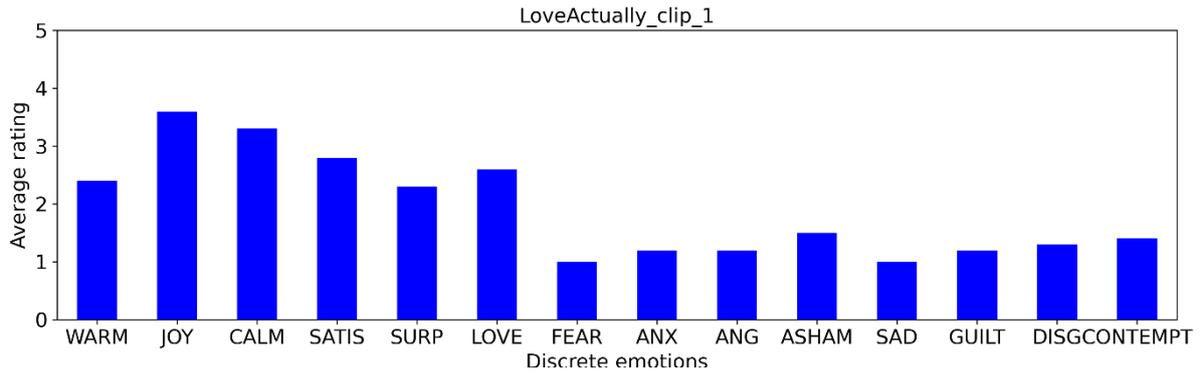
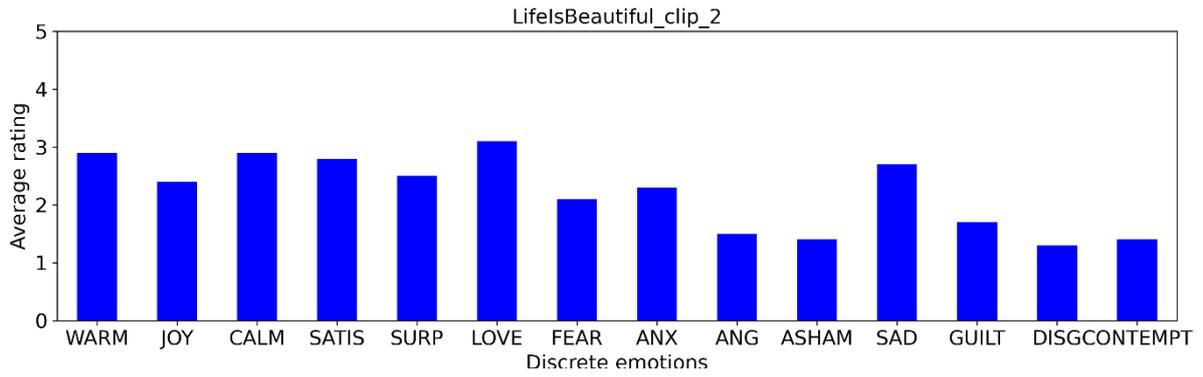
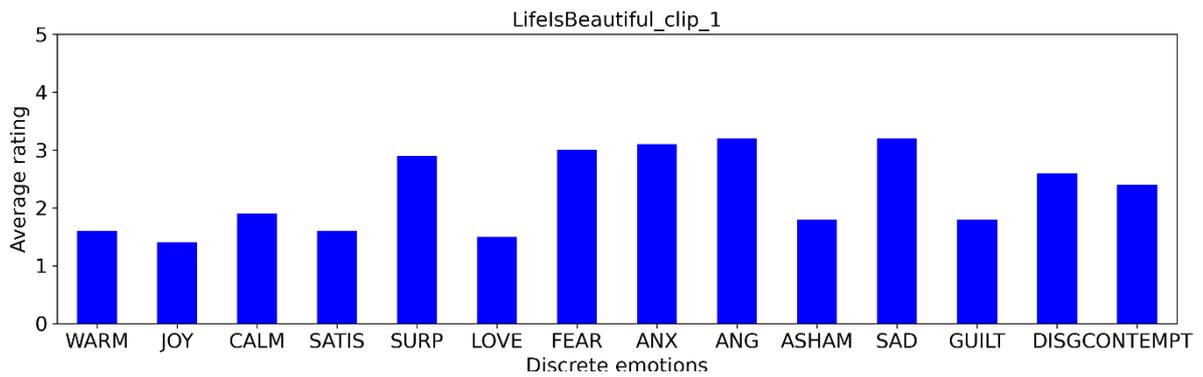

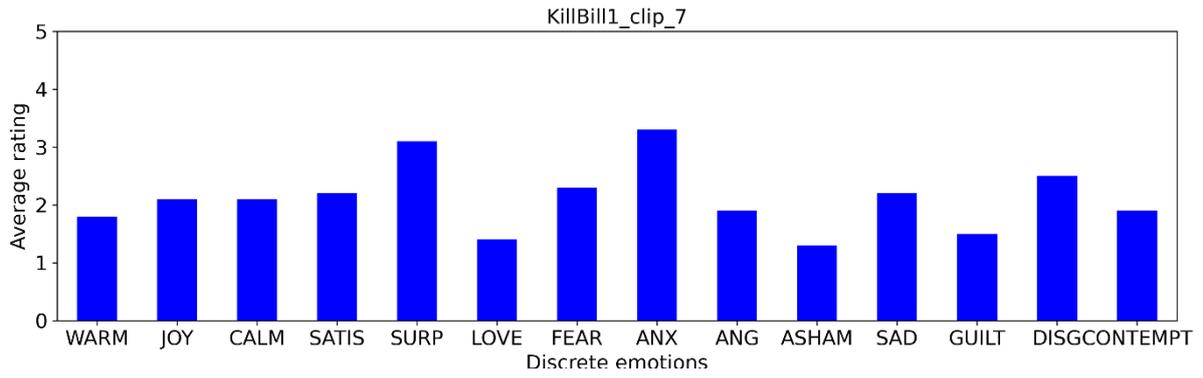
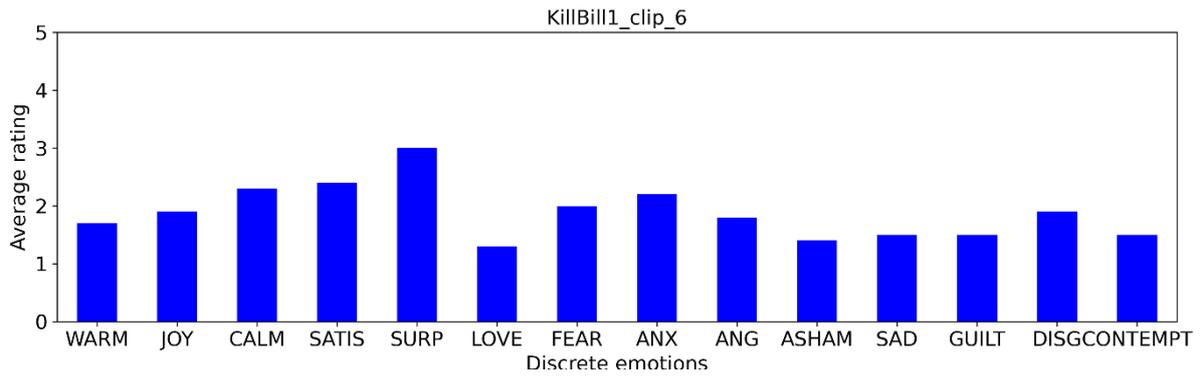
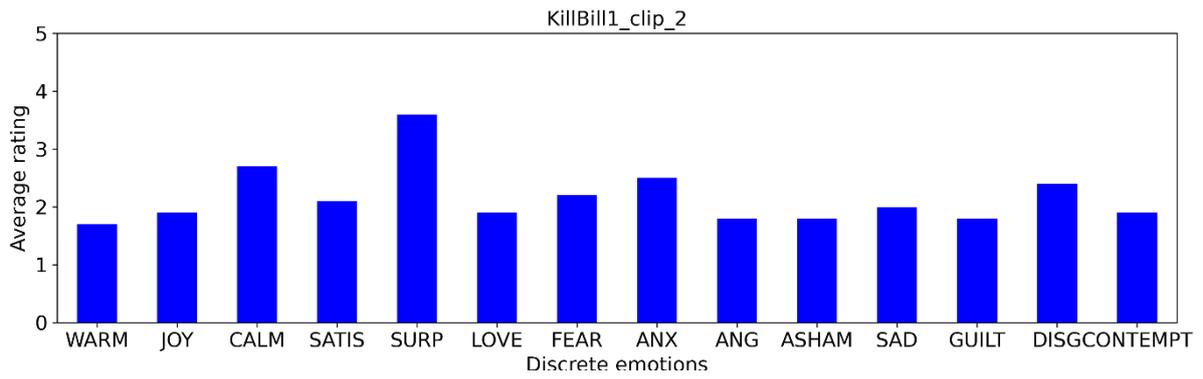
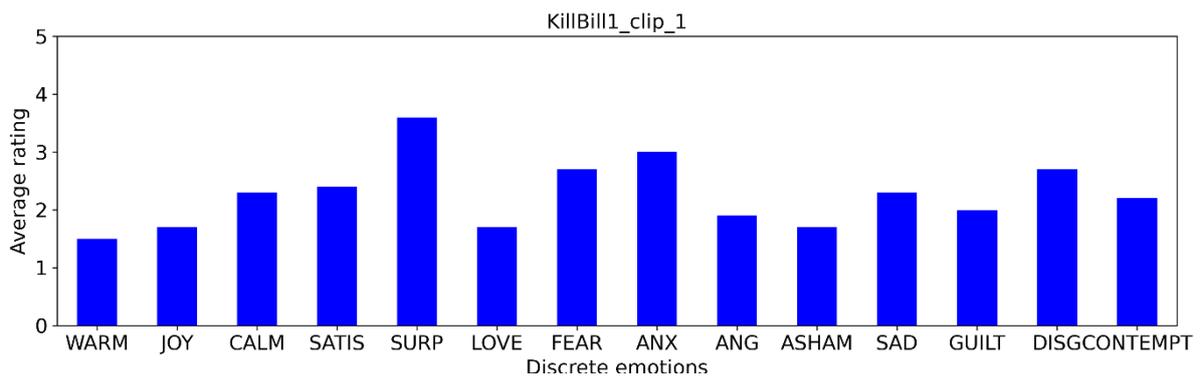

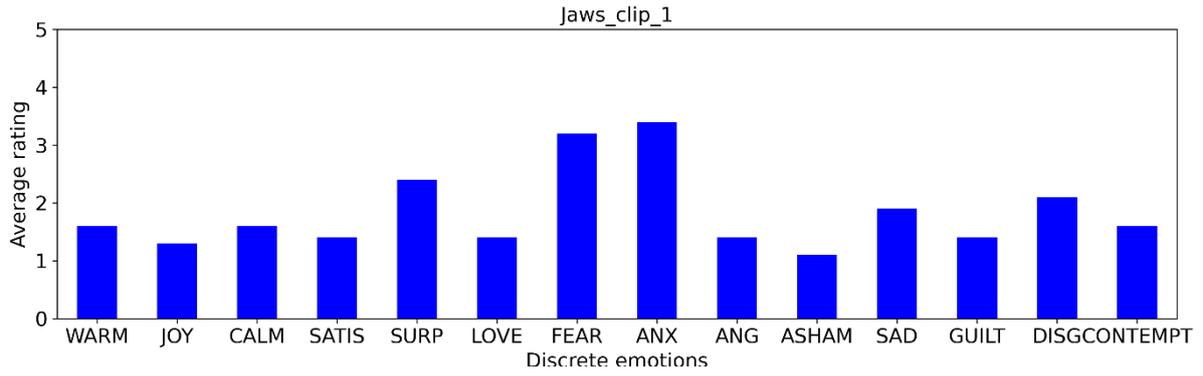
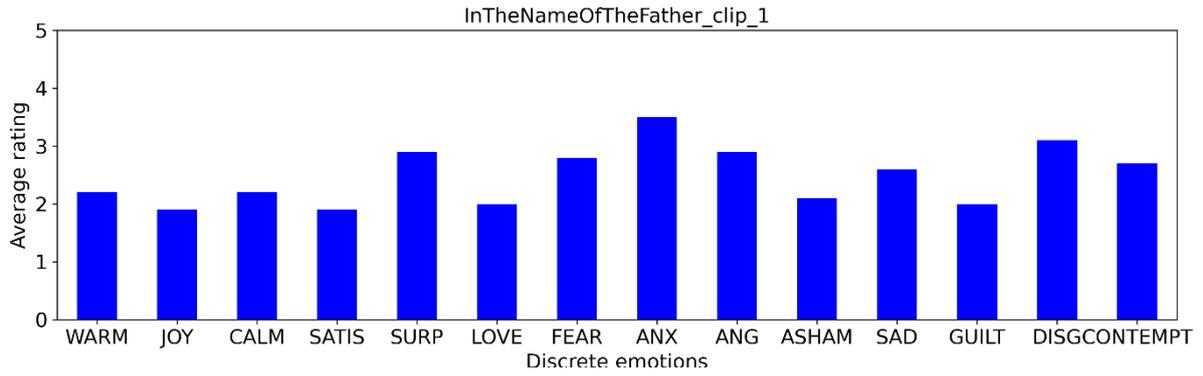
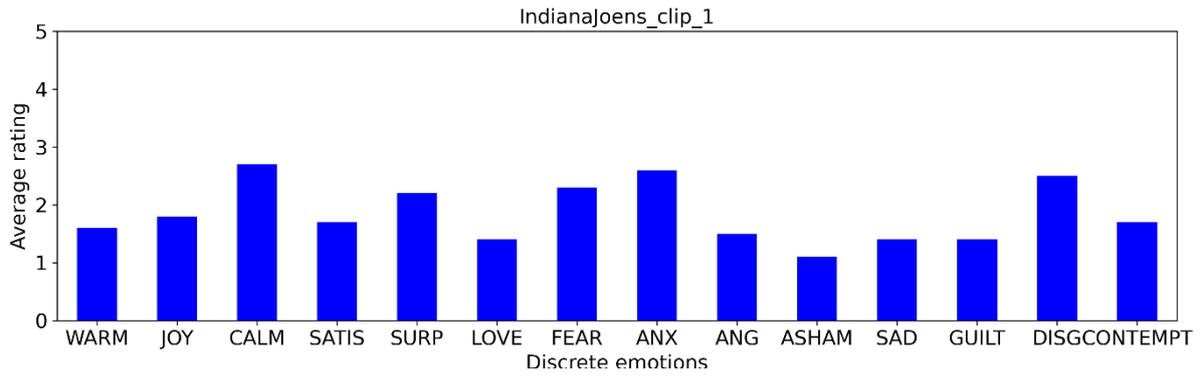
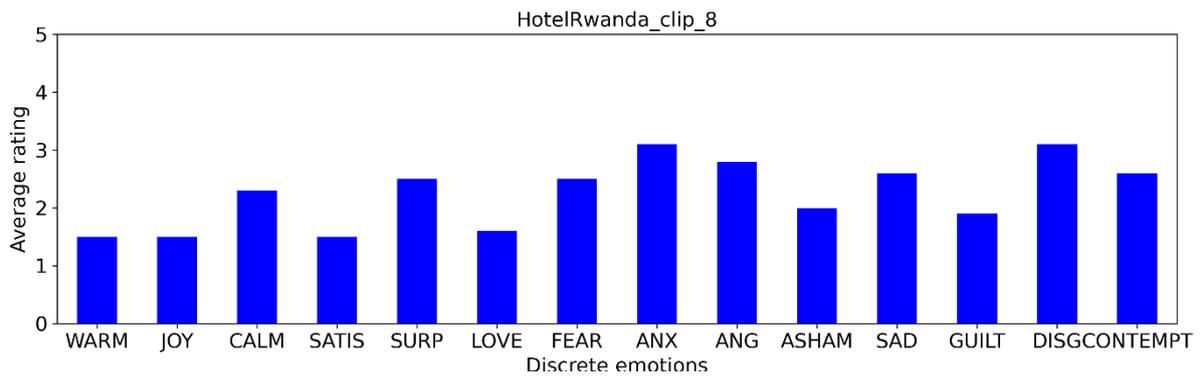

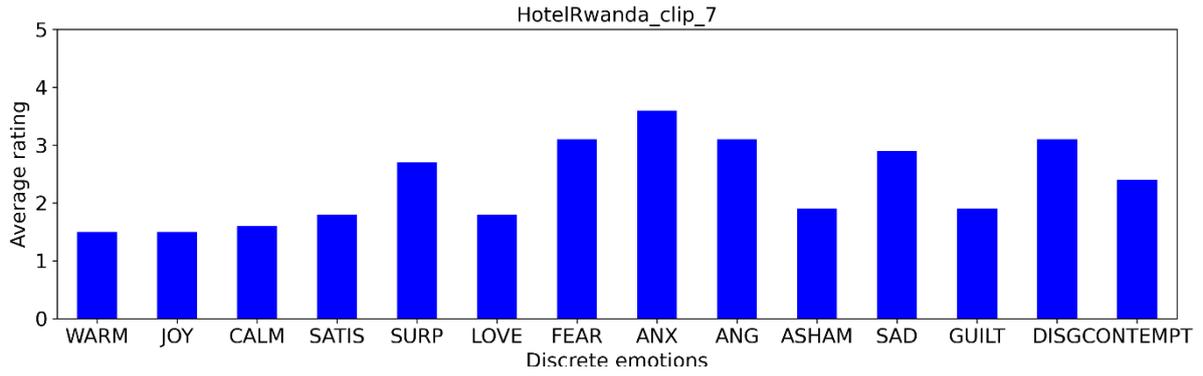
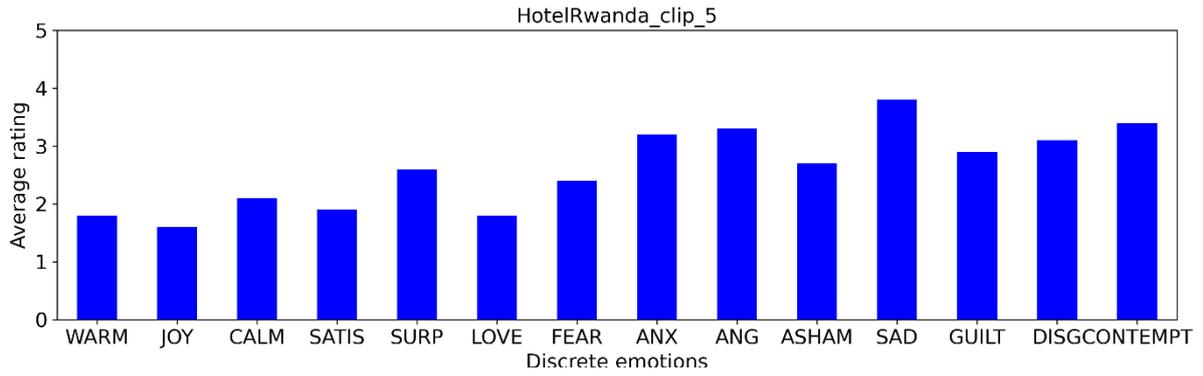
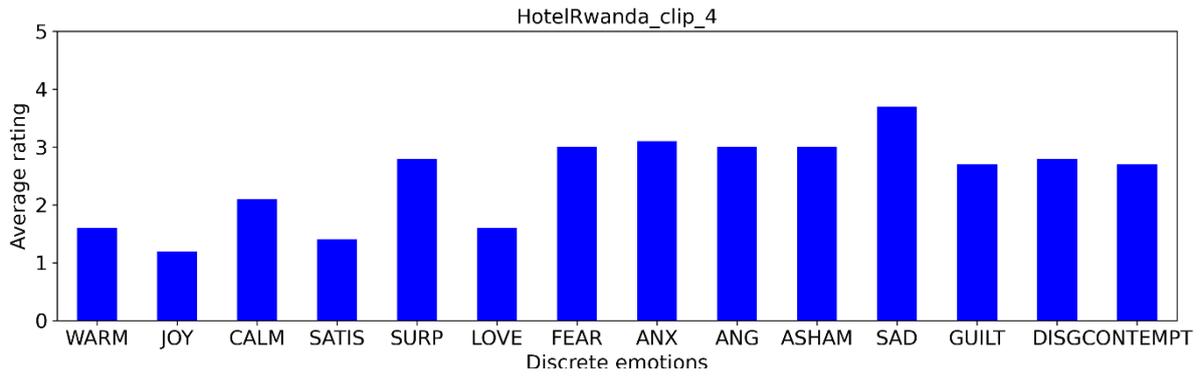
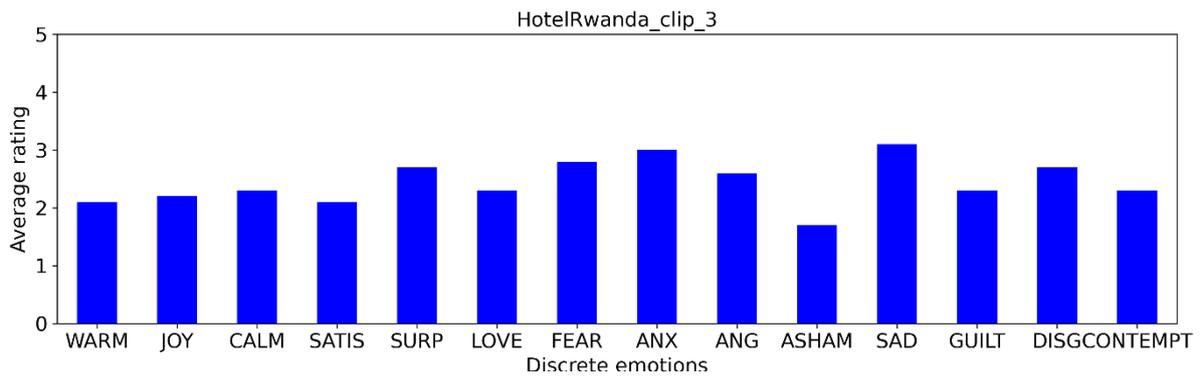

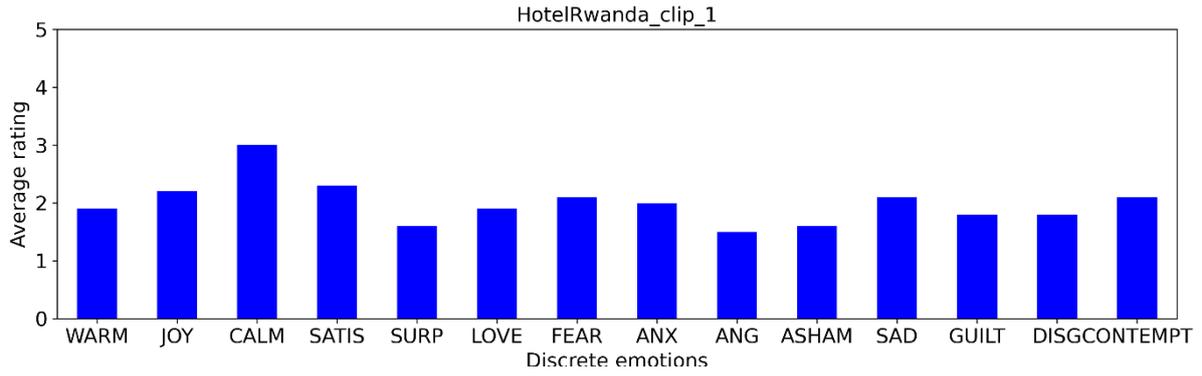
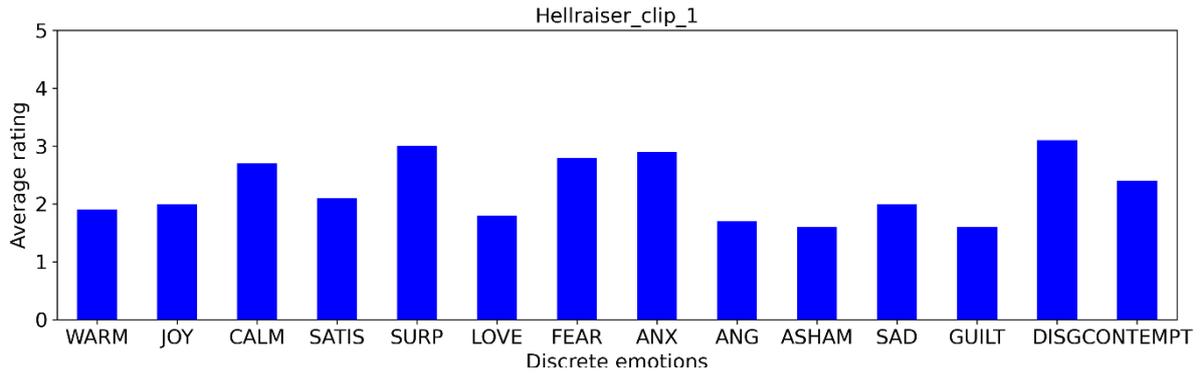
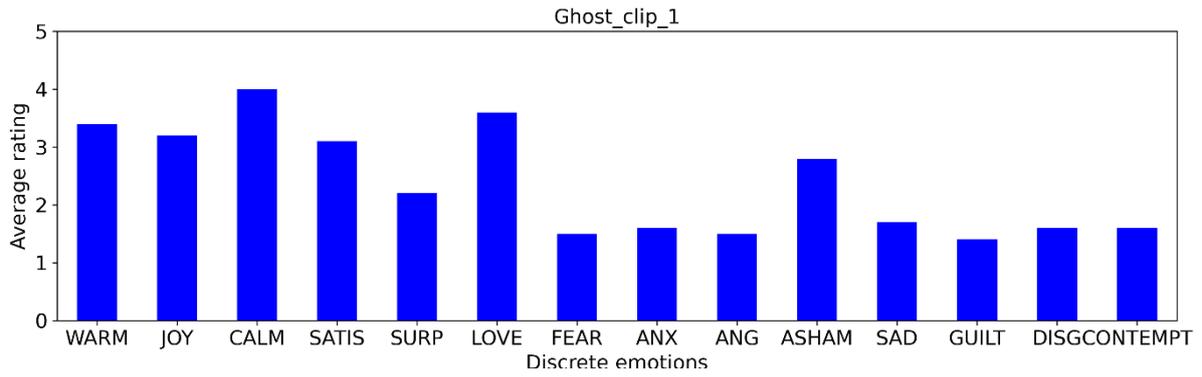
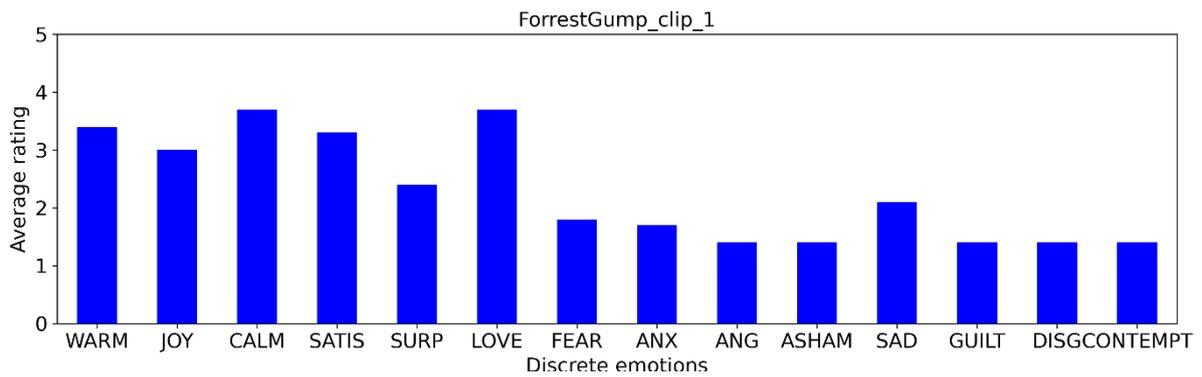

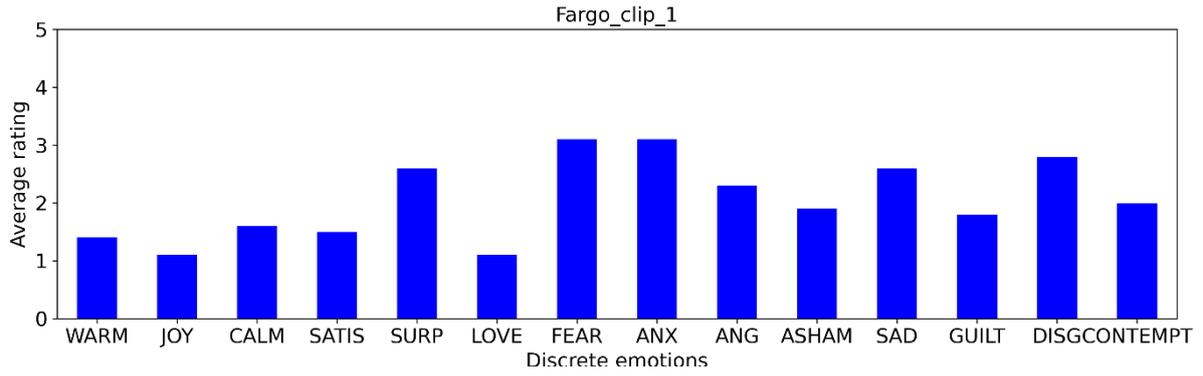
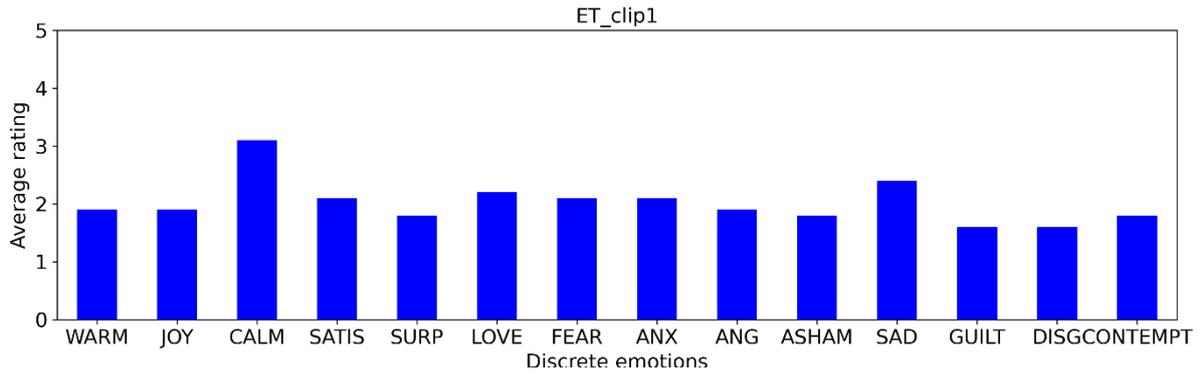
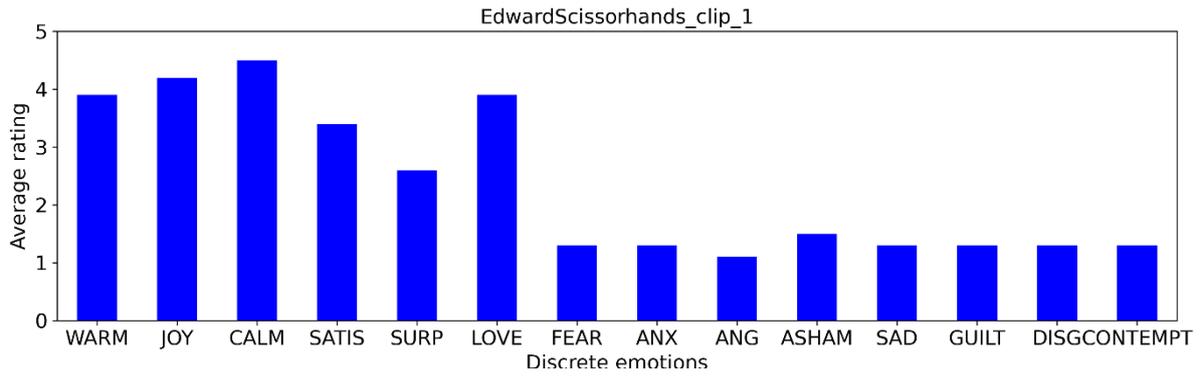
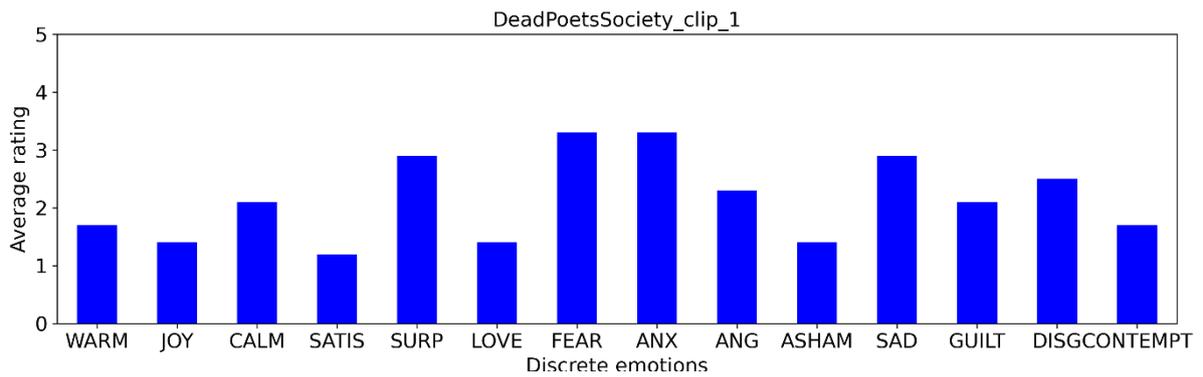

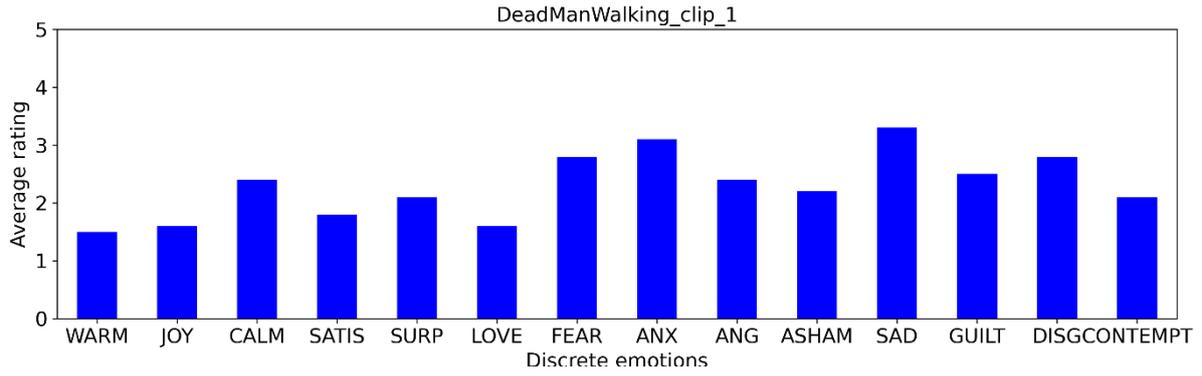
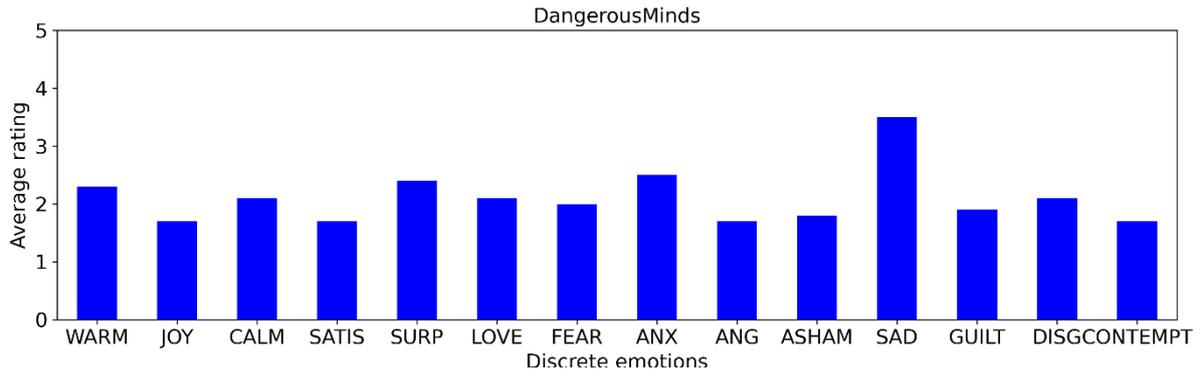
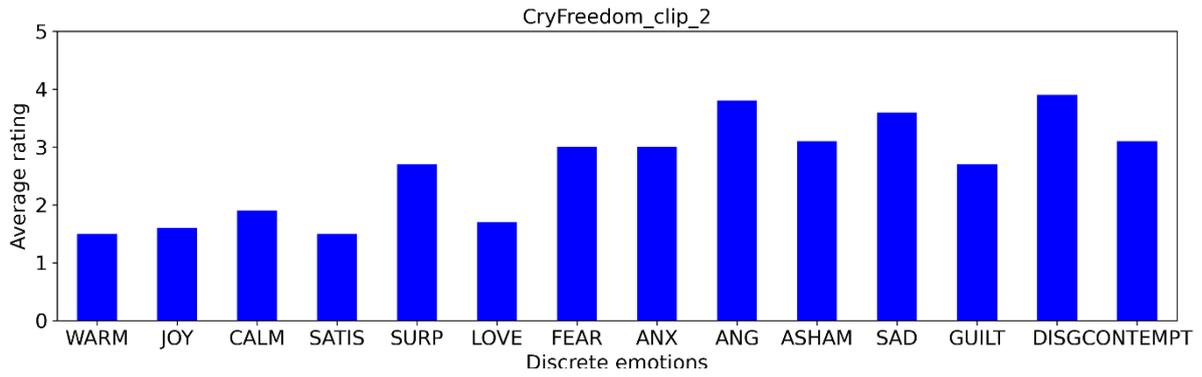
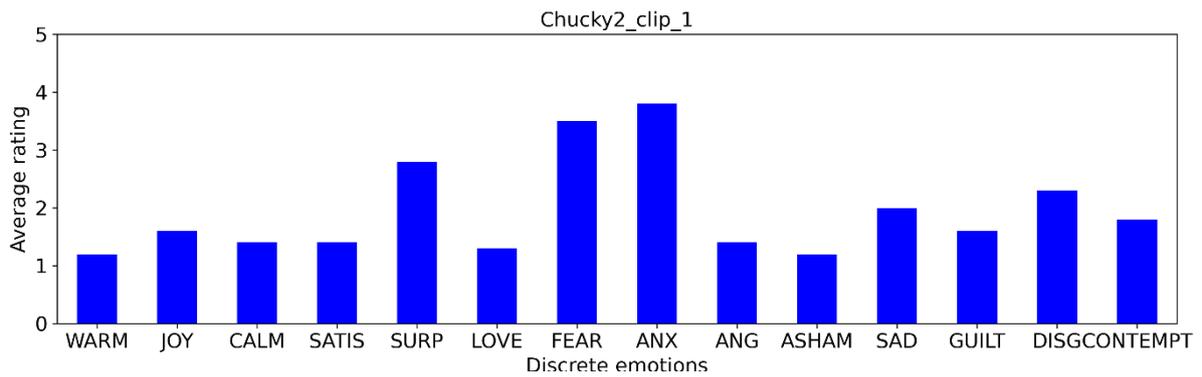

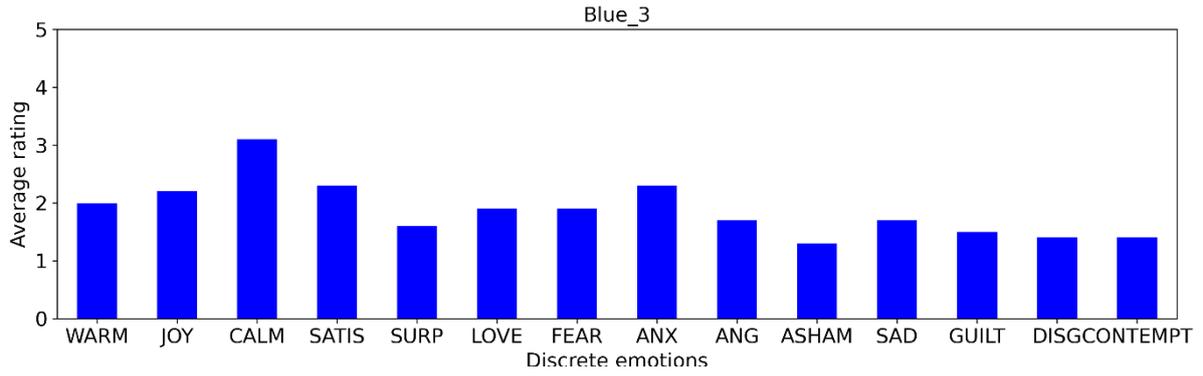
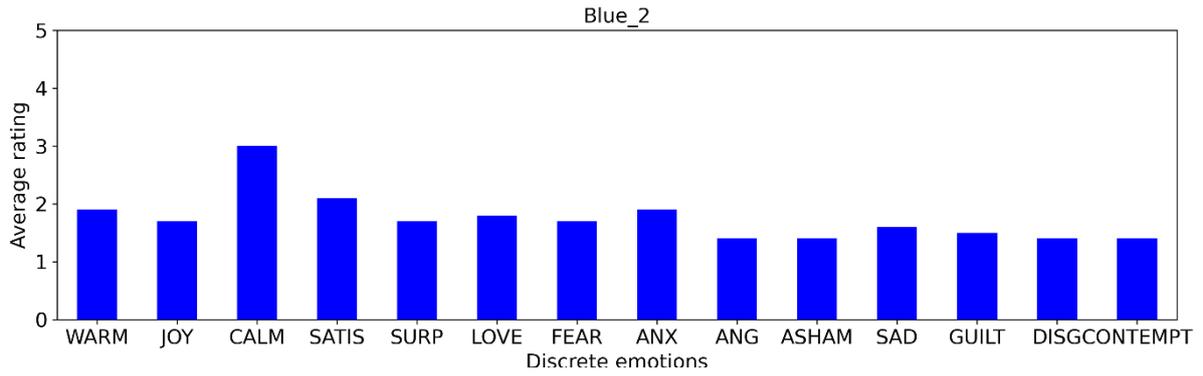
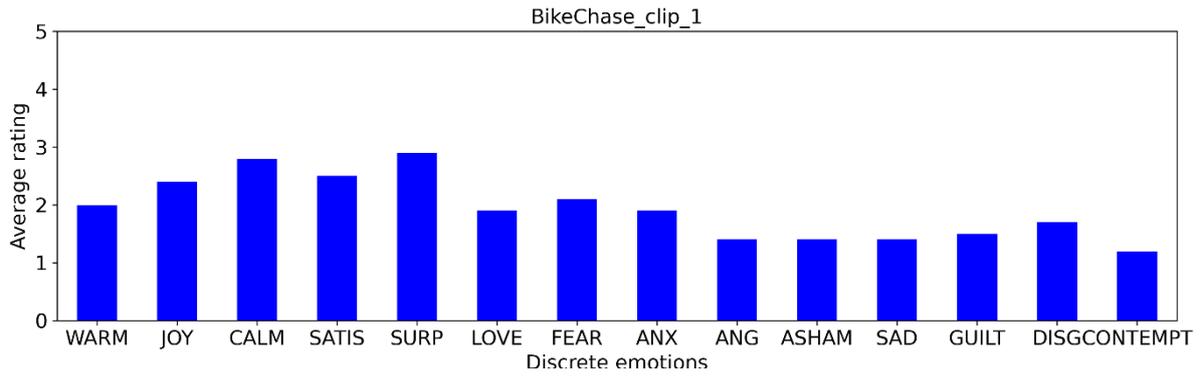
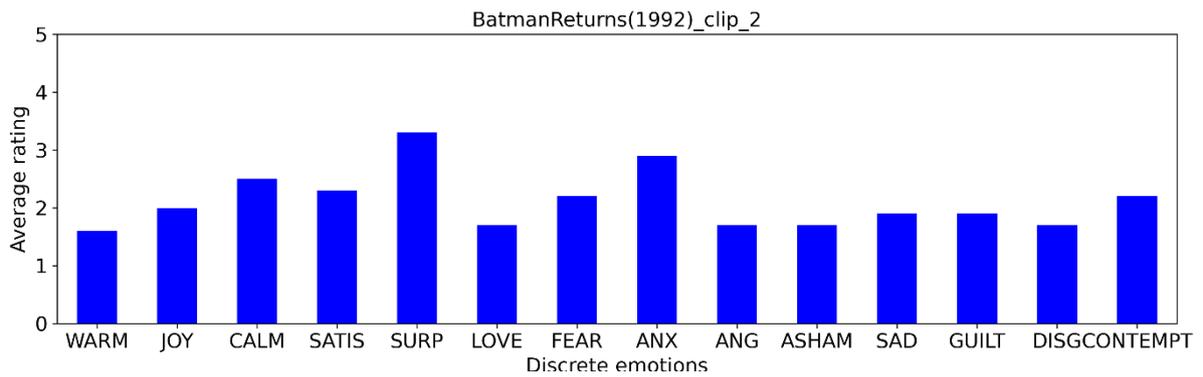

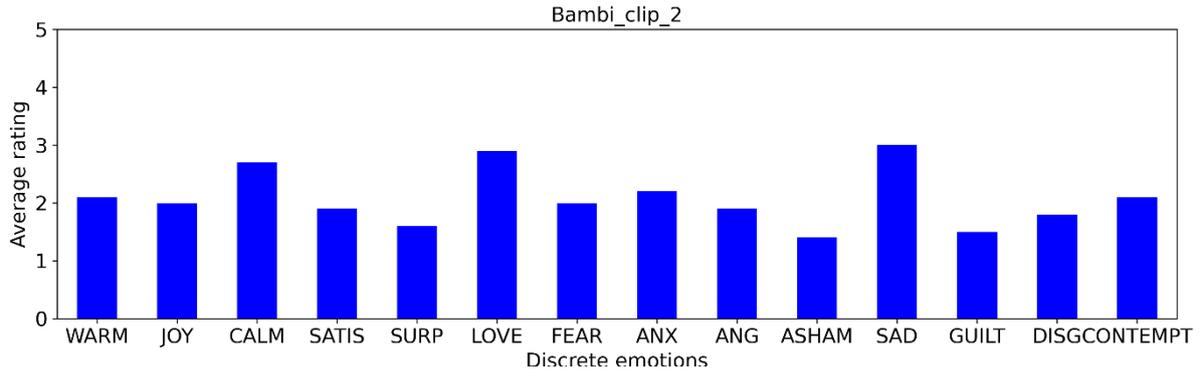
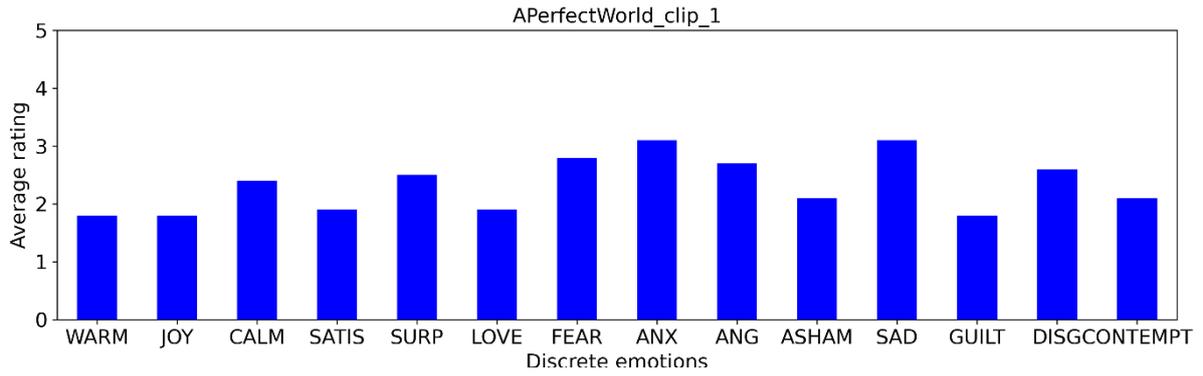
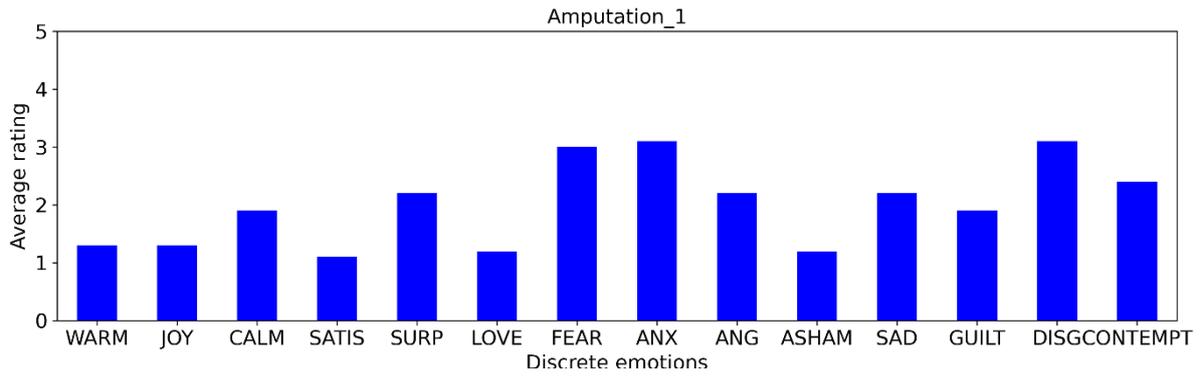
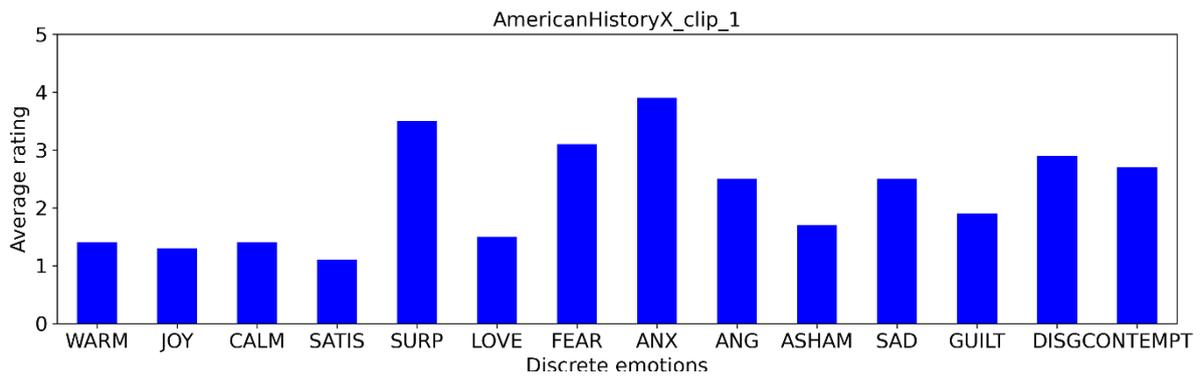

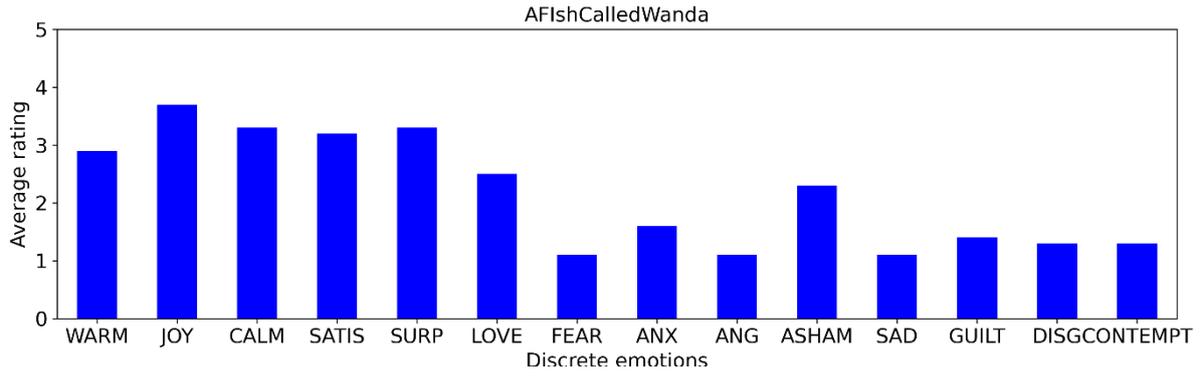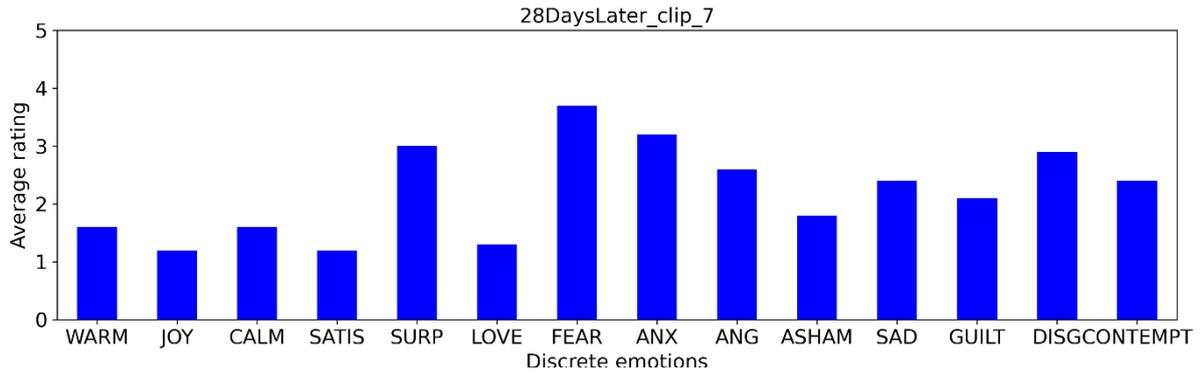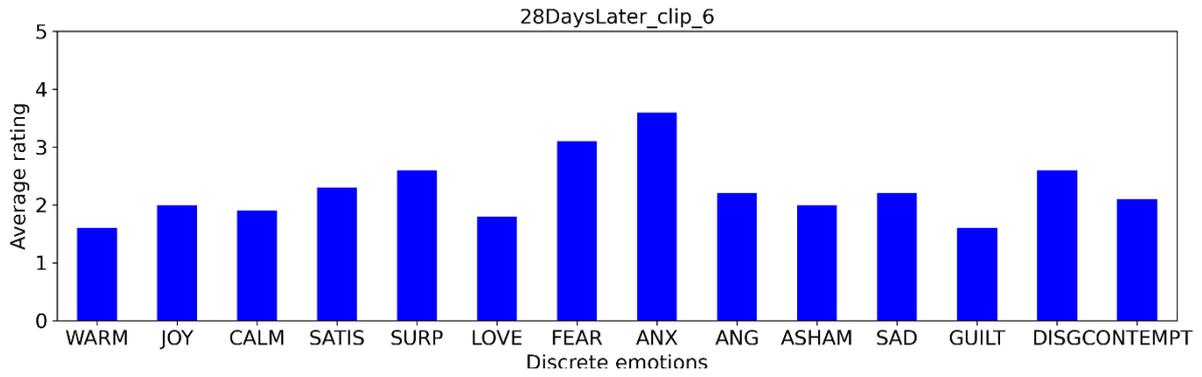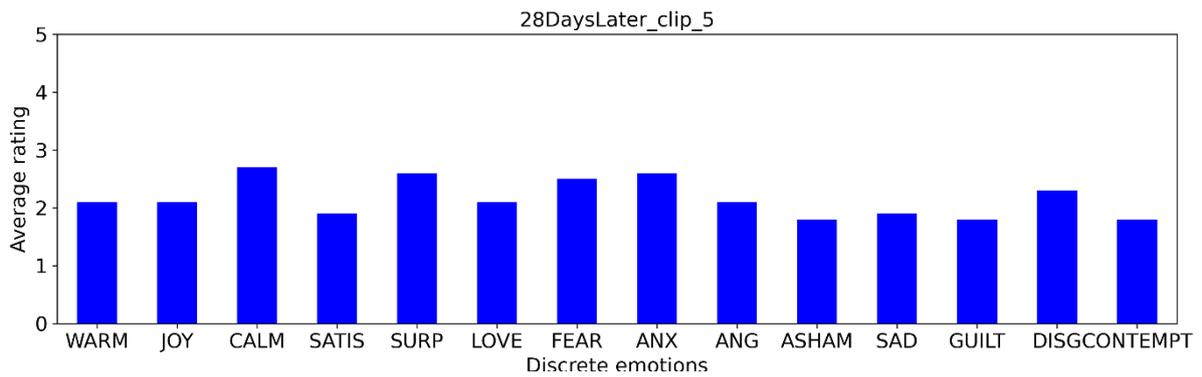

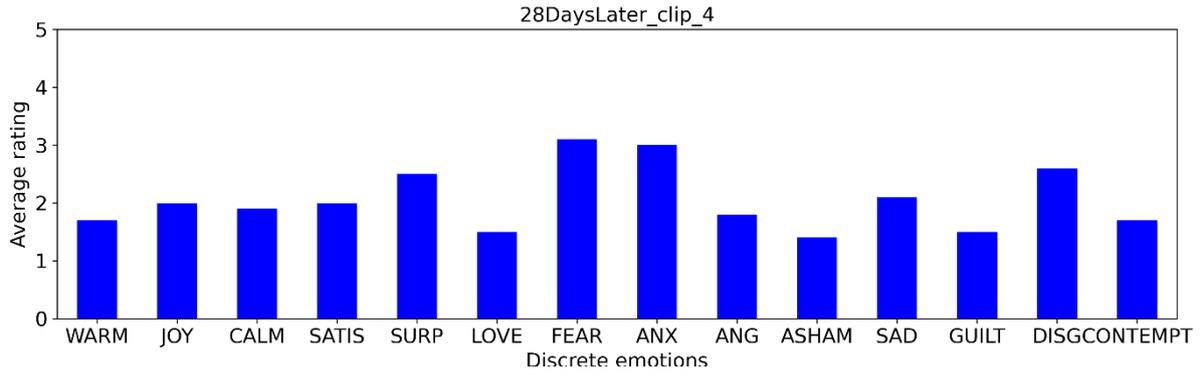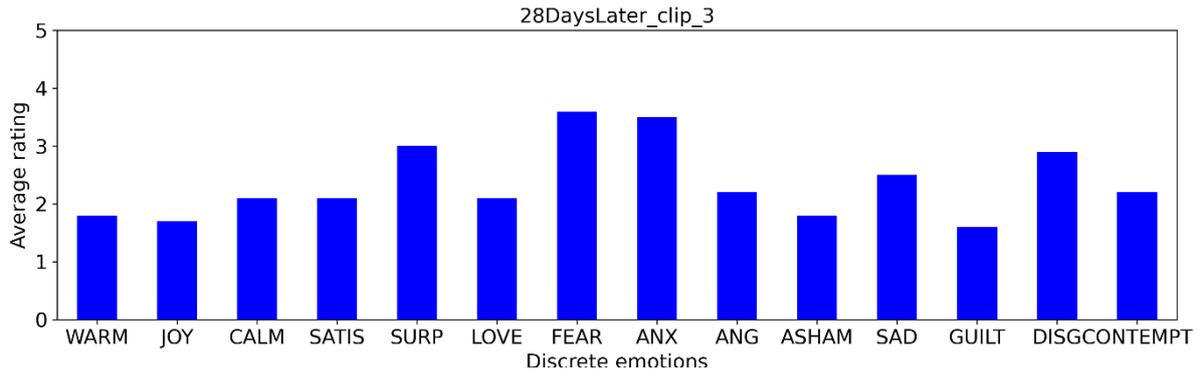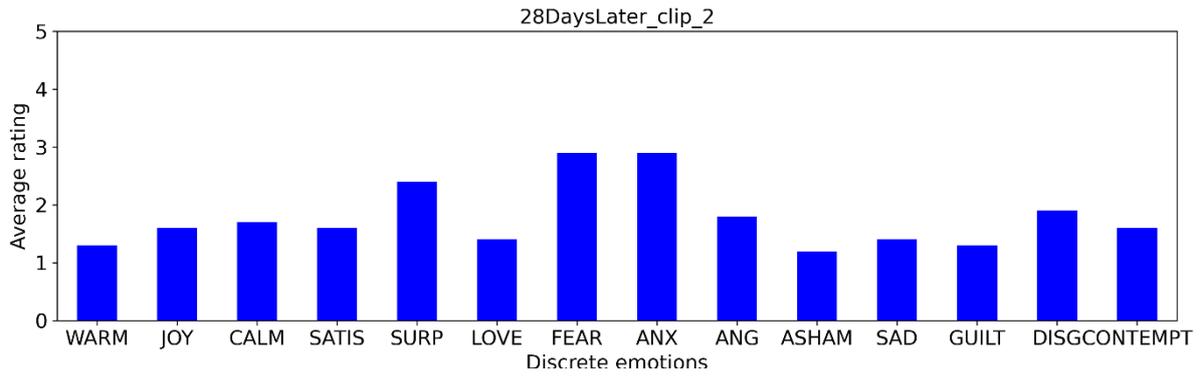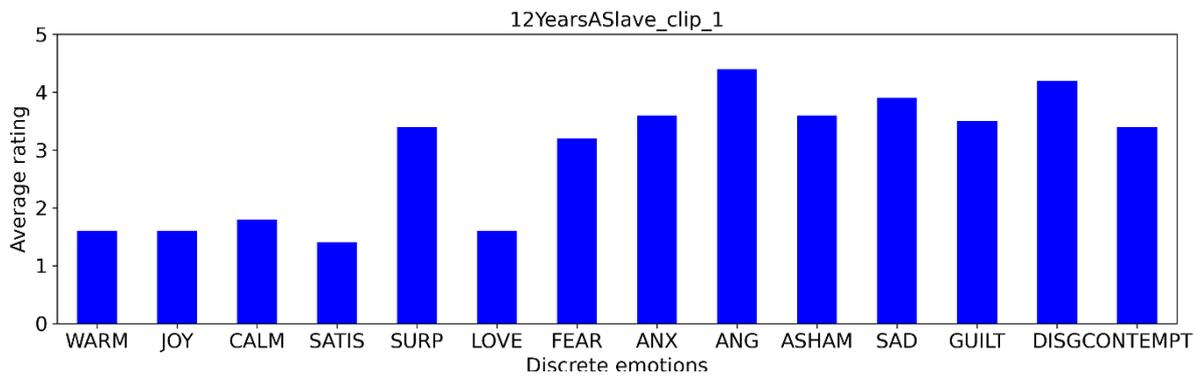